\documentclass[notitlepage,12pt]{gvbarticle}
\usepackage{amsmath}
\usepackage{times}
\usepackage{mhequ}			 
\usepackage{epsfig}
\usepackage{caption2}

\newtheorem{theorem}{Theorem}[section]
\newtheorem{condition}[theorem]{Condition}
\newtheorem{corollary}[theorem]{Corollary}
\newtheorem{definition}[theorem]{Definition}
\newtheorem{lemma}[theorem]{Lemma}
\newtheorem{proposition}[theorem]{Proposition}
\newtheorem{remark}[theorem]{Remark}

\newenvironment{proof}[1][Proof]{\textit{#1.} }{\ \rule{0.5em}{0.5em}\par}

\let\oldstackrel\stackrel
\renewcommand\stackrel[2]{
\oldstackrel{#1}{\scriptscriptstyle#2}
}

\newcommand{\class}{{\cal C}(K,L,\alpha,\hat{\epsilon},\hat{\epsilon}_0,c_{s_0},c_{\eta_0},\sigma)}
\newcommand{\supt}{{\displaystyle\sup_{t\geq0}}}
\newcommand{\bsp}{\hspace{-1mm}}

\newcommand{\tvert}{\vert\!\vert\!\vert}
\newcommand{\tvertb}{\Big\vert\!\Big\vert\!\Big\vert}
\newcommand{\Gepsilon}{G}
\newcommand{\Lepsilon }{{\cal L}_{\mu}}
\newcommand{\Lepsilonk}{{\cal L}_{\mu}(k)}
\newcommand{\Lepsilond}{{\cal L}_{\mu}(\delta)}

\newcommand{\Cinfty}{C}
\newcommand{\Cm}{C_m}
\newcommand{\balpha}{b(\alpha)}

\renewcommand{\epsilon}{\varepsilon}

\def\L{{\rm L}}

\newcommand{\ed}{e}

\newcommand{\essup}{\displaystyle\mathop{\rm ess~sup}}

\newcommand{\biblio}{

}

\makeatletter
\@addtoreset{equation}{section}
\def\theequation{\thesection.\the\c@equation}
\def\newappendix#1{%
        \let\@oldform\@seccntformat%
        \def\@seccntformat##1{Appendix~\csname the##1\endcsname:~}%
        \section{#1}%
        \let\@seccntformat\@oldform%
        }
\makeatother

\begin{document}

\title{Phase turbulence in the Complex Ginzburg--Landau equation
via Kuramoto--Sivashinsky phase dynamics.}
\author{Guillaume van Baalen\thanks{Supported in part by the Fonds National Suisse.}}
\institute{
D\'epartement de Physique Th\'eorique\\
Universit\'e de Gen\`eve\\
Switzerland\\
\email{guillaume.vanbaalen@physics.unige.ch}}

\maketitle
\tableofcontents

\abstract{We study the Complex Ginzburg--Landau initial
value problem
\begin{equs}   
\partial_t u=(1+i\alpha)~\partial_x^2 u + u - (1+i\beta)~u~|u|^2~,~~~
u(x,0)=u_0(x)~,
\label{eqn:cglabs}
\tag{CGL}
\end{equs}
for a complex field $u\in{\bf C}$, with $\alpha,\beta\in{\bf R}$. We
consider the Benjamin--Feir linear instability region
$1+\alpha\beta=-\epsilon^2$ with $\epsilon\ll1$ and $\alpha^2<1/2$. We show
that for all $\epsilon\leq{\cal O}(\sqrt{1-2\alpha^2}~L_0^{-32/37})$, and for all initial
data $u_0$ sufficiently close to $1$ (up to a global phase factor
$\ed^{i~\phi_0},~\phi_0\in{\bf R}$) in the
appropriate space, there exists a unique (spatially) periodic solution of
space period $L_0$. These solutions are small {\em even} perturbations of the
traveling wave solution,
$u=(1+\alpha^2~s)~\ed^{i~\phi_0-i\beta~t}~\ed^{i\alpha~\eta}$, and $s,\eta$
have bounded norms in various $\L^p$ and Sobolev spaces. We prove
that $s\approx-\frac{1}{2}~\eta''$ apart from ${\cal O}(\epsilon^2)$ corrections
whenever the initial data satisfy this condition, and that in the linear
instability range $L_0^{-1}\leq\epsilon\leq{\cal O}(L_0^{-32/37})$, the
dynamics is essentially determined by the motion of the phase alone, and so
exhibits `phase turbulence'. Indeed, we prove that the phase $\eta$
satisfies the Kuramoto--Sivashinsky equation
\begin{equs}
\partial_t\eta=
-\bigl({\textstyle\frac{1+\alpha^2}{2}}\bigr)~\triangle^2\eta
-\epsilon^2\triangle\eta
-{(1+\alpha^2)}~(\eta')^2
\label{eqn:KSabs}
\tag{KS}
\end{equs}
for times $t_0\leq{\cal O}(\epsilon^{-52/5}~L_0^{-32/5})$,
while the amplitude $1+\alpha^2~s$ is essentially constant.}

\section{Introduction}
\subsection{Generalities about the Ginzburg--Landau equation}\label{sec:facts}

The Complex Ginzburg--Landau equation (\ref{eqn:cglabs})
admits explicit traveling wave solutions of the form
\begin{equs}   
u(x,t)=c(p)~\exp\left(i(\phi_0+p~x-\omega(p)~t)\right)~,
\label{eqn:traveling}
\end{equs}
with $\phi_0\in{\bf R}$, $p\in[-1,1]$, $c(p)=\sqrt{1-p^2}$ and
$\omega(p)=\alpha~p^2+\beta~(1-p^2)$. For all $\alpha,\beta$ with
$1+\alpha~\beta>0$, there exists a parameter $p_E=p_{E}(\alpha,\beta)$, with
$p_{E}\to0$ as $1+\alpha~\beta\to0^{+}$ such that traveling wave solutions
(\ref{eqn:traveling}) with $|p|\geq p_{E}(\alpha,\beta)$ are linearly
unstable, a phenomenon called `sideband' or `Eckhaus' instability, while
those with $|p|\leq p_E$ are linearly stable (see e.g. \cite{Cross} and
the references therein). When $1+\alpha\beta<0$, all traveling wave
solutions are linearly unstable, a phenomenon called `Benjamin--Feir' or
`Benjamin--Feir--Newell' instability (see e.g. \cite{Benjamin} and
\cite{Newell}).

In this paper, we consider the case $1+\alpha~\beta=-\epsilon^2$. When
$\epsilon$ is small enough, numerical simulations on finite domains (see
e.g. \cite{Montagne} and the references therein) indicate that the dynamics
of the phase is turbulent, the phase evolving irregularly, (with
fluctuations of order $\epsilon^2$ around the global phase $\phi_0$), while
the amplitude of $u$ is constant up to ${\cal O}(\epsilon^4)$ corrections.
This type of behavior is called `phase turbulence'. The persistence of
phase turbulence on infinite domains is not known, while its existence on
finite domains is, to our knowledge, not proven rigorously.

As $\epsilon$ increases (or the domain is larger), `amplitude' or
`defect' turbulence occurs, the amplitude of $u$ vanishing at some instants
and places, called `defects' or `phase slips' (see also \cite{Eckmannslip}). 
Note that `phase' and `amplitude' turbulence may coexist at the same time
in the $\alpha,\beta$ parameter space, depending on initial conditions,
in which case one speaks of `bichaos'.

\begin{figure}[t]
\begin{center}  
\begin{picture}(0,0)%
\epsfig{file=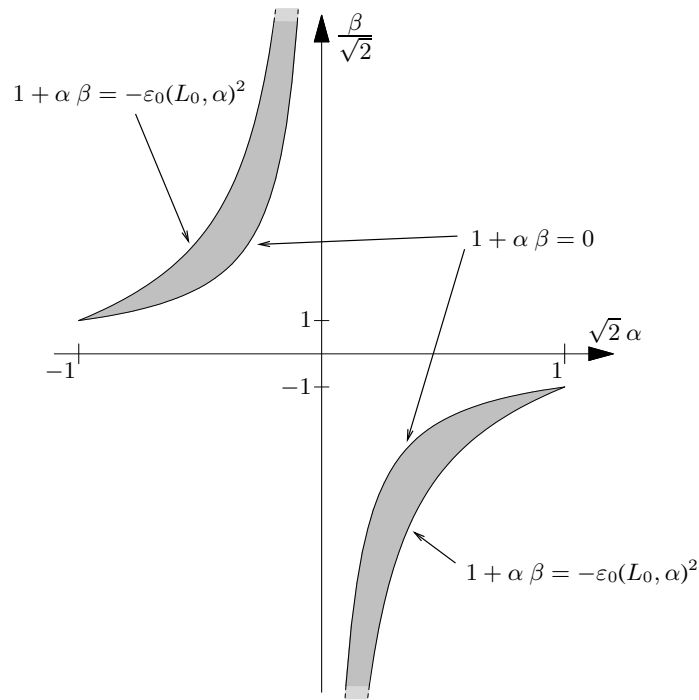}%
\end{picture}%
\setlength{\unitlength}{3947sp}%
\begingroup\makeatletter\ifx\SetFigFont\undefined%
\gdef\SetFigFont#1#2#3#4#5{%
  \reset@font\fontsize{#1}{#2pt}%
  \fontfamily{#3}\fontseries{#4}\fontshape{#5}%
  \selectfont}%
\fi\endgroup%
\begin{picture}(3787,4374)(5901,-3845)
\put(8776,-961){\makebox(0,0)[lb]{\smash{\SetFigFont{9}{10.8}{\familydefault}{\mddefault}{\updefault}$1+\alpha~\beta=0$}}}
\put(7702,-1469){\makebox(0,0)[lb]{\smash{\SetFigFont{9}{10.8}{\familydefault}{\mddefault}{\updefault}$1$}}}
\put(7581,-1895){\makebox(0,0)[lb]{\smash{\SetFigFont{9}{10.8}{\familydefault}{\mddefault}{\updefault}$-1$}}}
\put(6101,-1791){\makebox(0,0)[lb]{\smash{\SetFigFont{9}{10.8}{\familydefault}{\mddefault}{\updefault}$-1$}}}
\put(9281,-1791){\makebox(0,0)[lb]{\smash{\SetFigFont{9}{10.8}{\familydefault}{\mddefault}{\updefault}$1$}}}
\put(8744,-3061){\makebox(0,0)[lb]{\smash{\SetFigFont{9}{10.8}{\familydefault}{\mddefault}{\updefault}$1+\alpha~\beta=-\epsilon_0(L_0,\alpha)^2$}}}
\put(5901,-51){\makebox(0,0)[lb]{\smash{\SetFigFont{9}{10.8}{\familydefault}{\mddefault}{\updefault}$1+\alpha~\beta=-\epsilon_0(L_0,\alpha)^2$}}}
\put(9510,-1552){\makebox(0,0)[lb]{\smash{\SetFigFont{9}{10.8}{\familydefault}{\mddefault}{\updefault}$\sqrt{2}~\alpha$}}}
\put(7929,320){\makebox(0,0)[lb]{\smash{\SetFigFont{9}{10.8}{\familydefault}{\mddefault}{\updefault}$\frac{\textstyle\beta}{\textstyle\sqrt{2}}$}}}
\end{picture}
\setcaptionwidth{145mm}
\caption{
\small Parameter space for (\ref{eqn:cglabs}). Linear instability occurs 
for $1+\alpha~\beta<0$, and phase turbulence is shown in this paper
to occur in shaded region.}\label{fig:paramspace}
\end{center}
\end{figure}

The `amplitude' turbulence regime is technically difficult because the
phase is not well defined when the amplitude vanishes. In this paper, we
concentrate on the easier phase turbulence regime and prove that for the
particular case\footnote{The case $p\neq0$ should give a similar result but
is more challenging.} $p=0$, phase turbulence occurs for small initial
perturbations of the traveling wave $\ed^{i~\phi_0-i\beta~t}$ on domains of size
$L_0$ for all $\alpha^2<1/2$ and for all $\epsilon\leq\epsilon_0(L_0,\alpha)$ 
with $\epsilon_0(L_0,\alpha)\to0$ as $L_0\to\infty$ or $\alpha^2\to1/2$,
see figure \ref{fig:paramspace}. We restrict ourselves to {\em even}
perturbations for concision, though general perturbations could be treated
as well (see Remark \ref{rem:noteven} below). We believe the restriction
$\alpha^2<1/2$ to be an artifact of our technical treatment (see the
discussion at the end of Section \ref{sec:ampliphase}), though we expect
some restriction on the size of $\alpha$ to be necessary anyway, because
the large $\alpha$ limit of (\ref{eqn:cglabs}) is the so--called
Non--linear Schr\"odinger equation, whose dynamics is completely different
from the above picture.

\subsection{Setting}
We consider perturbations of $\ed^{i~\phi_0-i\beta~t}$, (this is a solution
of (\ref{eqn:cglabs})) which are of the form\footnote{The
$\alpha$ factors in front of $s$ and $\eta$ are only a convenient
normalization.}
\begin{equs}   
u(x,t)&=(1+\alpha^2~s(x,t))~\ed^{i~\phi_0-i\beta t}~\ed^{i\alpha~\eta(x,t)}~,
\label{eqn:ansatz}
\end{equs}
for (small) $s,\eta\in{\bf R}$. To state our results, we introduce the
following scalings\footnote{They will be justified in the next subsection.}
\begin{equs}
\eta(x,t)&=
{\scriptstyle\frac{1}{4}}
~\hat{\epsilon}^2~\hat{\eta}(\hat{x},\hat{t})~,
\label{eqn:firstcveta}\\
s(x,t)&=
~\hat{\epsilon}^4~\hat{s}(\hat{x},\hat{t})~,
\label{eqn:firstcvsintro}
\end{equs}
with
$\chi=\frac{4}{1+\alpha^2}$, 
$\hat{\epsilon}=\sqrt{\frac{\chi}{2}}~\epsilon$,
$\hat{x}=\hat{\epsilon}~x$ and
$\hat{t}=\frac{2}{\chi}~\hat{\epsilon}^4~t$.

We consider the initial value problem (\ref{eqn:cglabs}) with
$\eta(x,0)=\eta_0(x)$ and $s(x,0)=s_0(x)$, where $\eta_0$ and $s_0$ are
{\em even} periodic functions of period $L_0$, or equivalently, in terms of
the `hat' variables, $\hat{\eta}_0$ and $\hat{s}_0$ are {\em even} periodic
functions of period $L=\hat{\epsilon}~L_0$. To state our conditions on the initial data
$\hat{s}_0$ and $\hat{\eta}_0$, we introduce the Banach space ${\cal
W}_{0,\sigma}$ obtained by completing ${\cal C}^{\infty}_{\rm
per}([-L/2,L/2],{\bf R})$ under the norm
$\|\cdot\|_{\sigma}=\|\cdot\|_{\L^2([-L/2,L/2)}+\|\cdot\|_{{\cal
W},\sigma}$, where $\|\cdot\|_{{\cal W},\sigma}$ is a sup norm with
algebraic weight (going like $|k|^{\sigma}$ at infinity) on the Fourier
transform, see section \ref{sec:highfreq} for details. Essentially ${\cal
W}_{0,\sigma}$ consists of functions in $\L^2([-L/2,L/2])$, whose Fourier
transform decays (at least) like $|k|^{-\sigma}$ as $|k|\to\infty$ (this is
a regularity assumption). In the sequel we will often use the shorthand
notation $\L^2$ instead of $\L^2([-L/2,L/2])$, while we will always write
$\L^2([-L_0/2,L_0/2])$ to avoid confusion.
\begin{definition}\label{cond:thecondition}
We say that $\hat{\eta}_0$ and $\hat{s}_0$ are in the class
$\class\subset{\cal W}_{0,\sigma}\times{\cal W}_{0,\sigma-1}$ if 
\begin{equs}
\hat{\eta}_0(0)=0~,~~~\|\hat{\eta}_0'\|_{\sigma}\leq c_{\eta_0}~\rho~,~~~
\Bigl\|\hat{s}_0-\frac{\hat{\epsilon}^2~\hat{s}_0''}{2}\Bigr\|_{\sigma-1}&\leq c_{s_0}~\rho^3~,
\label{eqn:condiun}
\end{equs}
for $\rho=K~L^{8/5}$, and if
\begin{equs}
\left\|\hat{s}_0-\frac{\hat{\epsilon}^2~\hat{s}_0''}{2}
+\frac{\hat{\eta}_0''}{8}
+\frac{\hat{\epsilon}^2~(\hat{\eta}_0')^2}{32}
\right\|_{\sigma-1}
&<
2^{-8}~\min\Bigl(\frac{1}{3},\frac{1-2~\alpha^2}{1-\alpha^2}\Bigr)
\Bigl(\frac{\hat{\epsilon}}{\hat{\epsilon}_0}\Bigr)^2~\hat{\epsilon}^2~c_{\eta_0}~\rho~.
\label{eqn:condideux}
\end{equs}
\end{definition}
The parameter $L$ is the (space) period (in the scaled variables) of the
solution. The constant $K$ is essentially the same as that of \cite{Collet}
in their discussion of the Kuramoto--Sivashinsky equation
\begin{equs}
\partial_{\hat{t}}\hat{\eta}_c=-\triangle^2\hat{\eta}_c-\triangle\hat{\eta}_c
-\frac{1}{2}(\hat{\eta}_c')^2~,\label{eqn:KStt}
\end{equs}
where it appears in the bound
${\displaystyle\lim_{t\to\infty}}\|\hat{\eta}_c'(\cdot,t)\|_{\L^2}\leq
K~L^{8/5}$ for symmetric periodic solutions. Therefore, $K$ is independent
of $\alpha,\epsilon$ and $L$. The parameters $\alpha$ and $\hat{\epsilon}$
are those of (\ref{eqn:cglabs}), with
$\hat{\epsilon}^2=-2~\frac{1+\alpha~\beta}{1+\alpha^2}$, while
$\hat{\epsilon}_0$ is the maximal value of $\hat{\epsilon}$ for which our
results hold. The parameters $c_{\eta_0}$ and $c_{s_0}$ measure the size of
the initial perturbation. Note that only $\hat{\eta}_0'$ and
$\hat{\eta}_0(0)$ appear in the conditions. We can motivate this by noting
that (\ref{eqn:cglabs}) has a $U(1)$ symmetry (the global phase factor
$\ed^{i~\phi_0}$). Expressing all constraints in terms of $\hat{\eta}_0'$
and $\hat{\eta}_0(0)$ is a convenient way to take this invariance into
account. The condition $\eta_0(0)=0$ can always be satisfied, up to a
redefinition of the global phase $\phi_0$. Furthermore, this condition is
preserved by the evolution (see e.g. (\ref{eqn:mudef})). We will prove that
if $\hat{\eta}_0$ and $\hat{s}_0$ are in the class $\class$, the
(\ref{eqn:cglabs}) dynamics (which has a complex function as initial
condition) is increasingly well approximated as $\hat{\epsilon}\to0$ by the
Kuramoto--Sivashinsky dynamics (\ref{eqn:KStt}), which has a real function
as initial condition. For this to hold, $\hat{s}_0$ and $\hat{\eta}_0$ have
to be tightly related as $\hat{\epsilon}\to0$. This relation is quantified
by (\ref{eqn:condideux}), which says that, up to ${\cal O}(\epsilon^4)$
corrections, $\hat{s}_0$ and $\hat{\eta}_0$ are related by
\begin{equs}
\hat{s}_0=-\frac{1}{8}~\hat{G}~\hat{\eta}_0''
-\frac{\hat{\epsilon}^2}{32}~\hat{G}~(\hat{\eta}_0')^2~,
\end{equs}
where $\hat{G}$ is the operator with symbol
\begin{equs}
\hat{G}(k)=\frac{1}{1+\frac{\hat{\epsilon}^2}{2}~k^2}~.
\label{eqn:gedef}
\end{equs}
Note that $\hat{G}$ is the inverse of the (positive) operator
$1-\frac{\hat{\epsilon}^2}{2}~\partial_{\hat{x}}^2$.

\subsection{Main results and their physical discussion}
Our main results are twofold. We first have an existence and unicity
result for the solutions of (\ref{eqn:cglabs}), see Theorem
\ref{thm:existeunique} below, and then an approximation result in Theorem
\ref{thm:properties}. 

From now on, we will denote generic constants by the letters $C$ and $c$.
We will use the letter $c$ with different labels to recall the quantity on
which the bound is. By constants, we mean quantities which do not depend on
$\alpha,\hat{\epsilon},L$ and $\sigma$ in the ranges
\begin{equs}
0\leq\hat{\epsilon}\leq1~,~~\alpha^2<1/2~,~~L>2\pi~~~\mbox{and}~\sigma\leq\sigma_0
\end{equs}
for some finite $\sigma_0>\frac{11}{2}$.

\begin{theorem}\label{thm:existeunique}
Let $\alpha^2<1/2$, $\sigma>\frac{11}{2}$, $c_{s_0}>0$, $c_{\eta}>0$ and
$L>2~\pi$. There exist constants $K$ and $c_{\epsilon}$ such that for all
$m_{\epsilon}\geq4$, for any $\hat{\epsilon}\leq\hat{\epsilon}_0=
c_{\epsilon}~\sqrt{1-2\alpha^2}~\rho^{-m_{\epsilon}}$ and for all
$\hat{\eta}_0$ and $\hat{s}_0$ in the class $\class$,
the solution of (\ref{eqn:cglabs}) with parameters $\alpha$ and
$\beta=-\frac{2+(1+\alpha^2)~\hat{\epsilon}^2}{2~\alpha}$ 
exists for all times, is of the form (\ref{eqn:ansatz}) and satisfies
\begin{equs}
\sup_{\hat{t}\geq 0}
\|\hat{\eta}(\cdot,\hat{t})'\|_{\sigma}\leq c_{\eta}~\rho~,~~~
\sup_{\hat{t}\geq 0}
\|\hat{s}(\cdot,\hat{t})\|_{\sigma-1}\leq c_{s}~\rho^3,
\label{eqn:boundetasthm}\\
\sup_{\hat{t}\geq0}
\left\|\hat{s}(\cdot,\hat{t})
+\frac{1}{8}~\hat{G}~\hat{\eta}''(\cdot,\hat{t})
+\frac{\hat{\epsilon}^2}{32}~\hat{G}~(\hat{\eta}')^2(\cdot,\hat{t})
\right\|_{\L^2}\leq
\Bigl(\frac{\hat{\epsilon}}{\hat{\epsilon}_0}\Bigr)^2~c_{\eta}~\rho~,
\label{eqn:slaving}
\end{equs}
with $\rho=K~L^{8/5}$, $c_{\eta}>1+c_{\eta_0}$ and $c_{s}>c_{s_0}$. This
solution is unique among functions satisfying (\ref{eqn:boundetasthm}).
\end{theorem}
Our results are valid for any
$\hat{\epsilon}\leq\hat{\epsilon}_0=c_{\epsilon}~\sqrt{1-2\alpha^2}~\rho^{-
4}$ and for any $L>2~\pi$. Since $L=\hat{\epsilon}~L_0$ and
$\rho=K~L^{8/5}$, we see that the applicability range is
\begin{equs}
C~L_0^{-1}\leq\epsilon\leq C'\sqrt{1-2\alpha^2}~L_0^{-32/37}~.
\end{equs}
The lower bound is the linear instability condition. 

\begin{figure}   
\begin{center}
\begin{picture}(0,0)%
\includegraphics{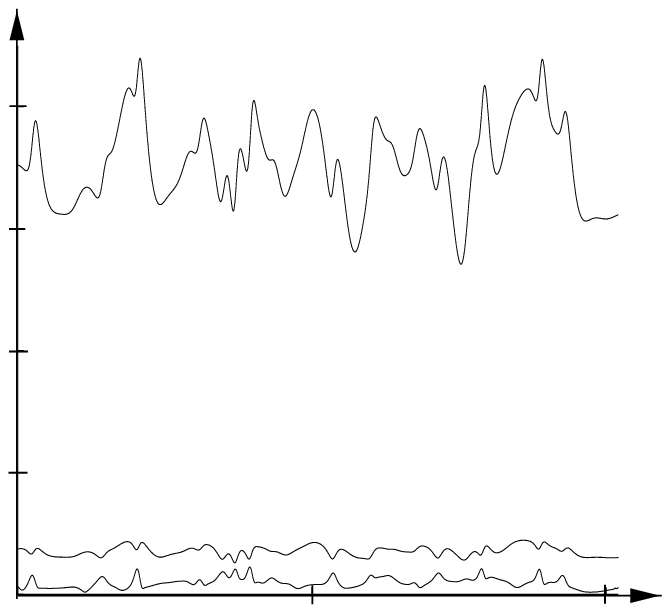}%
\end{picture}%
\setlength{\unitlength}{3947sp}%
\begingroup\makeatletter\ifx\SetFigFontx\undefined
\def\x#1#2#3#4#5#6#7\relax{\def\x{#1#2#3#4#5#6}}%
\expandafter\x\fmtname xxxxxx\relax \def\y{splain}%
\ifx\x\y   
\gdef\SetFigFontx#1#2#3{%
  \ifnum #1<17\tiny\else \ifnum #1<20\small\else
  \ifnum #1<24\normalsize\else \ifnum #1<29\large\else
  \ifnum #1<34\Large\else \ifnum #1<41\LARGE\else
     \huge\fi\fi\fi\fi\fi\fi
  \csname #3\endcsname}%
\else
\gdef\SetFigFontx#1#2#3{\begingroup
  \count@#1\relax \ifnum 25<\count@\count@25\fi
  \def\x{\endgroup\@setsize\SetFigFontx{#2pt}}%
  \expandafter\x
    \csname \romannumeral\the\count@ pt\expandafter\endcsname
    \csname @\romannumeral\the\count@ pt\endcsname
  \csname #3\endcsname}%
\fi
\fi\endgroup
\begin{picture}(3242,2987)(12331,-10309)
\put(12336,-7846){\makebox(0,0)[lb]{\smash{\SetFigFontx{8}{9.6}{rm}$4$}}}
\put(14664,-9757){\makebox(0,0)[lb]{\smash{\SetFigFontx{12}{14.4}{rm}$\|\hat{s}(\cdot,\hat{t})\|_{\L^2}$}}}
\put(12334,-8436){\makebox(0,0)[lb]{\smash{\SetFigFontx{8}{9.6}{rm}$3$}}}
\put(12336,-9026){\makebox(0,0)[lb]{\smash{\SetFigFontx{8}{9.6}{rm}$2$}}}
\put(12331,-10203){\makebox(0,0)[lb]{\smash{\SetFigFontx{8}{9.6}{rm}$0$}}}
\put(13741,-10299){\makebox(0,0)[lb]{\smash{\SetFigFontx{8}{9.6}{rm}$100$}}}
\put(15147,-10296){\makebox(0,0)[lb]{\smash{\SetFigFontx{8}{9.6}{rm}$200$}}}
\put(15447,-10309){\makebox(0,0)[lb]{\smash{\SetFigFontx{8}{9.6}{rm}$\hat{t}$}}}
\put(12332,-9599){\makebox(0,0)[lb]{\smash{\SetFigFontx{8}{9.6}{rm}$1$}}}
\put(14666,-8709){\makebox(0,0)[lb]{\smash{\SetFigFontx{12}{14.4}{rm}$\|\hat{\eta}'(\cdot,\hat{t})\|_{\L^2}$}}}
\end{picture}  
\begin{picture}(0,0)%
\includegraphics{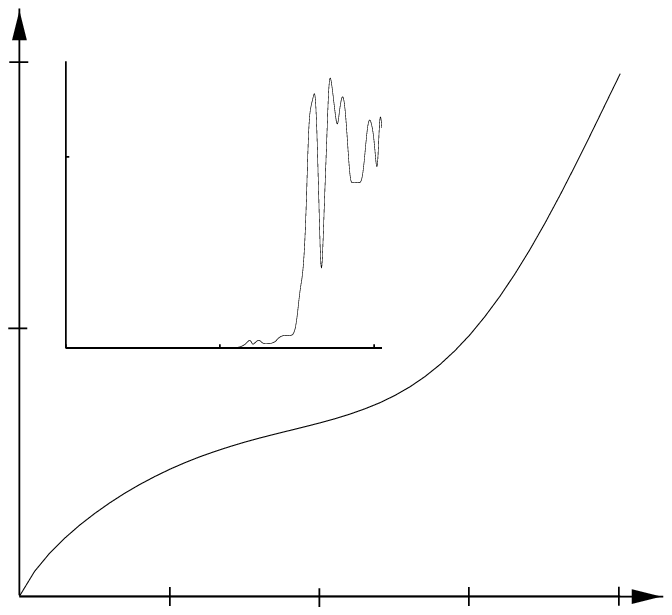}%
\end{picture}%
\setlength{\unitlength}{3947sp}%
\begingroup\makeatletter\ifx\SetFigFonta\undefined
\def\x#1#2#3#4#5#6#7\relax{\def\x{#1#2#3#4#5#6}}%
\expandafter\x\fmtname xxxxxx\relax \def\y{splain}%
\ifx\x\y   
\gdef\SetFigFonta#1#2#3{%
  \ifnum #1<17\tiny\else \ifnum #1<20\small\else
  \ifnum #1<24\normalsize\else \ifnum #1<29\large\else
  \ifnum #1<34\Large\else \ifnum #1<41\LARGE\else
     \huge\fi\fi\fi\fi\fi\fi
  \csname #3\endcsname}%
\else
\gdef\SetFigFonta#1#2#3{\begingroup
  \count@#1\relax \ifnum 25<\count@\count@25\fi
  \def\x{\endgroup\@setsize\SetFigFonta{#2pt}}%
  \expandafter\x
    \csname \romannumeral\the\count@ pt\expandafter\endcsname
    \csname @\romannumeral\the\count@ pt\endcsname
  \csname #3\endcsname}%
\fi
\fi\endgroup
\begin{picture}(3360,2991)(8821,-10348)
\put(11926,-10331){\makebox(0,0)[lb]{\smash{\SetFigFonta{8}{9.6}{rm}$4$}}}
\put(9201,-9046){\makebox(0,0)[lb]{\smash{\SetFigFonta{8}{9.6}{rm}$0$}}}
\put(8821,-7656){\makebox(0,0)[lb]{\smash{\SetFigFonta{8}{9.6}{rm}$\hat{\epsilon}^2$}}}
\put(9931,-9109){\makebox(0,0)[lb]{\smash{\SetFigFonta{8}{9.6}{rm}$100$}}}
\put(10681,-9109){\makebox(0,0)[lb]{\smash{\SetFigFonta{8}{9.6}{rm}$200$}}}
\put(9181,-8104){\makebox(0,0)[lb]{\smash{\SetFigFonta{8}{9.6}{rm}$1$}}}
\put(8956,-10201){\makebox(0,0)[lb]{\smash{\SetFigFonta{8}{9.6}{rm}$0$}}}
\put(9771,-10336){\makebox(0,0)[lb]{\smash{\SetFigFonta{8}{9.6}{rm}$1$}}}
\put(10486,-10336){\makebox(0,0)[lb]{\smash{\SetFigFonta{8}{9.6}{rm}$2$}}}
\put(11206,-10331){\makebox(0,0)[lb]{\smash{\SetFigFonta{8}{9.6}{rm}$3$}}}
\put(12057,-10348){\makebox(0,0)[lb]{\smash{\SetFigFonta{8}{9.6}{rm}$\hat{t}$}}}
\put(10018,-9838){\makebox(0,0)[lb]{\smash{\SetFigFonta{14}{16.8}{rm}$\frac{\|\hat{\eta}'(\cdot,\hat{t})-\hat{\eta}_c'(\cdot,\hat{t})\|_{\L^2}}{\|\hat{\eta}'(\cdot,\hat{t})\|_{\L^2}}$}}}
\end{picture}
\caption{
\small Numerical results for $\hat{\epsilon}=10^{-3}$, $\alpha=10^{-2}$ and
$L_0=10^{4}\cdot2\pi$.	     
	 }\label{fig:numerics}
\end{center}
\end{figure}

In terms of the original variables, Theorem \ref{thm:existeunique} shows
that solutions of (\ref{eqn:cglabs}) of the form (\ref{eqn:ansatz}) exist,
and that (see Appendix \ref{app:discussion} for details)
\begin{equs}
\sup_{t\geq 0}
\|\eta(\cdot,t)'\|_{\L^2([-L_0/2,L_0/2])}&\leq C~\epsilon^{5/2-1/m_{\epsilon}}
\label{eqn:boundetaphys}
~,\\
\sup_{t\geq 0}
\|s(\cdot,t)\|_{\L^2([-L_0/2,L_0/2])}&\leq
C~\epsilon^{7/2-3/m_{\epsilon}}~,\label{eqn:boundsphys}\\
\sup_{t\geq 0}~
\sup_{x\in[-L_0/2,L_0/2]}~
|\eta(x,t)|&\leq C~\epsilon^{2-13/(8~m_{\epsilon})}
\label{eqn:boundetaphysinf}
~,\\
\sup_{t\geq 0}~
\sup_{x\in[-L_0/2,L_0/2]}~
|s(x,t)|&\leq C~\epsilon^{4-4/m_{\epsilon}}
\label{eqn:boundsphysinf}~.
\end{equs}
The inequalities (\ref{eqn:boundetaphysinf}) and (\ref{eqn:boundsphysinf})
quantify the `physical intuition' $\eta={\cal O}(\epsilon^2)$ and
$s={\cal O}(\epsilon^4)$, see section \ref{sec:facts}.

Inequalities (\ref{eqn:boundetasthm}) or
(\ref{eqn:boundetaphys})--(\ref{eqn:boundsphysinf}) also show that the
solutions belongs to a (local) attractor, while (\ref{eqn:slaving}) shows
that on that attractor, the `amplitude' $s$ satisfies
$s=-\frac{1}{2}~\eta''+{\cal O}(\epsilon^2)$. The attractor is thus well
approximated by the graph $s=-\frac{1}{2}~\eta''$ in the $s,\eta$ space. 
This result was discovered at a heuristic level by Kuramoto and Tsuzuki in
\cite{Kur76}.

We do not expect the bounds (\ref{eqn:boundetasthm}) and
(\ref{eqn:slaving}) to be optimal. Numerical simulations show that
$\hat{\eta}'$ and $\hat{s}$ are uniformly bounded in space and time, at
least for a large range of $L=\epsilon~L_0$. This suggests that
$\|\hat{\eta}'\|_{\L^2}$ and $\|\hat{s}\|_{\L^2}$ should both scale with
$L$ like $\sqrt{L}$ and not like $L^{8/5}$ and $L^{24/5}$, hence
we should have $\rho\sim\sqrt{L}$. In the left panel of Figure
\ref{fig:numerics}, we display as a function of
$\hat{t}\in[0,200]$ (by decreasing size) the typical behavior of
$\|\hat{\eta}'(\cdot,\hat{t})\|_{\L^2}$,
$\|\hat{s}(\cdot,\hat{t})\|_{\L^2}$ and
$\hat{\epsilon}^{-4}\|\hat{s}(\cdot,\hat{t})+\frac{1}{8}~\hat{G}~\hat{\eta}''(\cdot,\hat{t})
+\frac{\hat{\epsilon}^2}{32}~\hat{G}~(\hat{\eta}')^2(\cdot,\hat{t})\|_{\L^2}$
in units proportional to $\sqrt{\epsilon L_0}$. We also see 
that $\hat{s}+\frac{1}{8}~\hat{G}~\hat{\eta}''
+\frac{\hat{\epsilon}^2}{32}~\hat{G}~(\hat{\eta}')^2$ is of order
$\hat{\epsilon}^4$ and not $\hat{\epsilon}^2$ as in (\ref{eqn:slaving}).

We now show that the dynamics of the phase on the attractor is well
approximated by the Kuramoto--Sivashinsky equation.
\begin{theorem}\label{thm:properties}
Under the assumptions of Theorem \ref{thm:existeunique}, there exists a constant $c_t$ 
such that if $\hat{t}_1\leq c_t~\rho^{-4}$, then for all $\hat{t}_0\geq0$,
\begin{equs}
\sup_{0\leq\hat{t}\leq\hat{t}_1}
\|\hat{\eta}(\cdot,\hat{t}_0+\hat{t})'-\hat{\eta}_c(\cdot,\hat{t})'\|_{\L^2}\leq 
\Bigl(\frac{\hat{\epsilon}}{\hat{\epsilon}_1}\Bigr)~c_{\eta}~\rho~,
\label{eqn:kuramapprox}
\end{equs}
where $\hat{\eta}_c$ satisfies the Kuramoto--Sivashinsky equation
(\ref{eqn:KStt}) with initial condition $\hat{\eta}_c(\hat{x},0)=\hat{\eta}(\hat{x},\hat{t}_0)$.
\end{theorem}
In physical terms, Theorem \ref{thm:properties} says that on each time
interval $[t_0,t_0+t_1]$, the distance between $\eta$ and the solution of
the Kuramoto--Sivashinsky equation with initial condition $\eta(t_0)$ is
small compared to the size of the attractor (see (\ref{eqn:kuramapprox})), at
least for time intervals of length $t_1$ of order
$\epsilon^{-4}~\rho^{-4}=\epsilon^{-52/5}~L_0^{-32/5}$.
This result gives a rigorous foundation to the heuristic derivation in
\cite{Kur76}) of the Kuramoto--Sivashinsky equation as a phase equation
for the Complex Ginzburg--Landau equation near the Benjamin--Feir line
(see also \cite{Manneville}). Furthermore, if $\epsilon$ is sufficiently
small, the amplitude $1+\alpha^2~s$ does not vanish by
(\ref{eqn:boundsphysinf}). This proves that the solution exhibits phase
turbulence for all times, the solutions of the Kuramoto--Sivashinsky
equation being believed to be chaotic.

The bound (\ref{eqn:kuramapprox})  for $\hat{t}_1\leq c_t~\rho^{-4}$ is
again certainly not optimal. Numerical simulations show that 
$\hat{t}_1$ scales like $L^{-2}$ (this is in agreement with
$\rho\sim \sqrt{L}$) with a bound of order $\hat{\epsilon}^2$ and not
$\hat{\epsilon}$ as in (\ref{eqn:kuramapprox}). In the right panel
of Figure \ref{fig:numerics}, we show in the large plot
$\frac{\|\hat{\eta}'(\cdot,\hat{t})-\hat{\eta}_c'(\cdot,\hat{t})\|_{\L^2}}{\|\hat{\eta}'(\cdot,\hat{t})\|_{\L^2}}$
in units of $\hat{\epsilon}^2$ for short times (large times are displayed in
small inserted plot in absolute units).

In the remainder of this section, we derive the dynamical equations for 
$\hat{s}$ and $\hat{\eta}$, then we discuss informally these equations to motivate
the analytical treatment that we will present in the next sections. In
particular, we will explain the particular choice of the scalings
(\ref{eqn:firstcveta}) and (\ref{eqn:firstcvsintro}). We will treat the phase 
dynamical equation in section \ref{sec:phase}, while the treatment of the 
dynamical equation for $s$ is postponed to section \ref{sec:ampli}, $s$ 
being `slaved' to $\eta$ by that equation.

\subsection{Derivation of the amplitude and phase equations}\label{sec:ampliphase}
The ansatz (\ref{eqn:ansatz}) leads, after separation of the real and imaginary
parts of equation (\ref{eqn:cglabs}), to
\begin{equs}[3]
\partial_t s&=
 s''-2~s -\eta''-(\eta')^2
&\hspace{1mm}-\hspace{1mm}&
\alpha^2~\left(
3~s^2+2~s'~\eta'+s~\eta''
+s~(\eta')^2+\alpha^2~s^3\right)~,
\label{eqn:fors}
\\
\partial_t\eta&=
 \eta''+\alpha^2~s''-2~\alpha~\beta~s
&\hspace{1mm}-\hspace{1mm}&
\alpha^2~
\left((\eta')^2+\alpha~\beta~s^2
-\frac{2s'~\eta'}{1+\alpha^2~s}+\frac{\alpha^2~s~s''}{1+\alpha^2~s}\right)
\label{eqn:foreta}~.
\end{equs}
Since these equations preserve the subspace of functions that are {\em
even} in the space variable, we restrict ourselves to that particular
case. We also use $\alpha,-\frac{1+\epsilon^2}{\alpha}$ as parameters
instead of $\alpha,\beta$ as it allows to emphasize the dependence on
the small parameter $\epsilon$. Finally, as the right hand sides of
(\ref{eqn:fors}) and (\ref{eqn:foreta}) contain only (space) derivatives of
the function $\eta$, we introduce the odd function $\mu$ (the phase
derivative) by 
\begin{equs}
\eta(x,t)=\int_0^{x}
\hspace{-2mm}
{\rm d}y~\mu(y,t)~,\label{eqn:mudef}
\end{equs}
and obtain
\begin{equs}
\partial_t s&=
 s''-2~s -\mu'-\mu^2\\
&\phantom{=}
-\alpha^2~\left(
3~s^2+2~s'~\mu+s~\mu'
+s~\mu^2+\alpha^2~s^3\right)~,
\label{eqn:forsmu}
\\
\partial_t\mu&=
 \mu''+\alpha^2~s'''+2~(1+\epsilon^2)~s'-\alpha^2~(\mu^2)'\\
&\phantom{=}+
\alpha^2~
\left((1+\epsilon^2)~s^2
+\frac{2s'~\mu}{1+\alpha^2~s}-\frac{\alpha^2~s~s''}{1+\alpha^2~s}\right)'
\label{eqn:formumu}~.
\end{equs}
We expect $\partial_t s,s'\ll s,\mu'\ll\mu\ll1$ when $\epsilon\ll1$.
We then have
\begin{equs}
\partial_t s&=
s''-2~s-\mu'-\mu^2
+f_s(s,\mu)~,
\label{eqn:forsmus}
\\
\partial_t\mu&=
-\Bigl(
s''-2~s-\mu'-\mu^2
\Bigr)'
+(1+\alpha^2)~s'''+2~\epsilon^2~s'
\\
&\phantom{=}
-
2~(1+\alpha^2)~\mu~\mu'+f_{\mu}(s,\mu)'
\label{eqn:formumus}~,
\end{equs}
where $f_s(s,\mu)$, respectively $f_{\mu}(s,\mu)$, is defined as the function
appearing in the second line of (\ref{eqn:forsmu}) resp.
(\ref{eqn:formumu}). The $-2s$ term in (\ref{eqn:forsmus}) strongly damps
$s$, which therefore is `slaved' to $\mu$. Indeed, as we will show in
Section \ref{sec:ampli}, for given $\mu$ satisfying appropriate bounds, the
map $\mu\mapsto s(\mu)$ defined by the (global and strong) solution of
(\ref{eqn:forsmus}) is well defined and Lipschitz in $\mu$. Furthermore, to
third order in $\epsilon$, the map is given by the solution $s_1$ of
$s_1''-2~s_1-\mu'-\mu^2=0$, which can be represented as
\begin{equs}
s_1(\mu)=-\frac{1}{2}~G~\left(\mu'+\mu^2\right)~,
\label{eqn:approxfors}
\end{equs}
where $G$ is the convolution operator with the fundamental solution ${\cal G}$ of
${\cal G}(x)-\frac{1}{2}{\cal G}''(x)=\delta(x)$. Note that $G$ acts
multiplicatively in Fourier space, with symbol $(1+\frac{k^2}{2})^{-1}$, in
particular, $G~ f$ has two more derivatives than $f$. As we will also show
in Section \ref{sec:ampli}, $s(\mu)$ will have the same structure as
$s_1(\mu)$, that is, the ${\cal G}$--convolution of another map with the same regularity as
$\mu'$. As such, $s(\mu)$ is once more differentiable than $\mu$, due to
the regularizing properties of $G$, and $s(\mu)=s(\eta')$ is as regular as
$\eta$. This is reasonable, since from
$u=(1+\alpha^2~s)~\ed^{-i\beta~t+i\alpha\eta}$ we see that $s$ and $\eta$
should have both the same degree of regularity as $u$.

Inserting (\ref{eqn:approxfors}) into (\ref{eqn:formumus}) and neglecting 
$f_{\mu}$ leads to the (modified) Kuramoto--Siva-shinsky equation for the phase
\begin{equs}
\partial_t\mu=
-{\scriptstyle\frac{1+\alpha^2}{2}}~G~\mu^{''''}
-\epsilon^2~G~\mu''
-
{\scriptstyle2(1+\alpha^2)}~
\mu\mu'
-\epsilon^2~G~(\mu^2)'
-{\scriptstyle\frac{1+\alpha^2}{2}}~G~
(\mu^2)'''~,
\label{eqn:kuramotounxxx}
\end{equs}
from which we recover the Benjamin--Feir linear instability criterion
$1+\alpha\beta<0$. Namely, linear stability analysis in Fourier space (set
$\mu=\epsilon_0~\ed^{ikx+\lambda(k)t}$ with $\epsilon_0\ll1$) gives the
dispersion relation
\begin{equs}
\lambda(k)=
\frac{\epsilon^2~k^2-k^4~
\bigl(\text{$\frac{1+\alpha^2}{2}$}\bigr)}{1+\frac{k^2}{2}}
=
\frac{-(1+\alpha\beta)~k^2-k^4~
\bigl(\text{$\frac{1+\alpha^2}{2}$}\bigr)}{1+\frac{k^2}{2}}~.
\end{equs}
This shows that there are linearly unstable modes for
$|k|\leq\epsilon\ll1$, growing at most like $\ed^{\epsilon^4 t}$. This
suggests that the dynamics of (\ref{eqn:kuramotounxxx}) should be dominated by
the dynamics of the Fourier modes in the small $|k|$ region, the high $|k|$
modes being slaved to them. For $|k|\ll1$, we have
$G\approx1$, and neglecting the last two terms of (\ref{eqn:kuramotounxxx}),
we get the Kuramoto--Sivashinsky equation in derivative form
\begin{equs}
\partial_t\mu\approx
-{\scriptstyle\frac{1+\alpha^2}{2}}~\mu^{''''}
-\epsilon^2\mu''-{\scriptstyle2(1+\alpha^2)}~\mu\mu'~.
\label{eqn:kuramotodeuxxxx}
\end{equs}
Defining
\begin{equs}
\mu(x,t)=
{\scriptstyle\frac{1}{4}}
~\hat{\epsilon}^3~\hat{\mu}(\hat{x},\hat{t})~,
\label{eqn:firstcv}
\end{equs}
with
\begin{equs}
\chi=\frac{4}{1+\alpha^2}~,~~
\hat{\epsilon}={\scriptstyle\sqrt{\frac{\chi}{2}}}~\epsilon~,~~
\hat{x}=\hat{\epsilon}~x~,~~
\hat{t}={\scriptstyle\frac{2}{\chi}}~\hat{\epsilon}^4~t~,
\end{equs}
we get from (\ref{eqn:kuramotodeuxxxx})
\begin{equs}
\partial_{t}\hat{\mu}=
-\hat{\mu}^{''''}-\hat{\mu}''-\hat{\mu}~\hat{\mu}'~,
\label{eqn:truekuramotoxxx}
\end{equs}
which is the original Kuramoto--Sivashinsky equation in derivative form.
This justifies the scalings (\ref{eqn:firstcveta}). Equation
(\ref{eqn:truekuramotoxxx}) possesses an universal attractor of finite radius
in $\L^2([-L/2,L/2])$ with periodic boundary conditions (see e.g.
\cite{Collet}), hence we can expect $\hat{\mu}$ to be of size $\epsilon^3$
times a typical solution in that attractor.

From (\ref{eqn:approxfors}), we get (the $\mu$--dependence of $s_1$ is
implicit here for concision) 
\begin{equs}
s_1(x,t)=-\frac{\hat{\epsilon}^4}{32}~\hat{G}~
\left( 
4\hat{\mu}'(\hat{x},\hat{t})+\hat{\epsilon}^2~\hat{\mu}(\hat{x},\hat{t})^2
\right)
\equiv\hat{\epsilon}^4~\hat{s}_1(\hat{x},\hat{t})
~,
\label{eqn:approxforsws}
\end{equs}
where $\hat{G}$ is the convolution operator with the fundamental
solution of $\hat{\cal G}(x)-\frac{\epsilon^2}{2}\hat{\cal G}''(x)=\delta(x)$.
As above, $\hat{G}$ acts multiplicatively in Fourier space, with
symbol $\hat{G}(k)=(1+\frac{\epsilon^2~k^2}{2})^{-1}$. Equation
(\ref{eqn:approxforsws}) motivates the scalings
\begin{equs}
s(x,t)=\hat{\epsilon}^4~\hat{s}(\hat{x},\hat{t})~,
\label{eqn:firstcvs}
\end{equs}
for $s$ (see also (\ref{eqn:firstcvsintro})).

We now apply (\ref{eqn:firstcv}) and (\ref{eqn:firstcvs}) to
(\ref{eqn:forsmus}) and (\ref{eqn:formumus}). From now on we drop the hats.
Then  $s$ and $\mu$ satisfy the following equations
\begin{equs}
\partial_t s&=
-\frac{\chi}{\epsilon^4}~{\cal L}_{s}~s
+\frac{\chi}{\epsilon^4}~r_1(\mu)
-{\textstyle\frac{\alpha^2~\chi}{8}}(2s'\mu+s\mu')+F_3(s,\mu)~,
\label{eqn:infinalforms}
\\
\partial_t\mu&=
-\Lepsilon~\mu
-\mu\mu'
+\epsilon^2~F_0(s,\mu)'+\epsilon^2~\chi~{\cal L}_{\mu,r}~r_2'~,
\label{eqn:infinalform}
\end{equs}
where ${\cal L}_{s},\Lepsilon $ and ${\cal L}_{\mu,r}$ are multiplicative operators
in Fourier space, with symbols given by
\begin{equs}
{\cal L}_{s}(k)&=1+\frac{\epsilon^2~k^2}{2}~,\\
\Lepsilonk&=
\frac{k^4-k^2}{1+\frac{\epsilon^2~k^2}{2}}~,\\
{\cal L}_{\mu,r}(k)&=
2~\frac{
 2+\epsilon^2~(1+\alpha^2)
-\alpha^2~\epsilon^2~k^2
}{1+\frac{\epsilon^2~k^2}{2}}
\label{eqn:lmudef}
~,
\end{equs}
while $r_1$, $r_2$, $F_3$ and $F_0$ are defined by
\begin{equs}
r_1(\mu)&=-\frac{1}{32}(4\mu'+\epsilon^2~\mu^2)~,\\
r_2&=\frac{1}{\epsilon^4}~\Bigl(s-\frac{\epsilon^2}{2}~s''\Bigr)-\frac{r_1(\mu)}{\epsilon^4}\\
F_3(s,\mu)&=
-\alpha^2~\chi~\left({\textstyle\frac{3}{2}}~s^2
+\epsilon^2~{\textstyle\frac{1}{32}}~s~\mu^2
+{\textstyle\frac{\alpha^2}{2}}~\epsilon^4~s^3\right)~,\\
F_0(s,\mu)&=\alpha^2~\chi
\left(
\bigl(2+\epsilon^2~(1+\alpha^2)\bigr)~s^2
+\frac{s'~\mu}{1+\epsilon^4~\alpha^2~s}
-\frac{2~\epsilon^2~\alpha^2~s~s''}{1+\epsilon^4~\alpha^2~s}\right)\\
&\phantom{=}~
-\frac{1}{4}~\Gepsilon~\mu^2-
\frac{1}{4}~\Gepsilon~(\mu^2)''~,
\end{equs}
where $\Gepsilon$ is the operator with symbol
\begin{equs}
\Gepsilon(k)&=\frac{1}{1+\frac{\epsilon^2~k^2}{2}}~.
\end{equs}
We will prove that (\ref{eqn:infinalforms}) defines a map $\mu\mapsto
s(\mu)$ for all $\mu$ in an open ball of ${\cal W}_{\sigma}$, and
that this map has indeed `the same properties' as $\Gepsilon~
r_1(\mu)$, e.g. in terms of regularity. This is so essentially because for
$\epsilon\ll1$, we have $\frac{\chi}{\epsilon^4}{\cal L}_{s}\gg1$, so that
by Duhamel's formula, $s\sim {\cal L}_s^{-1}~r_1(\mu)+{\cal O}(\epsilon^4)=
\Gepsilon~ r_1(\mu)+{\cal O}(\epsilon^4)$ (see Section
\ref{sec:ampli}). At the same time, as a dynamical variable,
$r_2$ satisfies
\begin{equs}
\partial_t r_2=-\frac{\chi}{\epsilon^4}~\Gepsilon ~{\cal L}_{r}~r_2
+\frac{\chi}{16}\mu~{\cal L}_{\mu,r}~r_2'+\frac{1}{\epsilon^4}~F_6(s,\mu)~,
\label{eqn:theeqforrdxxx}
\end{equs}
where ${\cal L}_{r}$ is the multiplicative operator in Fourier space with
symbol
\begin{equs}
{\cal
L}_{r}(k)=1+
\Bigl({\textstyle\frac{3}{2}+\epsilon^2\bigl(\frac{1+\alpha^2}{4}\bigr)}\Bigr)~\epsilon^2~k^2
+
\Bigl({\textstyle\frac{1-\alpha^2}{4}}\Bigr)~\epsilon^4~k^4~,
\end{equs}
and
\begin{equs}
F_6(s,\mu)&={\cal L}_s~\Bigl(F_3(s,\mu)+F_4(s,\mu)
\Bigr)+
F_7(s,\mu)+F_8(\mu)~,\\
F_4(s,\mu)&=-{\textstyle\frac{\alpha^2~\chi}{8}}(2s'\mu+s\mu')~,\\
F_7(s,\mu)&=
\frac{\epsilon^2}{8}~\bigl(
\partial_x+\frac{\epsilon^2~\mu}{2}
\bigr)
\Bigl(
F_0(s,\mu)'
\Bigr)~,\\
F_8(s,\mu)&=
-\frac{1}{8}~\bigl(
\partial_x+\frac{\epsilon^2~\mu}{2}
\bigr)
\Bigl(
\Lepsilon ~\mu+\mu~\mu'
\Bigr)~.
\end{equs}
Once $s$ is considered as a given map $\mu\mapsto s(\mu)$,
(\ref{eqn:theeqforrdxxx}) defines the map $\mu\mapsto r_2(\mu)$
through a {\em linear} equation for $r_2$. By the same mechanism as for
$s$, we have $r_2\sim(\Gepsilon~{\cal L}_r)^{-1}~
F_6(s,\mu)\sim\Gepsilon~ F_6(s,\mu)$ if $\alpha^2<1$ (see Section
\ref{sec:proprdeux}). The restriction $\alpha^2<1$ is necessary here to
make ${\cal L}_{r}$ positive definite. For technical reasons, we have in
fact to restrict $\alpha^2<1/2$ to prove Theorems \ref{thm:existeunique} and
\ref{thm:properties}. We believe that the results of these Theorems could
be extended to part of the $\alpha^2>1/2$ region by exploiting the
following argument. If $\alpha^2>1$, equation (\ref{eqn:theeqforrdxxx}) 
for $r_2$ is linearly unstable at high frequencies. However, the
linear coupling of $r_2$ to $\mu$ through (\ref{eqn:infinalform})
stabilizes $r_2$. To see this, we introduce the vector ${\bf v}=(\mu,r_2)$,
and consider (\ref{eqn:infinalform}) and
(\ref{eqn:theeqforrdxxx}) simultaneously, as a vector dynamical system
of the form
\begin{equs}
\partial_t {\bf v}={\cal L}_M{\bf v}+f({\bf v})~,
\label{eqn:matrix}
\end{equs}
for a (nonlinear) vector map $f$, where ${\cal L}_M$ is the operator with
(matrix) symbol
\begin{equs}
{\cal L}_{M}(k)=
\begin{pmatrix}   
-\Lepsilonk & \epsilon^2~\chi~{\cal L}_{\mu,r}(k)~ik\\
-\frac{1}{8~\epsilon^4}~\Lepsilonk~ik
& -\frac{\chi}{\epsilon^4}\Gepsilon{\cal L}_r(k)
\end{pmatrix}~.
\end{equs}
The stability of (\ref{eqn:matrix}) at high frequency is then determined by
the eigenvalues $\lambda_{\pm}(k)$ of ${\cal L}_M(k)$ for large $k$.
Since\footnote{This is the analogon of $(1+i\alpha)~u''$ in
(\ref{eqn:cglabs}).}
\begin{equs}
\lambda_{\pm}(k)\to-(1\pm i|\alpha|)~\frac{k^2}{\hat{\epsilon}^2}
\end{equs}
as $k\to\infty$, (\ref{eqn:matrix}) is stable at high frequency, the real
part of the eigenvalues $\lambda_{\pm}(k)$ of ${\cal L}(k)$ being negative
for large $k$. However to exploit this would force us to solve
(\ref{eqn:infinalform}) and (\ref{eqn:theeqforrdxxx}) simultaneously,
which is technically (and notationally) more difficult, see \cite{nswake}
for a similar problem. Instead, in our approach the system
(\ref{eqn:infinalforms}), (\ref{eqn:infinalform}) and
(\ref{eqn:theeqforrdxxx}) is considered as a `main' equation, 
(\ref{eqn:infinalform}), of the form
\begin{equs}
\partial_t\mu=
-\Lepsilon ~\mu
-\mu\mu'
+\epsilon^2~F(\mu)'~,
\label{eqn:fformtractpre}
\end{equs}
supplied with  two `auxiliary' equations, (\ref{eqn:infinalforms}) and
(\ref{eqn:theeqforrdxxx}), which can be solved independently.

We will first study (\ref{eqn:fformtractpre}) for a general class of map
$F(\mu)$ in Section \ref{sec:phase} below, because it explains the choice
of the functional space, and which properties of the solutions of
the amplitude equations (\ref{eqn:infinalforms}) and (\ref{eqn:theeqforrdxxx})
are needed. Then, in Sections \ref{sec:ampli} and \ref{sec:proprdeux}, we
will show that the solutions of the amplitude equations
(\ref{eqn:infinalforms}) and (\ref{eqn:theeqforrdxxx}) exist and satisfy
the `right' properties.

\section{The phase equation}\label{sec:phase}
\subsection{Strategy}
Having argued that $r_2=r_2(\mu)$, we rewrite (\ref{eqn:infinalform}) as
\begin{equs}
\partial_t\mu=-\Lepsilon ~\mu-\mu\mu'+\epsilon^2~F(\mu)'~,~~~~~
\mu(x,0)=\mu_0(x)~,
\label{eqn:fformtract}
\end{equs}
where $\mu_0$ is a given (odd) space periodic function of period $L$ for
some given $L$. Since (\ref{eqn:fformtract}) preserves the mean of $\mu$
over $[-L/2,L/2]$, and since $\mu_0$ is the space derivative of a space periodic
function, we restrict ourselves to $\mu_0$ which have zero mean over
$[-L/2,L/2]$.

We will show that the term $\epsilon^2~F(\mu)'$ is in some sense
negligible. If $\epsilon=0$, then
$\Lepsilon=\partial_x^4+\partial_x^2\equiv{\cal L}_{\mu,c}$,
and (\ref{eqn:fformtract}) is the Kuramoto--Sivashinsky equation. If $F=0$ and
$\epsilon>0$, (in this case, $\Lepsilon$ is of smaller order than
$\partial_x^4+\partial_x^2$), this situation can still be easily handled by the
techniques of \cite{Collet} or \cite{Nicolaenko}, which show that equation
(\ref{eqn:fformtract}) possesses a universal attractor of finite radius in
$\L^2([-L/2,L/2])$ if $F=0$. A key ingredient of that proof is the
observation that the trilinear form $\int{\rm d} x~\mu^2\mu'$ vanishes for
periodic functions. However, in general, $\epsilon^2\int{\rm d}
x~\mu~F(\mu)'$ will not vanish, and might even not exist at all for $\mu\in
\L^2$.

We will explain precisely below how we circumvent this, but the mechanism
is indeed quite simple. If the $n$--th Fourier coefficients of $\mu$ were
vanishing for all $n\geq\frac{\delta}{q}$ with $1\ll\delta\ll1/\epsilon$,
we would have e.g. $\|\mu''\|_{\L^2}\leq\delta^2~\|\mu\|_{\L^2}$, which would
(presumably) give $\epsilon^2\int{\rm d} x~\mu~F(\mu)'\sim\epsilon^2\delta^2~\|\mu\|_{\L^2}^2$.
For $\epsilon$ sufficiently small, this would only give a small blur to the
attractor of the true Kuramoto--Sivashinsky equation.

Evidently, we cannot expect the high--$n$ Fourier modes to vanish, so we
will have to treat them separately. On that matter, we want to point out
that contrary to the `true' Kuramoto--Sivashinsky equation
(\ref{eqn:truekuramotoxxx}), where the linear operator ${\cal L}_{\mu,c}$
acting on $\mu$ on the r.h.s.~is of fourth order, $\Lepsilon $ is only of
second order due to the regularizing properties of $\Gepsilon $. From the
point of view of derivatives of $\mu$, it is easy to see that
$\epsilon^2~\hat{s}_1'$ and $\epsilon^2~\hat{s}_1''$ contain at most first
derivatives of $\mu$, hence we expect $\epsilon^2~F(\mu)'$ to contain at
most second order derivatives of $\mu$, and we see that at high frequencies,
(\ref{eqn:fformtractpre}) is more similar to the well studied equation
$\dot{u}=u''+f(u,u',u'')$ (see e.g. \cite{Bricmont}) than to the
Kuramoto--Sivashinsky equation. 

Note that the term $F(\mu)'$ is `irrelevant' due to its prefactor
$\epsilon^2$, while $\mu~\mu'$ is certainly not. Indeed, it would be 
catastrophic to solve (\ref{eqn:fformtract}) by successive approximations,
beginning with the solution of the equation with
$-\mu\mu'+\epsilon^2~F(\mu)=0$, inserting that solution into the
nonlinear terms and solving again the linear inhomogeneous problem. This 
would lead to (apparently) exponentially growing modes, because the linear
operator $\Lepsilon $ is not positive definite at small
frequencies. Solving (\ref{eqn:fformtract}) iteratively as
\begin{equs}
\partial_t\mu_{n+1}=-\Lepsilon ~\mu_{n+1}-\mu_{n+1}\mu_{n+1}'+\epsilon^2~F(\mu_{n})'~,
\end{equs}
for $n\geq0$ is a much better choice. We therefore consider the
following class of equations
\begin{equs}
\partial_t\mu=-\Lepsilon ~\mu-\mu\mu'+\epsilon^2~g'~,~~~~~\mu(x,0)=\mu_0(x)~,
\label{eqn:inhomo}
\end{equs}
for some given time dependent and spatially periodic perturbation $g$ and
periodic initial data $\mu_0$.

From this (informal) discussion, we see that we should treat the small $n$
Fourier coefficients with an $\L^2$--like norm as in \cite{Collet} or
\cite{Nicolaenko}, and the high $n$ modes as in e.g. \cite{Bricmont}. In
the next three subsections, we implement this idea. We first show $\L^{2}$
estimates for (\ref{eqn:inhomo}) in Subsection \ref{sec:coercive}. Then in
Subsection \ref{sec:highfreq} we define functional spaces similar to those
of \cite{Bricmont}, and prove inequalities in these spaces, which will
allow us to prove the `high frequency estimates' in Subsection
\ref{sec:highkbound}. In subsection \ref{sec:nonlinear}, we will prove that
the full phase equation has a solution if $\mu\mapsto F(\mu)$ is a well
behaved Lipschitz map, and finally, in subsection \ref{sec:consequences},
we will show how the phase equation relates to the Kuramoto--Sivashinsky
equation.

\subsection{Coercive functional method, $\L^2$ estimates}\label{sec:coercive}
The initial value problem (\ref{eqn:inhomo}) is globally well posed in
$\L^2([-L/2,L/2])$ if the perturbation $g$ is periodic and satisfies
$\|g(\cdot,t)\|_{\L^2}<\infty$ for all $t\geq0$. The local
uniqueness/existence theory follows from standard techniques (see e.g.
\cite{Temam}), whereas the global existence follows from the a priori
estimate
\begin{equs}
\|\mu(\cdot,t)\|_{\L^2}^2\leq
\ed^{t}~\|\mu(\cdot,0)\|_{\L^2}^2
+2~\epsilon^4~\left(\ed^{t}-1\right)\sup_{0\leq s\leq t}\|g(\cdot,s)\|_{\L^2}^2~.
\label{eqn:apriori}
\end{equs}
This estimate can be obtained by multiplying (\ref{eqn:inhomo}) with $\mu$ and
averaging over $[-L/2,L/2]$. Denoting by $\int$ the integral over a (space)
period, we have, using integration by parts
\begin{equs}
\frac{1}{2}\partial_t\int\mu^2=
-\int\mu~\Lepsilon ~\mu
-\int\mu^2\mu'
-\epsilon^2\int\mu'~g~.
\end{equs}
Since $\mu$ is periodic, we have $\int\mu^2\mu'=0$. Using Young's
inequality, we have
\begin{equs}
\partial_t\int\mu^2\leq
-2\int\mu~\Lepsilon ~\mu
+\frac{1}{2}
\int(\mu')^2
+2~\epsilon^4\int g^2
\leq
\int\mu^2+2~\epsilon^4\int g^2~,
\end{equs}
from which (\ref{eqn:apriori}) follows immediately. As a much stronger
result, we can in fact prove that the $\L^2$--norm of the solution stays
bounded for all $t\geq0$. Introducing the operator ${\cal L}_v$,
whose symbol is
\begin{equs}
{\cal L}_{v}(k)=\sqrt{\frac{1}{3}\frac{1+k^4}{1+\frac{\epsilon^2~k^2}{2}}}~,
\label{eqn:lvdef}
\end{equs}
we have the following theorem.
\begin{theorem}\label{thm:ldeuxexiste}
There exist a constant $K$ such that the solution $\mu$ of 
\begin{equs}
\partial_t\mu=-\Lepsilon ~\mu-\mu\mu'+\epsilon^2~g'~,~~~~\mu(x,0)=\mu_0(x)
\end{equs}
is periodic, antisymmetric, and satisfies
\begin{equs}
\supt\|\mu(\cdot,t)\|_{\L^2}\leq
\rho+\|\mu_0\|_{\L^2}
+4~\epsilon^2~\supt\|{\cal L}_v^{-1}~g(\cdot,t)'\|_{\L^2}~,
\end{equs}
where $\rho=K~L^{8/5}$, if $\mu_0$ and $g'$ are antisymmetric (spatially)
periodic functions of period $L$, 
\end{theorem}
\begin{proof}
Note first that ${\cal L}_{v}^{-1}~\partial_x$ is a bounded operator on $\L^2$
with norm $\leq2$ (see Lemma \ref{lem:someproperties} in Appendix
\ref{app:therdeuxmap}), then local existence in $\L^2$ follows from the
above argument.

Following \cite{Nicolaenko} with the modifications of \cite{Collet}, we
write $\mu(x,t)=v(x,t)+\phi(x)$ for some constant periodic function $\phi$
to be chosen later on. Denoting by $\int$ the integral over $[-L/2,L/2]$,
we get from (\ref{eqn:inhomo})
\begin{equs}
\frac{1}{2}\partial_t
 \int\bsp v^2=
-\int\bsp v~\Lepsilon~v
-\int\bsp v~\Lepsilon~\phi
-\int\bsp v^2v'
-\frac{1}{2}
 \int\bsp v^2\phi'
-\int\bsp v\phi\phi'
+\epsilon^2\int\bsp v~g'~.
\label{eqn:coercun}
\end{equs}
The term $\int v^2v'$ vanishes because $v$ is periodic, giving a much
more compact form for (\ref{eqn:coercun}):
\begin{equs}
\frac{1}{2}\partial_t (v,v)=
-(v,v)_{\phi/2}
-(v,\phi)_{\phi}
+\epsilon^2~(v,g')~,
\label{eqn:coercdeux}
\end{equs}
where we have introduced the inner products
\begin{equs}
(v,w)&=\int vw~~~\text{and}~~~
(v,w)_{\gamma\phi}=\int v~(\Lepsilon +\gamma\phi')~w~.
\end{equs}
This decomposition is helpful because we have the following nondegeneracy result
which is proved in Appendix \ref{app:coercive}
\begin{proposition}\label{prop:coerciveness}
For all $L\geq2\pi$, there exist a constant $K$ and an antisymmetric
periodic function $\phi$ such that for all $\gamma\in[\frac{1}{4},1]$ and
$\epsilon\leq L^{-2/5}$, and for every antisymmetric periodic function $v$,
one has
\begin{equs}
\frac{3}{4}~({\cal L}_{v}~v,{\cal L}_{v}~v)\leq(v,v)_{\gamma\phi}&\leq
\|\phi'\|_{\infty}~(v,v)+(v'',v'')~,\\
(\phi,\phi)_{\gamma\phi}&\leq K~L^{16/5}~,\\
(\phi,\phi)&\leq{\textstyle\frac{4}{3}}~L^3~,
\end{equs}
where ${\cal L}_{v}$ is defined in (\ref{eqn:lvdef}).
\end{proposition}
We first note that
\begin{equs}
({\cal L}_{v}~v,{\cal L}_{v}~v)\geq
\frac{4}{3}\Bigl(
\frac{\sqrt{\epsilon^4+4}-2}{\epsilon^4}
\Bigr)~(v,v)\equiv c_v^2~(v,v)~.
\end{equs}
Using Young's inequality and Proposition
\ref{prop:coerciveness}, we get from (\ref{eqn:coercdeux}),
\begin{equs}
\partial_t (v,v)&\leq
-2~(v,v)_{\phi/2}
+{\scriptstyle\frac{2}{3}}(v,v)_{\phi}
+{\scriptstyle\frac{3}{2}}(\phi,\phi)_{\phi}
+2~\epsilon^2~(v,g')
\\
&\leq-{\scriptstyle\frac{4}{3}}~(v,v)_{\phi/4}
+{\scriptstyle\frac{3}{2}}(\phi,\phi)_{\phi}
+2~\epsilon^2~(v,g')\\
&\leq -({\cal L}_{v}~v,{\cal L}_{v}~v)
+{\scriptstyle\frac{3}{2}}(\phi,\phi)_{\phi}
+2~\epsilon^2~({\cal L}_{v}~v,{\cal L}_{v}^{-1}~g')
\\
&\leq -\frac{1}{2}~({\cal L}_{v}~v,{\cal L}_{v}~v)
+{\scriptstyle\frac{3}{2}}(\phi,\phi)_{\phi}
+2~\epsilon^4~\|{\cal L}_{v}^{-1}~g'\|_{\L^2}^2\\
&\leq -\frac{c_v^2}{2}~(v,v)
+{\scriptstyle\frac{3}{2}}(\phi,\phi)_{\phi}
+2~\epsilon^4~\|{\cal L}_{v}^{-1}~g'\|_{\L^2}^2~.
\label{eqn:coerctrois}
\end{equs}
Since $v(x,t)=\mu(x,t)-\phi(x)$ we conclude that
\begin{equs}
\|\mu(\cdot,t)-\phi(\cdot)\|_{\L^2}^2\leq
\|\mu_0-\phi\|_{\L^2}^2+
\frac{3}{c_v^2}(\phi,\phi)_{\phi}
+\frac{4~\epsilon^4}{c_v^2}~\supt~
\|{\cal L}_{v}^{-1}~g(\cdot,t)'\|_{\L^2}^2~.
\end{equs}
Finally, since $\frac{2}{c_v}\leq4$, we have
\begin{equs}
\supt\|\mu(\cdot,t)\|_{\L^2}\leq
\|\mu_0\|_{\L^2}+\rho+4~\epsilon^2~
\supt~\|{\cal L}_{v}^{-1}~g(\cdot,t)'\|_{\L^2}~,
\end{equs}
where
\begin{equs}
\rho=
2~\|\phi\|_{\L^2}+4~\sqrt{(\phi,\phi)_{\phi}}~.
\end{equs}
Furthermore, by Proposition \ref{prop:coerciveness}, we have
$\rho<\infty$, since $\|\phi\|_{\L^2}=\sqrt{(\phi,\phi)}<\infty$ and
$(\phi,\phi)_{\phi}<\infty$. This completes the proof of the theorem.
\end{proof}
\begin{corollary}\label{cor:ksoldresult}
The antisymmetric solution of the Kuramoto--Sivashinsky equation with 
periodic boundary conditions on $[-L/2,L/2]$
\begin{equs}
\partial_t\mu=-\mu^{''''}-\mu''-\mu~\mu'~,~~~~\mu(x,0)=\mu_0(x)~,
\label{eqn:truetrueks}
\end{equs}
stays in a ball of radius ${\cal O}(L^{8/5})$ in $\L^2$ as $L\to\infty$.
\end{corollary}
\begin{proof}
This result was already established in \cite{Nicolaenko} and \cite{Collet}.
To prove it, we only have to note that (\ref{eqn:truetrueks}) corresponds
to (\ref{eqn:inhomo}) with $\epsilon=0$, and that Theorem
\ref{thm:ldeuxexiste} is uniformly valid in $\epsilon\leq1$.
\end{proof}
\begin{remark}\label{rem:noteven}
The proof of Theorem \ref{thm:ldeuxexiste} is the only point in this paper
where we need $s$, respectively $\mu$, to be spatially even, resp.~odd,
functions. The theorem holds also in the general (non symmetric) case. The
proof can be obtained as a straightforward extension of the result of
\cite{Collet} for the Kuramoto--Sivashinsky equation in the non symmetric
case.
\end{remark}
If $\epsilon=0$, Theorem \ref{thm:ldeuxexiste} shows that the solution of
(\ref{eqn:inhomo}) stays in a ball in $\L^2$, centered on $0$ and of radius
$\|\mu_0\|_{\L^2}+\rho$ for all $t\geq0$, with $\rho={\cal O}(L^{8/5})$ as
$L\to\infty$. When $\epsilon\neq0$, the radius of the ball
widens to lowest order like $\epsilon^2~\supt\|g(\cdot,t)\|_{\L^2}$.

\subsection{Functional spaces, definitions and properties}\label{sec:highfreq}
In this section, we explain how to treat the high frequency part of the
solution of (\ref{eqn:inhomo}). This development is inspired by
\cite{Bricmont} (see also \cite{nswake} for similar definitions). We
will need some technical estimates which are proven in Appendix
\ref{app:highk}.

\subsubsection{Basic Definitions}\label{subsec:defs}
Let $L\geq2\pi$ and $q\equiv\frac{2\pi}{L}\leq1$. We define the Fourier
coefficients $f_n$ of a function $f:[-L/2,L/2]\to{\bf R}$ by
\begin{equs}
f_n=\frac{1}{L}\int_{-L/2}^{L/2}
\hspace{-4mm}
\text{d}x~\ed^{-iqnx}f(x)~,~~~~~\text{so that}~~~~~
f(x)=\sum_{n\in{\bf Z}}\ed^{iqnx}~f_n~,
\end{equs}
and $P_{<}$, $P_{>}$, the projectors on the small/high frequency part by
\begin{equs}   
P_{<}f(x)=\sum_{|n|\leq\frac{\delta}{q}}\ed^{iqnx}f_{n}~~~,~~~
P_{>}f(x)=\sum_{|n|>\frac{\delta}{q}}\ed^{iqnx}f_{n}~,
\end{equs}
where the parameter $\delta\geq2$ will be chosen later. We also define the $\L^p$ and
$l^p$ norms as
\begin{equs}[2]
\|f\|_{\L^p}&=\left(
\int_{-L/2}^{L/2}\hspace{-5mm}\text{d}x~
|f(x)|^{p}
\right)^{1/p}
~,~~~~
&\|f\|_{\L^{\infty}}&=\essup_{x\in[-L/2,L/2]}|f(x)|~,\\
\|f\|_{l^p}&=\left(
\sum_{n\in{\bf Z}}
|f_n|^{p}
\right)^{1/p}
~,~~~~
&\|f\|_{l^{\infty}}&=\sup_{n\in{\bf Z}}|f_n|~.
\end{equs}
We will use repeatedly Plancherel's equality without notice
\begin{equs}
\|f\|_{\L^2}=\sqrt{L}~\|f\|_{l^2}~.
\end{equs}
Finally, for $\sigma\geq0$, $\delta\geq 2$, we define the norm
$\|\cdot\|_{{\cal N},\sigma}$ by
\begin{equs}
\|f\|_{{\cal N},\sigma}&=
{\textstyle\frac{\sqrt{\delta}}{q}}
\sup_{n\in{\bf Z}}
\bigl(1+({\textstyle\frac{qn}{\delta}})^{\scriptscriptstyle2}\bigr)^{\frac{\sigma}{2}}
~|f_{n}|~.
\end{equs}
With a different normalization, the norm $\|\cdot\|_{{\cal
N},\sigma}$ was introduced in \cite{Bricmont} to study the long time
asymptotics of solutions of $\dot{u}=u''+f(u,u',u'')$, where $f$ is some
(polynomial) nonlinearity. From the point of view of the
nonlinearity, our situation is similar to the case treated there, but our
linear operator $\Lepsilon $ is not positive definite as $-\Delta$ was in
their case. The potentially exponentially growing modes correspond to
$|n|\leq\frac{1}{q}$, and we saw in Section \ref{sec:coercive} that their
$l^2$ norm was bounded. Since there are only a finite (but large)
number of linearly unstable modes, changing the definition of the $\|\cdot\|_{{\cal
N},\sigma}$--norm on these modes to an $l^2$--like norm will give an
equivalent norm which is better suited to our case. Thus we define the norms
$\|\cdot\|_{{\cal W},\sigma}$ and $\|\cdot\|_{\sigma}$ by
\begin{equs}
\|f\|_{{\cal W},\sigma}&=
{\textstyle\frac{\sqrt{\delta}}{q}}
\sup_{|n|>\frac{\delta}{q}}
\bigl(1+({\textstyle\frac{qn}{\delta}})^{\scriptscriptstyle2}\bigr)^{\frac{
\sigma}{2}}~|f_{n}|~,
\label{eqn:defnormhigh}
\\
\|f\|_{\sigma}&=\|f\|_{\L^2}+\|f\|_{{\cal W},\sigma}~.
\label{eqn:defnorm}
\end{equs}
While $\|\cdot\|_{{\cal W},\sigma}$ is clearly {\em not} a norm,
$\|\cdot\|_{\sigma}$ is a norm which is equivalent to $\|\cdot\|_{{\cal
N},\sigma}$ for $\sigma\geq1$. Indeed, easy calculations lead to
\begin{equs}[2]
\|f\|_{{\cal N},\sigma}&\leq 
\sqrt{2^{\sigma}~L~\delta}~\|f\|_{\L^2}+\|f\|_{{\cal W},\sigma}
\leq\bigl(1+\sqrt{2^{\sigma}~L~\delta}\bigr)~\|f\|_{\sigma}
~,
\label{eqn:equivalence}
\\
\|f\|_{\sigma}&\leq 
\Bigl(1+\pi~\sqrt{2}\Bigr)~\|f\|_{{\cal N},\sigma}~.
\label{eqn:equivalencedeux}
\end{equs}
We point out that if $\sigma>\frac{1}{2}$, the $\|\cdot\|_{{\cal
W},\sigma}$--semi--norm is a {\em decreasing} function of $\delta$. Indeed, we
have (here the norms carry an additional index to specify the value of $\delta$)
\begin{equs}
\|f\|_{{\cal W},\sigma,\delta_1}&\leq
\sqrt{\frac{\delta_1}{\delta_0}}~
\Bigl(
\frac{2}{1+\bigl(\frac{\delta_1}{\delta_0}\bigr)^2}
\Bigr)^{\frac{\sigma}{2}}
\|f\|_{{\cal W},\sigma,\delta_0}
\leq
2^{\frac{\sigma}{2}}~
\Bigl(
\frac{\delta_0}{\delta_1}
\Bigr)^{\sigma-\frac{1}{2}}~
\|f\|_{{\cal W},\sigma,\delta_0}~,
\label{eqn:decreasing}
\end{equs}
for all $\delta_1\geq\delta_0\geq2$.  As $\delta$ will be fixed later on,
the additional index is suppressed to simplify the notation. On the other
hand, $\|\cdot\|_{\sigma}$ is an {\em non--decreasing} function of
$\sigma$:
\begin{equs}
\|f\|_{\sigma_0}&\leq\|f\|_{\sigma_1}~,
\label{eqn:increasing}
\end{equs}
for all $\sigma_1\geq\sigma_0$.
\begin{definition}   
Denoting by ${\cal C}_{0,{\rm per}}^{\infty}([-L/2,L/2],{\bf R})$ the set
of infinitely differentiable periodic real valued functions on
$[-L/2,L/2]$, we define the (Banach) space ${\cal W}_{0,\sigma}$ as the
completion of ${\cal C}_{0,{\rm per}}^{\infty}([-L/2,L/2],{\bf R})$ under
the norm $\|\cdot\|_{\sigma}$, and ${\cal B}_{0,\sigma}(r)\subset{\cal
W}_{0,\sigma}$ the open ball of radius $r$ centered on $0\in{\cal
W}_{0,\sigma}$.
\end{definition}

Up to now, we considered functions depending on the space variable only. We
extend the definition (\ref{eqn:defnorm}) to functions
$f:[-L/2,L/2]\times[0,\infty)\to{\bf R}$ by
\begin{equs}
\tvert f\tvert_{\sigma}=\supt\|f(\cdot,t)\|_{\sigma}~.
\end{equs}
We will use the same convention for the $\L^p$ and $l^p$ norms, e.g.
\begin{equs}
\tvert f\tvert_{\L^2}=\supt\|f(\cdot,t)\|_{\L^2}~.
\end{equs}
Finally, we make the following definition.
\begin{definition}   
Let $\Omega=[-L/2,L/2]\times{\bf R}^{+}$ and ${\cal C}_{{\rm
per}}^{\infty}(\Omega,{\bf R})$ denote the set of infinitely differentiable
functions on $\Omega$ compactly supported on ${\bf R}^{+}$ and satisfying
$f(-L/2,t)=f(L/2,t)$ for all $t\in{\bf R}^{+}$. We define the (Banach)
space ${\cal W}_{\sigma}$ as the completion of ${\cal C}_{{\rm
per}}^{\infty}(\Omega,{\bf R})$ under the norm $\tvert\cdot\tvert_{\sigma}$, and
${\cal B}_{\sigma}(r)\subset{\cal W}_{\sigma}$ the open ball of radius
$r$ centered on $0\in{\cal W}_{\sigma}$.
\end{definition}
We then have the
\begin{proposition}   
For all $\sigma>\frac{5}{2}$, ${\cal W}_{\sigma}$ is a Banach
space included in the Banach space $\L^{\infty}({\bf
R}^{+},W_{2,2}([-L/2,L/2]))$ of functions (and their space derivatives up
to order $2$) on $\Omega$ uniformly (in time) bounded in $\L^2([-L/2,L/2])$.
\end{proposition}
\begin{proof}   
The proof follows directly from the Lemma \ref{lem:ldliestime} below.
\end{proof}
\begin{lemma}\label{lem:ldliestime}
Let $\sigma\geq\frac{3}{2}$. There exists a constant $C$ such that for all
$n\leq\sigma-\frac{3}{2}$ and for all $m\leq\sigma-1$, we have
\begin{equs}[3]
\|f^{(m)}\|_{\sigma-m}&+\|\Gepsilon~ f^{(m)}\|_{\sigma-m}
&~\leq~& C~\delta^{m}~&\|f\|_{\sigma}~,
\label{eqn:sdestim}
\\
\|f^{(m)}\|_{\L^{\infty}}&+
\|\Gepsilon~ f^{(n)}\|_{\L^\infty}
&~\leq~&
C~\delta^{n+\frac{1}{2}}~&\|f\|_{\sigma}~,
\label{eqn:Linfestim}
\end{equs}
where $f^{(m)}$ is the $m$--th order spatial
derivative of $f$.
\end{lemma}
\begin{proof}   
See Appendix \ref{app:highk}.
\end{proof}
Although the indices $m$ and $n$ make the reading of Lemma
\ref{lem:ldliestime} a bit cumbersome, it merely says that $\Gepsilon $ is
`transparent' for the norms, and that each derivative `cost' a factor
$\delta$.

\subsubsection{Properties}\label{sec:nonlinearities}
The $\|\cdot\|_{{\cal N},\sigma}$ and $\|\cdot\|_{{\cal W},\sigma}$ norms
can be used to control the nonlinear term $F_0(\mu)'$. For concision, all
the proofs of this section are relegated in Appendix \ref{app:highk}. The
map $F_0(\mu)$ admits the following decomposition (see Appendix
\ref{app:thefis})
\begin{equs}
F_0(\mu)=
F_{1}(\mu)
+\Gepsilon ~ F_{2}(\mu)~,
\label{eqn:decomposeF}
\end{equs}
where 
\begin{equs}
F_{1}(\mu)&=
(1+\epsilon^2)~\hat{s}^2
-\frac{\alpha^2~\hat{s}~(\epsilon^2\hat{s}'')}{1+\epsilon^4~\hat{s}}
-\frac{\alpha^2}{2}~\frac{\mu~\hat{s}~(\epsilon^4~\hat{s}')}{1+\epsilon^4~\hat{s}}~,\\
F_{2}(\mu)&=
-\frac{1}{4}\mu^2-\frac{1}{4}(\mu^2)''
+\frac{\alpha^2}{2}~\mu~\hat{r}'+\alpha^2~\mu'~\hat{r}-\alpha^2~\mu'~\hat{s}-
\frac{\alpha^2}{4}\mu''~(\epsilon^2~\hat{s}')~,\\
\hat{r}(\mu)&=\Bigl(1-\frac{{\epsilon}^2}{2}\partial_{x}^2\Bigr)
~\hat{s}(\mu)~.
\end{equs}
To bound the contribution of $F_0$ to $\mu$, we need a sequence of easy
Propositions and Lemmas. The first result concerns the various terms
appearing in Duhamel's formula.
\begin{proposition}\label{prop:propagation}
Let $\delta\geq2$, then
\begin{equs}
\left\|
\ed^{-\Lepsilon t}~
f(\cdot)
\right\|_{{\cal W},\sigma}
&\leq
\ed^{-4t}~
\|f(\cdot)\|_{{\cal W},\sigma}
~,\\
\left\|
\int_{0}^{t}
\hspace{-2mm}{\rm d}s~
\ed^{-\Lepsilon (t-s)}~
g'(\cdot,s)
\right\|_{{\cal W},\sigma}
&\leq \sup_{0\leq s\leq t}
\left\|\frac{g'(\cdot,s)}{\Lepsilon }\right\|_{{\cal W},\sigma}~,
\label{eqn:rhsdu}
\end{equs}
where $\ed^{-\Lepsilon t}$ is the propagation Kernel associated with
$\partial_t f=-\Lepsilon f$, and $\Lepsilon $ is defined in (\ref{eqn:fformtract}).
\end{proposition}
Then, on the r.h.s.~of (\ref{eqn:rhsdu}), we have the
\begin{lemma}\label{lem:unsurLe}
Let $\delta\geq2$, then
\begin{equs}
\left\|\frac{g'}{\Lepsilon }\right\|_{{\cal W},\sigma}
&\leq\frac{\sqrt{2}}{\delta}\left\|g\right\|_{{\cal W},\sigma-1}~,\\
\left\|\frac{\Gepsilon ~ g'}{\Lepsilon }\right\|_{{\cal W},\sigma}
&\leq\frac{2^{7/2}}{3~\delta^3}\left\|g\right\|_{{\cal W},\sigma-3}~.
\end{equs}
\end{lemma}
This shows that we need only control $\|F_1\|_{{\cal W},\sigma-1}$
and $\|F_2\|_{{\cal W},\sigma-3}$. These will be bounded using Propositions
\ref{prop:key} and \ref{prop:division} below, which show that
multiplication and division of functions are well defined in
${\cal W}_{\sigma}$.
\begin{proposition}\label{prop:key}
Let $\|u\|_{\sigma_1}<\infty$, $\|v\|_{\sigma_2}<\infty$ and
$\sigma=\min(\sigma_1,\sigma_2)\geq\frac{3}{2}$. Then there exists a
constant $\Cm$ depending only on $\sigma$ such that
\begin{equs}
\|uv\|_{\sigma}\leq
\Cm~\sqrt{\delta}~
\|u\|_{\sigma_1}~
\|v\|_{\sigma_2}~,
\end{equs}
and if $\sigma\leq1$, we have the two particular cases
\begin{equs}
\|uv\|_{{\cal W},\frac{1}{2}}&\leq
\Cm~\sqrt{\delta}~\|u\|_{1}~\|v\|_{1}~,\\
\|uv\|_{{\cal W},0}&\leq
\Cm~\sqrt{\delta}~\|u\|_{\L^2}~\|v\|_{\L^2}~.
\end{equs}
\end{proposition}
The following proposition shows that the $\|\cdot\|_{\sigma}$--norm
of $\frac{u}{1+v}$ is essentially given by $\|u\|_{\sigma}$ if 
$\|v\|_{\sigma}\ll1$.
\begin{proposition}\label{prop:division}
Let $\|u\|_{\sigma_1}<\infty$, $\|v\|_{\sigma_2}<\infty$ and
$\sigma=\min(\sigma_1,\sigma_2)\geq\frac{3}{2}$. Then
\begin{equs}
\left\|\frac{u}{1+v}\right\|_{\sigma}
\leq 
\frac{\|u\|_{\sigma_1}}{1-\Cm~\sqrt{\delta}~\|v\|_{\sigma_2}}
~.
\end{equs}
for all $v$ satisfying $\Cm~\sqrt{\delta}~\|v\|_{\sigma_2}<1$, where 
$\Cm$ is the constant of Proposition \ref{prop:key}.
\end{proposition}

\subsection{High frequency estimates}\label{sec:highkbound}
Here, we study the high frequency part of the solution of
\begin{equs}
\partial_t\mu=-\Lepsilon ~\mu-\mu\mu'+\epsilon^2~g'~,~~~~\mu(x,0)=\mu_0(x)~.
\end{equs}

The solution of this equation exists by Theorem \ref{thm:ldeuxexiste}, and
is bounded in $\L^2$ for all $t\geq0$ if $\|\mu_0\|_{\L^2}+\tvert{\cal
L}_v^{-1}~g'\tvert_{\L^2}<\infty$. We will now show that upon further
restrictions on $\mu_0$ and $g$, the solution has bounded
$\|\cdot\|_{\sigma}$--norm for all $t\geq0$.

We first need some definitions. We set
\begin{equs}
c_0&=1+\frac{\|\mu_0\|_{\sigma}}{\rho}
+\frac{4~\epsilon^2}{\rho}~\tvert {\cal L}_{v}^{-1}~g'\tvert_{\L^2}
+\frac{\epsilon^2}{\rho}~
\tvertb\frac{g'}{\Lepsilon }\tvertb_{{\cal W},\sigma}
~,\label{eqn:defczero}\\
\xi&=\frac{\Cm~c_0~\rho}{\sqrt{2~\delta}}~.
\end{equs}
Then we have
\begin{theorem}\label{thm:weirdball}
Assume that the initial condition $\mu_0$ and $g$ satisfy 
(\ref{eqn:defczero}) with $\xi<\frac{1}{4}$. Then the solution $\mu$ of
\begin{equs}
\partial_t\mu=-\Lepsilon ~\mu-\mu\mu'+\epsilon^2~g'~,~~~~
\mu(x,0)=\mu_0(x)
\label{eqn:inhomoundeux}
\end{equs}
satisfies
\begin{equs}
\tvert\mu\tvert_{\sigma}\leq
\left(\frac{1-\sqrt{1-4\xi}}{2\xi}\right)~
\Bigl(
\rho+\|\mu_0\|_{\sigma}
+4~\epsilon^2~\tvert {\cal L}_{v}^{-1}~g'\tvert_{\L^2}
+\epsilon^2~\tvertb\frac{g'}{\Lepsilon }\tvertb_{{\cal W},\sigma}
\Bigr)
~.
\label{eqn:borneweird}
\end{equs}
\end{theorem}
\begin{remark}   
Note that $c_0$ is implicitly dependent of $\delta$ (because the norm
$\|\cdot\|_{\sigma}$ is). If $\mu_0$ and $g$ are given, $c_0$ is a {\em non
increasing} function of $\delta$ (see (\ref{eqn:decreasing})). Hence we can
surely satisfy $\xi<\frac{1}{4}$ by taking $\delta$ sufficiently large.
\end{remark}
\begin{proof}[Proof of Theorem \ref{thm:weirdball}]
We first note that by Theorem \ref{thm:ldeuxexiste}, we have
\begin{equs}
\tvert\mu\tvert_{\L^2}\leq 
\rho+\|\mu_0\|_{L^2}+
4~\epsilon^2~\tvert {\cal L}_{v}^{-1}~g'\tvert_{\L^2}
\equiv c_1~\rho~.
\label{eqn:bldeux}
\end{equs}
To bound $\tvert\mu\tvert_{{\cal W},\sigma}$, we use Duhamel's formula for
the solution of (\ref{eqn:inhomoundeux})
\begin{equs}
\mu(x,t)=
\ed^{-\Lepsilon t}~\mu_0(x)
-\int_{0}^{t}\hspace{-2mm}{\rm d}s~
\ed^{-\Lepsilon (t-s)}~(\mu\mu')(x,s)
+\epsilon^2\bsp\int_{0}^{t}
\hspace{-2mm}{\rm d}s~
\ed^{-\Lepsilon (t-s)}~g'(x,s)~.
\label{eqn:reprep}
\end{equs}
Next, we define
\begin{equs}
T(\mu)(x,t)&=-\int_{0}^{t}\hspace{-2mm}{\rm d}s~
\ed^{-\Lepsilon (t-s)}~(\mu\mu')(x,s)~.
\end{equs}
Since $\mu~\mu'=\frac{1}{2}(\mu^2)'$, using Proposition
\ref{prop:propagation} and Lemma \ref{lem:unsurLe}, we get for all
$\sigma'\leq\sigma$, the bound
\begin{equs}
\tvert T(\mu)\tvert_{{\cal W},\sigma'}\leq
\frac{1}{\sqrt{2}~\delta}\tvert\mu^2\tvert_{{\cal W},\sigma'-1}~,
\end{equs}
and using again Proposition \ref{prop:propagation}, we get
\begin{equs}
\tvert\mu\tvert_{{\cal W},\sigma'}
\leq
\|\mu_0\|_{{\cal W},\sigma'}+
\frac{1}{\sqrt{2}~\delta}~
\tvert\mu^2\tvert_{{\cal W},\sigma'-1}
+\epsilon^2~\tvertb\frac{g'}{\Lepsilon }\tvertb_{{\cal W},\sigma'}~.
\label{eqn:inductionsdiv}
\end{equs}
Using that $\|f\|_{\sigma_1}\leq\|f\|_{\sigma_2}$ if $\sigma_1\leq\sigma_2$, 
inequality (\ref{eqn:bldeux}), and dividing 
(\ref{eqn:inductionsdiv}) by $c_0~\rho$ (note that $c_0\rho\geq\rho>0$), we get
\begin{equs}
\frac{1}{c_0~\rho}\tvert\mu\tvert_{\sigma'}
\leq
1+\frac{1}{\sqrt{2}~\delta~c_0\rho}~
\tvert\mu^2\tvert_{{\cal W},\sigma'-1}~.
\label{eqn:induction}
\end{equs}
We use this equation inductively in $\sigma'$ to show
(\ref{eqn:borneweird}). As a first step, notice that by Proposition
\ref{prop:key}, we have $\tvert\mu^2\tvert_{{\cal
W},0}\leq\Cm\sqrt{\delta}~\tvert\mu\tvert_{\L^{2}}^2$, so that from
(\ref{eqn:induction}) with $\sigma'=1$ and the definition of $c_1$
(see (\ref{eqn:bldeux})), we get
\begin{equs}
\frac{1}{c_0~\rho}
\tvert\mu\tvert_{1}
&\leq 1+\frac{\Cm    }{\sqrt{2~\delta}~c_0~\rho}~(c_1~\rho)^2
\leq 1+\frac{\Cm
~c_0~\rho}{\sqrt{2~\delta}}~\Bigl(\frac{c_1}{c_0}\Bigr)^2\\
&\leq 1+\xi~\Bigl(\frac{c_1}{c_0}\Bigr)^2
\equiv 
\frac{c_2}{c_0}~.
\label{eqn:defdeux}
\end{equs}
Using this inequality, (\ref{eqn:induction}) with
$\sigma'=\frac{3}{2}$ and Proposition \ref{prop:key}, we get
\begin{equs}
\frac{1}{c_0~\rho}
\tvert\mu\tvert_{\frac{3}{2}}
&\leq 1+
\frac{\Cm }{\sqrt{2~\delta}~c_0~\rho}
~\tvert\mu\tvert_{1}^2
\leq 1+\frac{\Cm~c_0~\rho}{\sqrt{2~\delta}}~\Bigl(\frac{c_2}{c_0}\Bigr)^2\\
&\leq 1+\xi~\Bigl(\frac{c_2}{c_0}\Bigr)^2
\equiv\frac{c_3}{c_0}~.
\label{eqn:deftrois}
\end{equs}
Let now $\sigma_3=\frac{3}{2}$, $\sigma_n=\sigma_{n-1}+1$ for all $4\leq
n\leq n_0$, where the integer $n_0$ is defined by $\sigma-1\leq\sigma_{n_0}<\sigma$.
Using (\ref{eqn:induction}) and Proposition \ref{prop:key}, we have
\begin{equs}
\frac{1}{c_0~\rho}
\tvert\mu\tvert_{\sigma_n}
&\leq
1+
\frac{\Cm }{\sqrt{2~\delta}~c_0~\rho}
~(\tvert\mu\tvert_{\sigma_{n-1}})^2
\leq
1+\frac{\Cm ~c_0~\rho}{\sqrt{2~\delta}}
~\Bigl(\frac{c_{n-1}}{c_0}\Bigr)^2\\
&\leq
1+\xi~\Bigl(\frac{c_{n-1}}{c_0}\Bigr)^2
\equiv\frac{c_{n}}{c_0}~,
\label{eqn:defnnnn}
\end{equs}
for all $4\leq n\leq n_0$. Let now $\tilde{c}_{n}=\frac{c_n}{c_0}$ for
$n\geq2$. We can write (\ref{eqn:defdeux})--(\ref{eqn:defnnnn}) as
$\tilde{c}_{n+1}=1+\xi~\tilde{c}_n^2$ for $n\geq2$. Furthermore, since
$\frac{c_1}{c_0}\leq1$, if we set $\tilde{c}_1=1$, we will also get an
upper bound for $\|\mu\|_{\sigma}$. And now, since $\xi<\frac{1}{4}$, the
(infinite) sequence $\tilde{c}_{n+1}=1+\xi~\tilde{c}_n^2$, $\tilde{c}_1=1$,
is increasing and satisfies
$\tilde{c}_n\leq{\displaystyle\lim_{n\to\infty}}\tilde{c}_n=\tilde{c}_{\infty}
\equiv\frac{1-\sqrt{1-4\xi}}{2\xi}$, hence
\begin{equs}
\tvert\mu\tvert_{\sigma}
&\leq
\tilde{c}_{\infty}~c_0~\rho
\\&=\left(\frac{1-\sqrt{1-4\xi}}{2\xi}\right)~
\Bigl(
\rho+\|\mu_0\|_{\sigma}
+4~\epsilon^2~\tvert {\cal L}_{v}^{-1}~g'\tvert_{\L^2}
+\epsilon^2~\tvertb\frac{g'}{\Lepsilon }\tvertb_{{\cal W},\sigma}
\Bigr)
~.
\label{eqn:truebound}
\end{equs}
This completes the proof.
\end{proof}

\subsection{Existence and unicity of the solution of the phase equation}\label{sec:nonlinear}
Let $\tilde{\mu}\in{\cal W}_{\sigma}$ and $\mu_0\in{\cal
B}_{0,\sigma}(c_{\mu_0}~\rho)\subset{\cal W}_{0,\sigma}$. We consider the
equation
\begin{equs}   
\partial_t f=-{\cal
L}_{\epsilon}f-f~f'+\epsilon^2~F(\tilde{\mu})'~,~~~~f(x,0)=\mu_0(x)~.
\label{eqn:uneeq}
\end{equs}
By Theorem \ref{thm:weirdball}, $f$ exists if $\|\mu_0\|_{\sigma}
+\tvert{\cal L}_v^{-1}~F(\tilde{\mu})'\tvert_{\L^2} +\tvert
\Lepsilon^{-1}~F(\tilde{\mu})'\tvert_{{\cal W},\sigma}<\infty$, in which
case, we define the map $(\tilde{\mu},\mu_0)\mapsto{\cal
F}(\tilde{\mu},\mu_0)$, by ${\cal F}(\tilde{\mu},\mu_0)\equiv f$. We will
show that for fixed $\mu_0$, $\tilde{\mu}\mapsto{\cal
F}(\tilde{\mu},\mu_0)$ is a contraction in the ball ${\cal
B}_{\sigma}(c_{\mu}~\rho)$ if the following condition holds.
\begin{condition}\label{cond:condonF}
There exists a constant $\lambda_1<1$ such that for all
$c_{\mu}>\frac{c_{\mu_0}+1}{1-\lambda_1}$, there exists a constant $\epsilon_0$ such
that for all $\epsilon\leq\epsilon_0$ and for all $\mu_1,\mu_2\in{\cal
B}_{\sigma}(c_{\mu}~\rho)$ the following bounds hold
\begin{equs}
4~\epsilon^2~\tvert {\cal L}_{v}^{-1}~F(\mu_i)'\tvert_{\L^2}
+\epsilon^2~
\tvertb\frac{F(\mu_i)'}{\Lepsilon }\tvertb_{{\cal W},\sigma}
&\leq 
\lambda_1~c_{\mu}~\rho~,
\label{eqn:lundef}
\\
\epsilon^2~\tvert{\cal L}_{v}^{-1}~\bigl(F(\mu_1)-F(\mu_2)\bigr)'\tvert_{\L^2}
+\epsilon^2~
\tvertb\frac{F(\mu_1)'-F(\mu_2)'}{\Lepsilon }\tvertb_{{\cal W},\sigma}
&\leq
\lambda_1~
\tvert\mu_1-\mu_2\tvert_{\sigma}~,
\label{eqn:lundefdiff}\\
\epsilon^4~\tvert r_2(\mu_i)\tvert_{\L^2}+
\epsilon^2~\tvert {\cal L}_{v}^{-1}~F(\mu_i)'\tvert_{\L^2}
&\leq 
\Bigl(\frac{\epsilon}{\epsilon_0}\Bigr)^2~c_{\mu}~\rho~,
\label{eqn:lundefw}
\end{equs}
where ${\cal L}_v$ is defined in (\ref{eqn:lmudef}).
\end{condition}
We prove that this condition holds in Section \ref{sec:proprdeux}. The
proof requires bounds on $s$ and $r_2$. At this point, we note that if
$s=s_1(\mu)$, or equivalently $r_2=0$, we have $F(\mu)=F_0(s_1(\mu),\mu)$,
and that (see Appendix \ref{app:thefis} or the beginning of Section
\ref{sec:proprdeux}) we can satisfy Condition
\ref{cond:condonF} for any $\lambda_1<1$ and
$\epsilon_0=c_{\epsilon}~\delta^{-5/4}~\rho^{-1/2}$ if $c_{\epsilon}$ is
sufficiently small (depending on $\lambda_1$). To apply Theorem
\ref{thm:weirdball}, we need
$\xi=\frac{\Cm ~c_{0}~\rho}{\sqrt{2~\delta}}<\frac{1}{4}$, and from
(\ref{eqn:lundef}), we have $c_0<c_{\mu}$, hence we can satisfy
$\xi<\frac{1}{4}$ by choosing $\delta=c_{\delta}~\rho^2$ for some constant
$c_{\delta}$. This implies also that we should take (at least)
$\epsilon_0=c_{\epsilon}~\rho^{-m_{\epsilon}}$ with $m_{\epsilon}\geq3$.

We then have the following Proposition
\begin{proposition}\label{prop:contraun}
Let $c_{\mu}>\frac{1+c_{\mu_0}}{1-\lambda_1}$, and assume that Condition
\ref{cond:condonF} holds with $\epsilon_0$ sufficiently small. Then there
exists a constant $c_{\delta}$ such that if $\delta=c_{\delta}~\rho^2$ and
$\epsilon\leq\epsilon_0$, then
\begin{equs}
\tvert{\cal F}(\tilde{\mu}_i,\mu_0)\tvert_{\sigma}&< c_{\mu}~\rho
\label{eqn:concor}
\end{equs}
for all $\mu_0\in{\cal B}_{0,\sigma}(c_{\mu_0}~\rho)$.
\end{proposition}
\begin{proof}   
Let 
\begin{equs}
c_0(\tilde{\mu})&=1+c_{\mu_0}
+4~\epsilon^2~\frac{\tvert{\cal L}_{v}^{-1}~F(\tilde{\mu})'\tvert_{\L^2}}{\rho}
+\frac{\epsilon^2}{\rho}~
\tvertb\frac{F(\tilde{\mu})'}{\Lepsilon }\tvertb_{{\cal W},\sigma}~,\\
\xi(\tilde{\mu})&=\frac{\Cm~c_0(\tilde{\mu})~\rho}{\sqrt{2~\delta}}~.
\end{equs}
for all $\mu\in{\cal B}_{\sigma}(c_{\mu}~\rho)$ and $\mu_0\in{\cal
B}_{0,\sigma}(c_{\mu_0}~\rho)$, we have $c_0(\tilde{\mu})<\lambda~c_{\mu}$
with $\lambda=\lambda_1+\frac{1+c_{\mu_0}}{c_{\mu}}<1$. Choosing
\begin{equs}
c_{\delta}>\frac{\Cm^2~c_{\mu}^2}{2}\max\Bigl(4~,~\frac{1}{1-\lambda^2}\Bigr)^2~,
\end{equs}
we have $\xi(\tilde{\mu})<\frac{1}{4}$ and 
$\left(\frac{1-\sqrt{1-4\xi(\tilde{\mu})}}{2\xi(\tilde{\mu})}\right)\lambda<1$, so that
by Theorem \ref{thm:weirdball}, we have
\begin{equs}
\tvert{\cal F}(\tilde{\mu},\mu_0)\tvert_{\sigma}\leq
\left(\frac{1-\sqrt{1-4\xi(\tilde{\mu})}}{2\xi(\tilde{\mu})}\right)~c_0(\tilde{\mu})~\rho
<c_{\mu}~\rho~.
\end{equs}
This completes the proof.
\end{proof}

\begin{proposition}\label{prop:contradeux}
Let $c_{\mu}$, $c_{\delta}$ and $\epsilon_0$ be given by Proposition
\ref{prop:contraun}, and assume that for all
$\tilde{\mu}_1,\tilde{\mu}_2\in{\cal B}_{\sigma}(c_{\mu}~\rho)$ the following bound
holds
\begin{equs}
\tvert{\cal F}(\tilde{\mu}_i,\mu_0)\tvert_{\sigma}&< c_{\mu}~\rho
\label{eqn:concorrap}
\end{equs}
for all $\mu_0\in{\cal B}_{0,\sigma}(c_{\mu}~\rho)$.
Then there exists a time $t_0$ such that
\begin{equs}
\sup_{0\leq t\leq t_0}~
\|{\cal F}(\tilde{\mu}_1,\mu_{0})(\cdot,t)
 -{\cal F}(\tilde{\mu}_2,\mu_{0})(\cdot,t)\|_{\sigma}
&<\sup_{0\leq t\leq t_0}
\|\tilde{\mu}_1(\cdot,t)-\tilde{\mu}_2(\cdot,t)\|_{\sigma}~.
\label{eqn:difftzero}
\end{equs}
\end{proposition}
\begin{proof}
The proof of (\ref{eqn:difftzero}), being very similar to the estimates leading
to (\ref{eqn:concor}), can be found in Appendix \ref{app:contra}. Note that
here we only asked for $\mu_0\in{\cal B}_{0,\sigma}(c_{\mu}~\rho)$ and not for
$\mu_0\in{\cal B}_{0,\sigma}(c_{\mu_0}~\rho)$.
\end{proof}
We now deduce from Propositions \ref{prop:contraun} and
\ref{prop:contradeux} existence, unicity, and estimates
for the solution of the phase equation.
\begin{theorem}\label{thm:principal}
Let $c_{\mu}$, $c_{\delta}$ and $\epsilon_0$ be given by Proposition
\ref{prop:contraun}. Then for all $T\geq 0$, the solution $\mu_{\star}$ of
\begin{equs}   		     
\partial_t\mu_{\star}=-\Lepsilon ~\mu_{\star}-\mu_{\star}~\mu_{\star}'+\epsilon^2~F(\mu_{\star})'~,
~~~~\mu(x,0)=\mu_0(x)
\label{eqn:complete}
\end{equs}
exists for all $0\leq t\leq T$ and satisfies
\begin{equs}
\sup_{0\leq t\leq
T}\|\mu_{\star}(\cdot,t)\|_{\sigma}\leq c_{\mu}~\rho~,
\label{eqn:bornestar}
\end{equs}
for all $\mu_0\in{\cal B}_{0,\sigma}(c_{\mu_0}~\rho)$.
\end{theorem}
\begin{proof}
Let ${\cal F}(\tilde{\mu},\mu_0)$ be the solution of
\begin{equs}   
\partial_t\mu=-\Lepsilon ~\mu-\mu~\mu'+\epsilon^2~F(\tilde{\mu})'~,~~~~\mu(x,0)=\mu_0(x)~.
\end{equs}
By Proposition \ref{prop:contraun}, we know that 
$\tvert{\cal F}(\tilde{\mu})\tvert_{\sigma}< c_{\mu}~\rho$ if
$\tvert\tilde{\mu}\tvert_{\sigma}\leq c_{\mu}~\rho$. Hence, we can apply 
Proposition \ref{prop:contradeux} and get that $\tilde{\mu}\mapsto{\cal
F}(\tilde{\mu},\mu_0)$ is a contraction for $0\leq t\leq t_0$ in the ball
of radius $c_{\mu}~\rho$. Thus
$\tilde{\mu}\mapsto{\cal F}(\tilde{\mu},\mu_0)$ has a unique fixed point
$\mu_{\star}$ in that ball. By easy arguments (see e.g. \cite{nswake}),
this fixed point is the unique {\em strong} solution of
(\ref{eqn:complete}) for $0\leq t\leq t_0$. Furthermore, since the image of
$\tilde{\mu}\mapsto{\cal F}(\tilde{\mu},\mu_0)$ is in a ball of radius
$c_{\mu}~\rho$, $\mu_{\star}$ satisfies (\ref{eqn:bornestar}) with $T=t_0$.

We can now show inductively that $\mu_{\star}$ exists for all $t\geq0$ and
satisfies (\ref{eqn:bornestar}) for all $T\geq0$. Define $t_n=(n+1)t_0$
for $n\geq1$, and suppose that $\mu_{\star}$ exists on $0\leq t\leq
t_{n-1}$ and satisfies (\ref{eqn:bornestar}) with $T=t_{n-1}$. By
Proposition \ref{prop:contraun}, we know that for $t_{n-1}\leq t\leq
t_{n}$, the solution ${\cal F}(\tilde{\mu},\mu_{\star}(\cdot,t_{n-1}))$ of
\begin{equs}   
\partial_t\mu=-\Lepsilon ~\mu-\mu~\mu'+\epsilon^2~F(\tilde{\mu})'~,~~~~\mu(x,t_0)=\mu_{\star}(x,t_{n-1})
\label{eqn:phasedecale}
\end{equs}
is in a ball of size $c_{\mu}~\rho$ if $\tilde{\mu}$ is in a ball of size
$c_{\mu}~\rho$ for $t_{n-1}\leq t\leq t_{n}$, because it is the {\em
continuation} of a solution of (\ref{eqn:uneeq}), beginning with $\mu_0$ in
$t=0$, with $\tilde{\mu}(x,t)=\mu_{\star}(x,t)$ for $0\leq t\leq t_{n-1}$.
Shifting the origin of time to $t_{n-1}$ and replacing $\mu_0$ by
$\mu_{\star}(\cdot,t_{n-1})$, we see that the conditions of Proposition
\ref{prop:contradeux} are satisfied, hence 
$\tilde{\mu}\mapsto{\cal F}(\tilde{\mu},\mu_{\star}(\cdot,t_{n-1}))$ is a
contraction for $t_{n-1}\leq t\leq t_{n}$. As above, this implies that
there exists an unique fixed point $\mu_{\star}$ which is the unique {\em
strong} solution of \ref{eqn:complete} on $0\leq t\leq t_{n}$ and satisfies
(\ref{eqn:bornestar}) with $T=t_n$.
\end{proof}
\subsection{Consequences}\label{sec:consequences}
Up to now, we did not use (\ref{eqn:lundefw}) of Condition
\ref{cond:condonF}. This inequality has two important consequences. The
first one is that $s$ (if it exists) and $\eta$ are related by
$s=-\frac{1}{8}~\eta''=-\frac{1}{8}~\mu'$ up to corrections of order
$\epsilon^2$ (see Theorem \ref{thm:graph} below), and the second one
concerns the relation with the Kuramoto--Sivashinsky equation (see Theorem
\ref{thm:comparison} below). Once these theorems are proved, we will only have
to prove the bound on $\hat{s}$ to complete the proof of Theorems
\ref{thm:existeunique} and \ref{thm:properties}.

\begin{theorem}\label{thm:graph}
There exists a constant $c_{\epsilon}>0$ such that if Condition
\ref{cond:condonF} holds with $\epsilon_0\leq c_{\epsilon}~\rho^{-3}$, and
$\delta$ is given by Proposition \ref{prop:contraun}, then
\begin{equs}
\frac{\tvert s
+\frac{1}{8}~G~\mu'
+\frac{\epsilon^2}{32}~G~(\mu)^2
\tvert_{\L^2}}{c_{\mu}~\rho}
\leq\Bigl(\frac{\epsilon}{\epsilon_0}\Bigr)^2
\end{equs}
if $\epsilon\leq\epsilon_0$.
\end{theorem}
\begin{proof}
The proof is very simple. We use that
$s+\frac{1}{32}\Gepsilon ~(4\mu'+\epsilon^2~\mu^2)=
\epsilon^4~\Gepsilon ~ r_2$, and that by assumption (see
(\ref{eqn:lundefw})), we have
$\epsilon^4~\|r_2\|_{\L^2}\leq\bigl(\frac{\epsilon}{\epsilon_0}\bigr)^2~c_{
\mu}~\rho$.
\end{proof}
We next show that the solution $\mu_c$ of the Kuramoto--Sivashinsky
equation (in derivative form) captures the dynamics of the (derivative of
the) phase for short times (then $-\frac{1}{8}~\mu_c'$ captures the
dynamics of the amplitude by Theorem \ref{thm:graph}). To state the result,
we introduce the operator ${\cal L}_{\mu,c}=\partial_x^4+\partial_x^2$. We
have the following Theorem.
\begin{theorem}\label{thm:comparison}
Let $\mu$ and $\mu_c$ be the solutions of 
\begin{equs}[3]
\partial_t\mu  &=-\Lepsilon~\mu-\mu\mu'+\epsilon^2~F(\mu)'~,~~~~
&\mu(x,0)&=\mu_0(x)~,\\
\partial_t\mu_c&=-{\cal L}_{\mu,c}~\mu_c       -\mu_c\mu_c'
~,~~~~
&\mu_c(x,0)&=\mu_0(x)~,
\end{equs}
There exist constants $c_{\epsilon}$ and $c_t$ such that if Condition
\ref{cond:condonF} holds with $\epsilon_0\leq c_{\epsilon}~\rho^{-4}$, then
\begin{equs}
\sup_{0\leq t\leq t_0}
\frac{\|\mu(\cdot,t)-\mu_c(\cdot,t)\|_{\L^2}}{c_{\mu}~\rho}
\leq\frac{\epsilon}{\epsilon_0}~,
\label{eqn:comparison}
\end{equs}
for all $t_0\leq c_t~\rho^{-4}$ and for all
$\epsilon\leq\epsilon_0$.
\end{theorem}
This theorem implies directly Theorem \ref{thm:properties}.

\noindent\begin{proof}[Proof of Theorem \ref{thm:comparison}]
Let $\mu_{\pm}=\mu\pm\mu_{c}$ and ${\cal L}_{\pm}={\cal L}_{\mu,c}\pm{\cal
L}_{\mu}$. Note that $\mu_c$ exists and satisfies
$\tvert\mu_c\tvert_{\sigma}\leq c_{\mu}~\rho$. Furthermore, we have
\begin{equs}
\partial_{t}\mu_{-}=-\frac{{\cal L}_{+}}{2}\mu_{-}
+{\cal L}_{-}\mu_{+}-\frac{1}{2}(\mu_{+}\mu_{-})'
+\epsilon^2~F(\mu)'
~.
\end{equs}
Multiplying this equation by $\mu_{-}$ and integrating over $[-L/2,L/2]$,
we get
\begin{equs}
\frac{1}{2}~\partial_t~(\mu_{-},\mu_{-})
=
-\frac{1}{2}(\mu_{-},{\cal L}_{+}\mu_{-})
+(\mu_{-},{\cal L}_{-}\mu_{+})
-\frac{1}{4}(\mu_{-},\mu_{+}'\mu_{-})
+\epsilon^2(\mu_{-},F(\mu)')
~.
\end{equs}
Next, we use the Cauchy--Schwartz inequality, the identity $({\cal L}_{+}-{\cal
L}_{-})=2\Lepsilon $ and ${\cal L}_{-}\geq0$, (this follows from ${\cal
L}_{\mu,c}(k)=k^4-k^2$), to get
\begin{equs}
\partial_t(\mu_{-},\mu_{-})&\leq
-(\mu_{-},(2\Lepsilon -{\cal L}_{v}^2)\mu_{-})+
(\mu_{+},{\cal L}_{-}\mu_{+})
\\&\phantom{=}~+\frac{\|\mu_{+}'\|_{\L^{\infty}}}{2}~(\mu_{-},\mu_{-})
+\epsilon^4~\|{\cal L}_{v}^{-1}~F(\mu)'\|_{\L^2}^2
\\
&\leq
(1+\frac{\|\mu_{+}'\|_{\L^{\infty}}}{2})(\mu_{-},\mu_{-})
+(\mu_{+},{\cal L}_{-}\mu_{+})
+\epsilon^4~\|{\cal L}_{v}^{-1}~F(\mu)'\|_{\L^2}^2
~.
\end{equs}
Since the Fourier coefficients of ${\cal L}_{-}$ satisfy $({\cal
L}_{-})_n\leq\epsilon^2~(qn)^6$ we get, using Lemma
\ref{lem:ldliestime}
\begin{equs}
(\mu_{+},{\cal L}_{-}\mu_{+})
\leq\epsilon^2
~\|\mu_{+}'''\|_{\L^2}^2\leq
C~\epsilon^2~\delta^6~\|\mu_{+}\|_{\sigma}^2
\leq C~\epsilon^2~\delta^6~(c_{\mu}~\rho)^2
~.
\end{equs}
Let $\zeta=1+\Cinfty ~c_{\mu}~\rho~\delta^{3/2}=c_{\zeta}~\rho^4$. We have
\begin{equs}
\partial_t(\mu_{-},\mu_{-})&\leq
\zeta~(\mu_{-},\mu_{-})
+C~\epsilon^2~\delta^6~(c_{\mu}~\rho)^2
+\epsilon^4~\|{\cal L}_{v}^{-1}~F(\mu)'\|_{\L^2}^2\\
&\leq
\zeta~(\mu_{-},\mu_{-})
+C~\epsilon^2~\delta^6~(c_{\mu}~\rho)^2
+\Bigl(
\frac{\epsilon}{\epsilon_0}
\Bigr)^4~
(c_{\mu}~\rho)^2~,
\end{equs}
where we used (\ref{eqn:lundefw}). Let $\epsilon_0\leq
c_{\epsilon}~\rho^{-4}$, and $t_0\leq c_t~\rho^{-4}$. We have
\begin{equs}
\sup_{0\leq t\leq t_0}
\frac{\|\mu(\cdot,t)-\mu_{c}(\cdot,t)\|_{\L^2}}{c_{\mu}~\rho}
&\leq
\Bigl(
\frac{\epsilon}{\epsilon_0}
\Bigr)~
\Bigl(\frac{2~\Cm~c_{\delta}^3~c_{\epsilon}}{\sqrt{c_{\zeta}}}
+
\frac{1}{\sqrt{c_{\zeta}}~\rho^2}
\Bigr)~
\sqrt{\ed^{c_{\zeta}~c_{t}}-1}\leq
\frac{\epsilon}{\epsilon_0}~,
\label{eqn:commonun}
\end{equs}
if $c_{\epsilon}$ and $c_t$ are sufficiently small.
\end{proof}

\section{The amplitude equation}\label{sec:ampli}
In (\ref{eqn:infinalforms}), we showed that in terms of the amplitude
$s$ and the (derivative of the) phase $\mu$ of the perturbation of
$\ed^{i~\phi_0-i\beta t}$, the `amplitude' part of the Complex Ginzburg
Landau equation becomes
\begin{equs}
\partial_t s=
-\frac{\chi}{\epsilon^4}
\bigl(s-\frac{\epsilon^2}{2}s''\bigr)
+\frac{\chi}{\epsilon^4}~r_1(\mu)
-{\textstyle\frac{\alpha^2~\chi}{8}}(2s'\mu+s\mu')+F_3(s,\mu)~,
\label{eqn:ampliscal}
\end{equs}
with $s(x,0)=s_0(x)$ and
\begin{equs}
r_1(\mu)&=-\frac{1}{32}(4\mu'+\epsilon^2~\mu^2)~,\\
F_3(s,\mu)&=
-\alpha^2~\chi~\left({\textstyle\frac{3}{2}}~s^2
+\epsilon^2~{\textstyle\frac{1}{32}}~s~\mu^2
+{\textstyle\frac{\alpha^2}{2}}~\epsilon^4~s^3\right)~.
\end{equs}
We now show that for given $\mu$ with
$\tvert\mu\tvert_{\sigma}$ not too large, the solution of 
(\ref{eqn:ampliscal}) is determined by a well defined Lipschitz map of $\mu$.
We will use the definitions and properties of the norms
$\tvert\cdot\tvert_{\sigma}$ of Subsections \ref{subsec:defs} and
\ref{sec:nonlinearities}. We proceed as we did for the phase equation,
that is, we first show $\L^2$ estimates, and then $\|\cdot\|_{\sigma}$
estimates. Equation (\ref{eqn:ampliscal}) suggests that we study
\begin{equs}
\partial_t s=
-\frac{\chi}{\epsilon^4}
\bigl(s-\frac{\epsilon^2}{2}s''\bigr)
-{\textstyle\frac{\alpha^2~\chi}{8}}(2s'\nu+s\nu')+f~,~~~~~s(x,0)=s_0(x)~,
\label{eqn:swithin}
\end{equs}
for given $s_0$, $\nu$ and $f$. If $\tvert\nu\tvert_{\L^{\infty}}+\tvert\nu'\tvert_{\L^{\infty}}$
is sufficiently small, this equation is a linear (in $s$) inhomogeneous
damped heat equation, hence the local existence and unicity of the solution in 
$\L^2$ is known by classical arguments (see e.g. \cite{Temam}). Furthermore,
the solution satisfies the
\begin{lemma}\label{lem:ldeuxfors}
If $s$ is the solution of (\ref{eqn:swithin}) then
\begin{equs}
\tvert s\tvert_{\L^2}\leq\|s_0\|_{\L^2}
+\frac{\epsilon^4}{\chi}~\tvert f\tvert_{\L^2}~.
\label{eqn:ldeuxfors}
\end{equs}
\end{lemma}
\begin{proof}
Multiplying (\ref{eqn:swithin}) with $s$, integrating over one period,
using Young's inequality, and using that $\int s(2s'\nu+s\nu')=\int(s^2\nu)'=0$ 
because $s$ and $\nu$ are periodic, we get
\begin{equs}
\partial_t\int s^2&\leq-\frac{2\chi}{\epsilon^4}\int s^2
+\int s~f
\leq-\frac{\chi}{\epsilon^4}\int s^2+\frac{\epsilon^4}{\chi}\int f^2~,
\end{equs}
from which (\ref{eqn:ldeuxfors}) follows immediately.
\end{proof}
\begin{proposition}\label{prop:bs}
If $s$ is the solution of (\ref{eqn:swithin}), then
\begin{equs}
\tvert s\tvert_{\sigma-1}\leq2\|s_0\|_{\sigma-1}
+\frac{2~\epsilon^4}{\chi}\tvert f\tvert_{\sigma-1}
\label{eqn:ssmundelta}
\end{equs}
for all $\nu$ satisfying
$\epsilon^3~\sqrt{\delta}~\alpha^2~\Cm~(4+\epsilon\delta)~\tvert\nu\tvert_{\sigma}\leq4$.
\end{proposition}
\begin{proof}
The idea is again to use Duhamel's representation formula for $s$. Let
${\cal L}$ be the operator with symbol
\begin{equs}
{\cal L}(k)=\frac{\chi}{\epsilon^4}\Bigl(1+\frac{\epsilon^2~k^2}{2}\Bigr)~,
\end{equs}
then $s$ satisfies
\begin{equs}
s(x,t)&=\ed^{-{\cal L}t}s_0(x)+{\cal T}(s,f)(s,t)~,
\end{equs}
with 
\newcommand{\tzero}{\tau}
\begin{equs}
{\cal T}(s,f)(s,t)&=-
\int_{0}^{t}{\rm d}\tzero~\ed^{-{\cal L}(t-\tzero)}~
\Bigl({\textstyle\frac{\alpha^2~\chi}{8}}
\bigl(2~\nu(x,\tzero)~\partial_x
s(x,\tzero)+s(x,\tzero)~\partial_x\nu(x,\tzero)\bigr)-f(x,\tzero)\Bigr)\\
&=
\int_{0}^{t}{\rm d}\tzero~\ed^{-{\cal L}(t-\tzero)}~
\Bigl({\textstyle\frac{\alpha^2~\chi}{8}}
\bigl(2~\partial_x\bigl(s(x,\tzero)~\nu(x,\tzero)\bigr)-s(x,\tzero)~
\partial_x\nu(x,\tzero)\bigr)-f(x,t_0)\Bigr)~.
\end{equs}
Using the inequality
\begin{equs}
\left|\!\left|\!\left|
\int_{0}^{t}{\rm d}\tzero~\ed^{-{\cal L}(t-\tzero)}~f(x,\tzero)
\right|\!\right|\!\right|_{{\cal W},\sigma}
\leq \frac{\epsilon^4}{\chi}\tvert \Gepsilon ~ f\tvert_{{\cal W},\sigma}~,
\end{equs}
we get for any $\sigma'\leq\sigma-1$, 
\begin{equs}
\tvert s\tvert_{{\cal W},\sigma'}&\leq
\|s_0\|_{{\cal W},\sigma-1}
+\frac{\epsilon^4}{\chi}\tvert f\tvert_{{\cal W},\sigma-1}
+\frac{\epsilon^4~\alpha^2}{4}\tvert \Gepsilon ~(s~\nu)'\tvert_{{\cal W},\sigma'}
+\frac{\epsilon^4~\alpha^2}{8}\tvert \Gepsilon ~ (s~\nu')\tvert_{{\cal W},\sigma'}
\label{eqn:preescalier}
\\
&\leq
\|s_0\|_{{\cal W},\sigma-1}
+\frac{\epsilon^4}{\chi}\tvert f\tvert_{{\cal W},\sigma-1}
+\frac{\epsilon^2~\alpha^2}{2}\tvert s~\nu\tvert_{{\cal W},\sigma'-1}
+\frac{\epsilon^3~\alpha^2}{4}\tvert s~\nu'\tvert_{{\cal W},\sigma'-1}~.
\label{eqn:escalier}
\end{equs}
We now use (\ref{eqn:escalier}) inductively in $\sigma'$ to show that
$\tvert s\tvert_{{\cal W},\sigma-1}$ is bounded. Then we will use 
(\ref{eqn:preescalier}) to show that $\tvert s\tvert_{{\cal
W},\sigma-1}$ satisfies the bound (\ref{eqn:ssmundelta}).

Using (\ref{eqn:escalier}) with $\sigma'=1$, we get 
$\tvert s\tvert_{{\cal W},1}<\infty$, and
$\tvert s\tvert_{1}<\infty$, because
\begin{equs}
\tvert s~\nu\tvert_{{\cal W},0}\leq
\Cm ~\sqrt{\delta}\tvert s\tvert_{\L^2}~\tvert\nu\tvert_{\L^2}
\leq
\Cm ~\sqrt{\delta}\tvert s\tvert_{\L^2}~\tvert\nu\tvert_{\sigma}
~,\\
\tvert s~\nu'\tvert_{{\cal W},0}\leq
\Cm ~\sqrt{\delta}\tvert s\tvert_{\L^2}~\tvert\nu'\tvert_{\L^2}
\leq
C~\delta^{3/2}\tvert s\tvert_{\L^2}~\tvert\nu\tvert_{\sigma}~.
\end{equs}
Then, using (\ref{eqn:escalier}) with $\sigma'=\frac{3}{2}$, we get
$\tvert s\tvert_{{\cal W},\frac{3}{2}}<\infty$ because
\begin{equs}
\tvert s\tvert_{{\cal W},\frac{3}{2}}&\leq
\|s_0\|_{{\cal W},\sigma-1}
+\frac{\epsilon^4}{\chi}\tvert f\tvert_{{\cal W},\sigma-1}
+\frac{\epsilon^2~\alpha^2}{2}\tvert s~\nu\tvert_{{\cal W},1}
+\frac{\epsilon^3~\alpha^2}{4}\tvert s~\nu'\tvert_{{\cal W},1}\\
&\leq
\|s_0\|_{{\cal W},\sigma-1}
+\frac{\epsilon^4}{\chi}\tvert f\tvert_{{\cal W},\sigma-1}
+\frac{\epsilon^2~\alpha^2~\Cm~\delta^{3/2}}{2}
\Bigl(\frac{1}{\delta}+\frac{1}{2}\Bigr)
\tvert s\tvert_{1}~\tvert\nu\tvert_{\sigma}~.
\end{equs}
Then, for any $\sigma'\geq\frac{5}{2}$, we get
\begin{equs}
\tvert s\tvert_{{\cal W},\sigma'}&\leq
\|s_0\|_{{\cal W},\sigma-1}
+\frac{\epsilon^4}{\chi}\tvert f\tvert_{{\cal W},\sigma-1}
+\frac{\epsilon^2~\alpha^2}{2}\tvert s~\nu\tvert_{{\cal W},\sigma'-1}
+\frac{\epsilon^3~\alpha^2}{4}\tvert s~\nu'\tvert_{{\cal W},\sigma'-1}\\
&\leq
\|s_0\|_{{\cal W},\sigma-1}
+\frac{\epsilon^4}{\chi}\tvert f\tvert_{{\cal W},\sigma-1}
+\frac{\epsilon^3~\alpha^2~\Cm~\delta^{3/2}}{2}
\Bigl(\frac{1}{\delta}+\frac{1}{2}\Bigr)
\tvert s\tvert_{\sigma'-1}~\tvert\nu\tvert_{\sigma}~.
\end{equs}
Using this last inequality with $\sigma'=\frac{5}{2},\frac{7}{2},\ldots$ until 
we reach $\sigma-1$ shows that $\tvert s\tvert_{{\cal W},\sigma-1}<\infty$.
Then, from (\ref{eqn:preescalier}) and Lemma \ref{lem:ldeuxfors}, we also get
\begin{equs}
\tvert s\tvert_{\sigma-1}&\leq
\|s_0\|_{\sigma-1}
+\frac{\epsilon^4}{\chi}\tvert f\tvert_{\sigma-1}
+\frac{\epsilon^4~\alpha^2}{4}\tvert \Gepsilon ~(s~\nu)'\tvert_{{\cal W},\sigma-1}
+\frac{\epsilon^4~\alpha^2}{8}\tvert \Gepsilon ~(s~\nu')\tvert_{{\cal W},\sigma-1}\\
&\leq
\|s_0\|_{\sigma-1}
+\frac{\epsilon^4}{\chi}\tvert f\tvert_{\sigma-1}
+\frac{\epsilon^3~\alpha^2}{2}\tvert s~\nu\tvert_{{\cal W},\sigma-1}
+\frac{\epsilon^4~\alpha^2}{8}\tvert s~\nu'\tvert_{{\cal W},\sigma-1}\\
&\leq
\|s_0\|_{\sigma-1}
+\frac{\epsilon^4}{\chi}\tvert f\tvert_{\sigma-1}
+\tvert s\tvert_{\sigma-1}
\frac{\epsilon^3~\alpha^2~\sqrt{\delta}~\Cm~(4+\epsilon\delta)~\tvert\nu\tvert_{\sigma}}{8}~.
\end{equs}
Since by hypothesis 
$(\epsilon^3~\alpha^2~\sqrt{\delta}~\Cm~(4+\epsilon\delta)~\tvert\nu\tvert_{\sigma})/8\leq1/2$,
the proof is completed.
\end{proof}
\begin{corollary}\label{cor:bds}
Let $s_1$, resp. $s_2$ be the solution of (\ref{eqn:swithin}) with $f=f_1$
resp. $f_2$, and assume that $s_1(x,0)=s_2(x,0)$. Then
\begin{equs}
\tvert s_1-s_2\tvert_{\sigma-1}\leq
\frac{2~\epsilon^4}{\chi}~\tvert f_1-f_2\tvert_{\sigma-1}~.
\label{eqn:ldeuxforsdiff}
\end{equs}
\end{corollary}
\begin{proof}
Since (\ref{eqn:swithin}) is linear in $s$, $s_1-s_2$ satisfies
(\ref{eqn:swithin}) with $s_0=0$ and $f\equiv f_1-f_2$.
\end{proof}
\begin{corollary}\label{cor:forlipchi}
Let $s(\nu)$ be the solution of (\ref{eqn:swithin}) with $f=f(\nu)$
satisfying $\tvert f(\nu)\tvert_{\sigma-1}<\infty$. Then
\begin{equs}
\tvert s(\nu_1)-s(\nu_2)\tvert_{\sigma-1}\leq
\frac{2~\epsilon^4}{\chi}~\tvert f(\nu_1)-f(\nu_2)\tvert_{\sigma-1}~.
\label{eqn:ldeuxforslip}
\end{equs}
\end{corollary}
\begin{proof}   
Since (\ref{eqn:swithin}) is linear in $s$, $s(\nu_1)-s(\nu_2)$ satisfies
(\ref{eqn:swithin}) with $s_0=0$ and $f\equiv f(\nu_1)-f(\nu_2)$.
\end{proof}
We are now in position to prove that the solution of (\ref{eqn:ampliscal})
exists if $\epsilon_0$ is sufficiently small.
\begin{theorem}\label{thm:onr}
Let $c_{r_1}$ and $c_{\mu}$ be given by Proposition \ref{prop:run} and
Theorem \ref{thm:principal}, and $c_{s_0}>0$. Let $c_s>2(c_{r_1}+c_{s_0})$.
There exists a constant $c_\epsilon$ such that for all $\epsilon\leq
c_{\epsilon}~\delta^{-5/4}~\rho^{-1/2}$, for all
$\mu\in{\cal B}_{\sigma}(c_{\mu}~\rho)$ and for all
$s_0\in{\cal B}_{0,\sigma-1}(c_{s_0}~\delta~\rho)$, the solution
$s$ of
\begin{equs}
\partial_t s=
-\frac{\chi}{\epsilon^4}
\bigl(s-\frac{\epsilon^2}{2}s''\bigr)
+\frac{\chi}{\epsilon^4}~r_1(\mu)
-{\textstyle\frac{\alpha^2~\chi}{8}}(2s'\mu+s\mu')+F_3(s,\mu)~,~~~~~
s(x,0)=s_0(x)~,
\label{eqn:ampliscalrap}
\end{equs}
exists and is unique in ${\cal B}_{\sigma-1}(c_s~\delta~\rho)$. As such, it
defines the map $\mu\mapsto s(\mu)$, which satisfies
\begin{equs}
\tvert s(\mu_i)\tvert_{\sigma-1}&\leq c_s~\delta~\rho~,
\label{eqn:existesb}
\\
\tvert s(\mu_1)-s(\mu_2)\tvert_{\sigma-1}&\leq 
c_s~\delta~
\tvert\mu_1-\mu_2\tvert_{\sigma}~,
\label{eqn:existesbdiff}
\end{equs}
for all $\mu_i\in{\cal B}_{\sigma}(c_{\mu}~\rho)$.
\end{theorem}
\begin{proof}
Fixing $\tilde{s}\in{\cal W}_{\sigma-1}$, we consider the equation
\begin{equs}
\partial_t f=
-\frac{\chi}{\epsilon^4}
\bigl(f-\frac{\epsilon^2}{2}f''\bigr)
+\frac{\chi}{\epsilon^4}~r_1(\mu)
-{\textstyle\frac{\alpha^2~\chi}{8}}(2f'\mu+f\mu')+F_3(\tilde{s},\mu)~,~~~~~
f(x,0)=s_0(x)~.
\end{equs}
By Proposition \ref{prop:bs}, $f$ exists if $\|s_0\|_{\sigma-1} +\tvert
r_1(\mu)\|_{\sigma-1} +\tvert F_3(\tilde{s},\mu)\|_{\sigma-1}<\infty $, in
which case we define the map $\tilde{s}\mapsto T(\tilde{s},\mu)$ by
$T(\tilde{s},\mu)\equiv f$. To show that $s(\mu)$ exists, is unique and
satisfies (\ref{eqn:existesb}), we only have to show that if $\epsilon$ is
sufficiently small, $\tilde{s}\mapsto T(\tilde{s},\mu)$ is a contraction in
${\cal B}_{\sigma-1}(c_s~\delta~\rho)\subset{\cal W}_{\sigma-1}$. Using
Propositions \ref{prop:bs} and \ref{prop:run} and the assumption on $s_0$,
we have
\begin{equs}
\tvert T(s,\mu)\tvert_{\sigma-1}&\leq
2\tvert r_1(\mu)\tvert_{\sigma-1}+
2\| s_0\|_{\sigma-1}+
\frac{2~\epsilon^4}{\chi}~
\tvert F_3(s,\mu)\tvert_{\sigma-1}\\
&\leq
2~(c_{r_1}+c_{s_0})~\delta~\rho
+\frac{2~\epsilon^4}{\chi}~\tvert F_3(s,\mu)\tvert_{\sigma-1}~.
\label{eqn:formap}
\end{equs}
Similarly, using Corollary \ref{cor:bds}, we have
\begin{equs}
\tvert T(s_1,\mu)-T(s_2,\mu)\tvert_{\sigma-1}&\leq
\frac{2~\epsilon^4}{\chi}~
\tvert F_3(s_1,\mu)-F_3(s_2,\mu)\tvert_{\sigma-1}~.
\label{eqn:forcontr}
\end{equs}
By Proposition \ref{prop:onFtrois}, there exists a constant
$c_\epsilon$ such that for all $\epsilon\leq
c_{\epsilon}~\delta^{-5/4}~\rho^{-1/2}$, for all
$\mu\in{\cal B}_{\sigma}(c_{\mu}~\rho)$, and for all
$s_i\in{\cal B}_{\sigma-1}(c_{s}~\delta~\rho)$, we have
\begin{equs}
\frac{2~\epsilon^4}{\chi}~\tvert F_3(s_i,\mu)\tvert_{\sigma-1}
&<
\Bigl(
1-\frac{2(c_{r_1}+c_{s_0})}{c_s}
\Bigr)~c_s~\delta~\rho~,
\label{eqn:formapF}
\\
\frac{2~\epsilon^4}{\chi}~
\tvert F_3(s_1,\mu)-F_3(s_2,\mu)\tvert_{\sigma-1}
&\leq
\Bigl(
1-\frac{2(c_{r_1}+c_{s_0})}{c_s}
\Bigr)
\tvert s_1-s_2\tvert_{\sigma-1}~.
\label{eqn:forcontrF}
\end{equs}
From (\ref{eqn:formap}) and (\ref{eqn:formapF}), and since
$c_s>2(c_{r_1}+c_{s_0})$, we conclude that $s\mapsto
T(s,\mu)$ maps the ball ${\cal B}_{\sigma-1}(c_s~\delta~\rho)$ (strictly)
inside itself, whereas from (\ref{eqn:forcontr}) and (\ref{eqn:forcontrF}) we see
that it is a contraction in that ball. Hence, the map $s\mapsto T(s,\mu)$
has a unique fixed point $s^{\star}(\mu)$. This fixed point satisfies
(\ref{eqn:existesb}) and is a strong solution of (\ref{eqn:ampliscalrap})
(see also \cite{nswake}).

For (\ref{eqn:existesbdiff}), using Corollary \ref{cor:forlipchi}, we have
\begin{equs}
\tvert s^{\star}(\mu_1)-s^{\star}(\mu_2)\tvert_{\sigma-1}
&\leq 2
\tvert r_1(\mu_1)-r_1(\mu_2)\tvert_{\sigma-1}
\\&\leq\phantom{=}~
+\frac{2\epsilon^4}{\chi}~
\tvert F_3(s^{\star}(\mu_1),\mu_1)-F_3(s^{\star}(\mu_2),\mu_2)\tvert_{\sigma-1}~.
\end{equs}
Using that
\begin{equs}
F_3(s_1,\mu_1)-F_3(s_2,\mu_2)
&=
F_3(s_1,\mu_1)-F_3(s_1,\mu_2)
+
F_3(s_1,\mu_2)-F_3(s_2,\mu_2)
\end{equs}
and that by Proposition \ref{prop:onFtrois} we have
\begin{equs}
\frac{2\epsilon^4}{\chi}~
\tvert F_3(s_1,\mu_1)-F_3(s_1,\mu_2)\tvert_{\sigma-1}&\leq
2~c_{s_0}~\delta~\tvert\mu_1-\mu_2\tvert_{\sigma}
~,\\
\frac{2\epsilon^4}{\chi}~
\tvert F_3(s_1,\mu_2)-F_3(s_2,\mu_2)\tvert_{\sigma-1}&\leq
\Bigl(
1-\frac{2(c_{r_1}+c_{s_0})}{c_s}
\Bigr)~\tvert s_1-s_2\tvert_{\sigma-1}~,
\end{equs}
we conclude, using Proposition \ref{prop:run}, that
\begin{equs}
\tvert s^{\star}(\mu_1)-s^{\star}(\mu_2)\tvert_{\sigma-1}
&\leq 2~(c_{r_1}+c_{s_0})~\delta~\tvert\mu_1-\mu_2\tvert_{\sigma}
\\&\phantom{=}~+\frac{2\epsilon^4}{\chi}~
\tvert F_3(s^{\star}(\mu_1),\mu_1)-F_3(s^{\star}(\mu_2),\mu_2)\tvert_{\sigma-1}\\
&\leq 2~(c_{r_1}+c_{s_0})~\delta~\tvert\mu_1-\mu_2\tvert_{\sigma}
\\&\phantom{=}~+
\Bigl(
1-\frac{2(c_{r_1}+c_{s_0})}{c_s}
\Bigr)
~\tvert s^{\star}(\mu_1)-s^{\star}(\mu_2)\tvert_{\sigma-1}~.
\end{equs}
Since $c_s>2(c_{r_1}+c_{s_0})$, we have
\begin{equs}
\frac{2(c_{r_1}+c_{s_0})}{c_s}
\tvert s^{\star}(\mu_1)-s^{\star}(\mu_2)\tvert_{\sigma-1}
&\leq 
2(c_{r_1}+c_{s_0})
~\delta~\tvert\mu_1-\mu_2\tvert_{\sigma}~.
\end{equs}
Therefore, $\tilde{s}\mapsto T(\tilde{s},\mu)$ is a contraction.
\end{proof}
We define $r(\mu)=s(\mu)-\frac{\epsilon^2}{2}s(\mu)''$ and
$r_0=s_0-\frac{\epsilon^2}{2}s_0''$, and prove that $\mu\mapsto
r(\mu)$ satisfies essentially the same bounds as $\mu\mapsto s(\mu)$.
\begin{corollary}\label{cor:bonr}
Assume that $r_0\in{\cal B}_{0,\sigma-1}(c_{r_1}~\delta~\rho)$. Then
$\mu\mapsto r(\mu)$ satisfies
\begin{equs}
\tvert r(\mu)\tvert_{\sigma-1}&\leq 8~c_{r_1}~\delta~\rho~,
\label{eqn:existerb}
\\
\tvert r(\mu_1)-r(\mu_2)\tvert_{\sigma-1}&\leq 8~c_{r_1}~\delta~
\tvert\mu_1-\mu_2\tvert_{\sigma}~,
\label{eqn:existerbdiff}
\end{equs}
if the conditions of Theorem \ref{thm:onr} are satisfied.
\end{corollary}
\begin{proof}
The proof, being very similar to the ones of Proposition \ref{prop:bs}
and Theorem \ref{thm:onr} is outlined in Appendix \ref{app:amplitude}.
\end{proof}

\section{The Condition \ref{cond:condonF}, properties of $\mu\mapsto
F(\mu)$ and $\mu\mapsto r_2(\mu)$}\label{sec:proprdeux}
We recall that
\begin{equs}
F(\mu)=F_0(s(\mu),\mu)+\chi~{\cal L}_{\mu,r}~r_2(\mu)~,
\end{equs}
where
\begin{equs}
r_2(\mu)&=\frac{r(\mu)}{\epsilon^4}-\frac{r_1(\mu)}{\epsilon^4}~.
\label{eqn:encorerdeuxdiff}
\end{equs}
For Condition \ref{cond:condonF} to hold, we need to show that there exists
a constant $\lambda_1<1$ such that for all $\mu_1,\mu_2\in{\cal
B}_{\sigma}(c_{\mu}~\rho)$,
\begin{equs}
\epsilon^4~\tvert r_2(\mu_i)\tvert_{\L^2}+\epsilon^2~\tvert{\cal L}_{v}^{-1}~F(\mu_i)'\tvert_{\L^2}
&\leq 
\Bigl(\frac{\epsilon}{\epsilon_0}\Bigr)^2~c_{\mu}~\rho~,
\label{eqn:lundefrap}
\\
4~\epsilon^2~\tvert{\cal L}_{v}^{-1}~F(\mu_i)'\tvert_{\L^2}
+\epsilon^2~
\tvertb\frac{F(\mu_i)'}{\Lepsilon }\tvertb_{{\cal W},\sigma}
&\leq 
\lambda_1~c_{\mu}~\rho~,
\label{eqn:lundefwrap}
\\
\epsilon^2~\tvert{\cal L}_{v}^{-1}~\bigl(F(\mu_1)-F(\mu_2)\bigr)'\tvert_{\L^2}
+\epsilon^2~
\tvertb\frac{F(\mu_1)'-F(\mu_2)'}{\Lepsilon }\tvertb_{{\cal W},\sigma}
&\leq
\lambda_1~
\tvert\mu_1-\mu_2\tvert_{\sigma}~,
\label{eqn:lundefdiffrap}
\end{equs}
for all $\epsilon\leq\epsilon_0$ if $\epsilon_0$ is sufficiently small. If
$r_2=0$, these conditions can be satisfied if $\epsilon_0\leq
c_{\epsilon}~\rho^{-3}$ with $c_{\epsilon}$ sufficiently small. Namely,
from Theorem \ref{thm:lesfonctions}, Appendix \ref{app:thefis}, using also
$\|{\cal L}_{v}^{-1}~f'\|_{\L^2}\leq 2~\|f\|_{\L^2}$, we have
\begin{equs}
\epsilon^2~\tvert{\cal L}_{v}^{-1}~F_0(\mu_i)'\tvert_{\L^2}\leq
2~\epsilon^2~\tvert F_0(\mu_i)\tvert_{\L^2}
&\leq 
2~\epsilon^2~c_{F_0}~\delta^{5/2}~\rho^2~,
\label{eqn:lundefrapzero}
\\
4~\epsilon^2~\tvert{\cal L}_{v}^{-1}~F_0(\mu_i)'\tvert_{\L^2}
+\epsilon^2~
\tvertb\frac{F_0(\mu_i)'}{\Lepsilon }\tvertb_{{\cal W},\sigma}
&\leq 
9~\epsilon^2~c_{F_0}~\delta^{5/2}~\rho^2~,
\label{eqn:lundefwrapzero}
\\
\epsilon^2~\tvert{\cal L}_{v}^{-1}~\Delta F_0(\mu_1,\mu_2)'\tvert_{\L^2}
+\epsilon^2~
\tvertb\frac{\Delta F_0(\mu_1,\mu_2)'}{\Lepsilon }\tvertb_{{\cal W},\sigma}
&\leq
5~\epsilon^2~c_{F_0}~\delta^{5/2}~\rho~
\tvert\mu_1-\mu_2\tvert_{\sigma}~,
\label{eqn:lundefdiffrapzero}
\end{equs}
where $\Delta F_0(\mu_1,\mu_2)=F_0(\mu_1)-F_0(\mu_2)$. Since
$\delta=c_{\delta}~\rho^2$, we see that for
$\epsilon_0=c_{\epsilon}~\rho^{-3}$, the contribution of $F_0$ to the
bounds (\ref{eqn:lundefrap})--(\ref{eqn:lundefdiffrap})
can be made arbitrarily small, choosing $c_{\epsilon}$ sufficiently small,
independently of $\rho$, or of the size of the system $L$. So what we need
is more detailed information on $r_2$. Note that $r_2$ inherits the bounds
of $r$ and $r_1$, but with a factor $\epsilon^{-4}$, so that we have to
work a little more to show that the bounds on $r_2$ are finite as
$\epsilon\to0$, and that (\ref{eqn:lundefrap})--(\ref{eqn:lundefdiffrap})
are also satisfied when the contribution of $r_2$ is taken into account.
The essential input will be (\ref{eqn:theeqforrdxxx}), where we showed
that, as a dynamical variable, $r_2$ satisfies
\begin{equs}
\partial_t r_2=-\frac{\chi}{\epsilon^4}~\Gepsilon ~{\cal L}_{r}~r_2
+\frac{\chi}{16}\mu~{\cal L}_{\mu,r}~r_2'+\frac{1}{\epsilon^4}~F_6(s,\mu)~,~~~
r_2(x,0)=r_{2,0}(x)~,
\label{eqn:therdeqrapun}
\end{equs}
with $r_{2,0}=\frac{r_0}{\epsilon^4}-\frac{r_1(\mu_0)}{\epsilon^4}$. Since
we know that $s(\mu)$ exists, we can view this equation as a linear
inhomogeneous equation for $r_2$ and derive bounds from it.
We first have the following lemma:
\begin{lemma}\label{lem:ldrd}
If $r_2$ solves (\ref{eqn:therdeqrapun}), one has
\begin{equs}
\tvert r_2(\mu)\tvert_{\L^2}&\leq
\|r_{2,0}\|_{\L^2}+\tvert F_6(s(\mu),\mu)\tvert_{\L^2}~,
\label{eqn:rdeuxsansdiff}\\
\tvert\chi{\cal L}_{\mu,r}~{\cal L}_{v}^{-1}~r_2(\mu)'\tvert_{\L^2}&\leq
64~\|r_{2,0}\|_{\L^2}+
\sqrt{2}~\tvert{\cal L}_{\mu,r}~{\cal L}_{v}^{-1}~F_6(s(\mu),\mu)'\tvert_{\L^2}
\label{eqn:rdeuxsansdiffw}~,
\end{equs}
for all $\mu$ with $\tvert\mu\tvert_{\sigma}\leq c_{\mu}~\rho$.
\end{lemma}
\begin{remark}   
Without the term $\mu~{\cal L}_{\mu,r}~r_2'$ in
(\ref{eqn:therdeqrapun}), the proof of (\ref{eqn:rdeuxsansdiff}) and
(\ref{eqn:rdeuxsansdiffw}) would follow immediately by positivity of
$\Gepsilon ~{\cal L}_r$, and we would have the same estimates for
$r_2(\mu_1)-r_2(\mu_2)$, but with $r_{2,0}=0$ and $F_6(\mu)$ replaced by
$F_6(\mu_1)-F_6(\mu_2)$ . The term $\mu~{\cal L}_{\mu,r}~r_2'$ cannot
destroy the positivity of $\Gepsilon ~{\cal L}_r$ if $\epsilon$ is
sufficiently small (see (\ref{eqn:harmless}) below or Proposition
\ref{prop:coercrdeux}), and it will add a (small) correction to the norm of
the difference.
\end{remark}
\begin{proof}
From (\ref{eqn:therdeqrapun}), using Young's inequality and Proposition
\ref{prop:coercrdeux} (see subsection \ref{app:coercrdeux} of Appendix
\ref{app:therdeuxmap}), we get
\begin{equs}
\partial_t \int r_2^2&=
-\frac{2~\chi}{\epsilon^4}\left(
\int r_2~\Gepsilon ~{\cal L}_{r}~r_2
-\frac{\epsilon^4}{16}
\int r_2~\mu~{\cal L}_{\mu,r}~r_2'\right)
+\frac{2}{\epsilon^4}\int r_2~F_6(s,\mu)~,
\label{eqn:harmless}
\\
&\leq
-\frac{\chi}{\epsilon^4}
\int r_2^2+\frac{2}{\chi~\epsilon^4}\int F_6(s,\mu)^2~.
\end{equs}
The proof of (\ref{eqn:rdeuxsansdiff}) is completed integrating this
differential inequality and noting that $\frac{2}{\chi^2}\leq1$. The proof
of (\ref{eqn:rdeuxsansdiffw}) can be found in
subsection \ref{sec:varboundsrd}.
\end{proof}
\begin{corollary}\label{cor:firstpartrd}
Assume that $256~\epsilon^2~\|r_{2,0}\|_{\L^2}\leq
\lambda_{2,2}~c_{\mu}~\rho$. Then there exists a constant $C$ such that
\begin{equs}
\epsilon^4~\tvert r_2(\mu)\tvert_{\L^2}
&\leq
\epsilon^2~\lambda_{2,2}~c_{\mu}~\rho+
 \epsilon^2~C~\bigl(\delta^{5/2}~\rho^{2}+\delta^{4}~\rho\bigr)~,\\
4~\epsilon^2~\tvert\chi~{\cal L}_{\mu,r}~{\cal L}_{v}^{-1}~r_2(\mu)'\tvert_{\L^2}
&\leq
\lambda_{2,2}~c_{\mu}~\rho+
 \epsilon^2~C~\bigl(\delta^{5/2}~\rho^{2}+\delta^{4}~\rho\bigr)~,
\label{eqn:stronger}
\end{equs}
for all $\mu\in{\cal B}_{\sigma}(c_{\mu}~\rho)$.
\end{corollary}
\begin{proof}
This is really a statement on $F_6$. For the proof, see Section
\ref{sec:boundfsix}.
\end{proof}
Choosing $\epsilon_0=c_{\epsilon}~\rho^{-4}$ with $c_{\epsilon}$
sufficiently small and a $\lambda_{2,2}$ sufficiently small, the r.h.s. of
(\ref{eqn:stronger}) is bounded by $c_{\mu}~\rho$. 
Thus the hypotheses of the following lemma can be fulfilled.
\begin{lemma}\label{lem:brdeuxmain}
Let $r_2$ be the solution of (\ref{eqn:therdeqrapun}) with
$\mu\in{\cal B}_{\sigma}(c_{\mu}~\rho)$, and let
$\balpha=\max(2,\frac{\alpha^2}{1-\alpha^2})$. Assume that $r_2$
satisfies
\begin{equs}
\epsilon^2~\tvert\chi~{\cal L}_{\mu,r}~{\cal L}_{v}^{-1}~r_2(\mu)'\tvert_{\L^2}
\leq c_{\mu}~\rho~.
\label{eqn:oneassumption}
\end{equs}
Then there exists a constant $C$ such that
\begin{equs}
\tvertb
\frac{\chi~{\cal L}_{\mu,r}}{\Lepsilon }~r_2(\mu)'
\tvertb_{{\cal W},\sigma}&\leq
\frac{
\frac{4~\chi}{\delta}~
\|r_{2,0}\|_{{\cal W},\sigma-1}+
\tvertb
\frac{{\cal L}_{\mu,r}}{\Lepsilon }~
\frac{F_6(s(\mu),\mu)'}{\Gepsilon ~{\cal L}_r}
\tvertb_{{\cal W},\sigma}+
C~\balpha~\epsilon^2~c_{\mu}~\rho}
{1-\epsilon^2~2~\balpha~\Cm ~c_{\mu}~c_{\delta}^{-1/2}}~,
\end{equs}
for all $\epsilon$ satisfying
$\epsilon^2~2~\balpha~\Cm ~c_{\mu}~c_{\delta}^{-1/2}<1$.
\end{lemma}
\begin{proof}   
See Section \ref{sec:varboundsrd}.
\end{proof}
\begin{corollary}\label{cor:onprogresse}
Assume that $256~\epsilon^2~\|r_{2,0}\|_{{\cal W},\sigma-1}\leq
\lambda_{2,{\cal W}}~c_{\mu}~\rho$. Then there exists a constant $C$ such that
\begin{equs}
\epsilon^2~
\tvertb\frac{\chi{\cal L}_{\mu,r}}{\Lepsilon }~r_2(\mu)'\tvertb_{{\cal W},\sigma}
&\leq 
\frac{
\max\Bigl(\frac{1}{3},\frac{\alpha^2}{1-\alpha^2}\Bigr)
\left((1+C~\epsilon^2)~
c_{\mu}~\rho
+\epsilon^2~C~\delta^{5/2}~\rho^{2}
\right)
}{1-\epsilon^2~2~\balpha~\Cm ~c_{\mu}~c_{\delta}^{-1/2}}\\
&\phantom{=}~+
\frac{
\lambda_{2,{\cal W}}~c_{\mu}~\rho
}{1-\epsilon^2~2~\balpha~\Cm ~c_{\mu}~c_{\delta}^{-1/2}}
~,
\label{eqn:onprogresse}
\end{equs}
for all $\mu\in{\cal B}_{\sigma}(c_{\mu}~\rho)$, and for all
$\epsilon$ satisfying
$\epsilon^2~2~\balpha~\Cm ~c_{\mu}~c_{\delta}^{-1/2}<1$.
\end{corollary}
\begin{proof}
This is also a statement on $F_6$. For the proof, see Section
\ref{sec:boundfsix}.
\end{proof}
Choosing $\lambda_{2,{\cal W}}$ and $c_{\epsilon}$ sufficiently small, we
bound the r.h.s. of (\ref{eqn:onprogresse}) by $c_{\mu}~\rho$.
Thus, the hypotheses of the next Lemma can also be fulfilled.
\begin{lemma}\label{lem:brdeuxdiffmain}
Let $r_2(\mu_i)$ be the solution of (\ref{eqn:therdeqrapun}) with
$\mu=\mu_i$, and define $\Delta r_2=r_2(\mu_1)-r_2(\mu_2)$ and
$\Delta F_6=F_6(s(\mu_1),\mu_1)-F_6(s(\mu_2),\mu_2)$. Assume that
$\mu_i\in{\cal B}_{\sigma}(c_{\mu}~\rho)$, and that 
\begin{equs}
\epsilon^2~\tvert\chi~{\cal L}_{\mu,r}~{\cal L}_{v}^{-1}~r_2(\mu_i)'\tvert_{\L^2}+
\epsilon^2~\tvertb
\frac{\chi~{\cal L}_{\mu,r}}{\Lepsilon }~{r_2(\mu_i)}'
\tvertb_{{\cal W},\sigma}\leq c_{\mu}~\rho~.
\label{eqn:assumebrdeuxmain}
\end{equs}
Then there exists a constant $C$ such that
\begin{equs}
\tvert\chi~{\cal L}_{\mu,r}~{\cal L}_{v}^{-1}~\Delta r_2'\tvert_{\L^2}&\leq
\tvert\chi~{\cal L}_{\mu,r}~{\cal L}_{v}^{-1}~\Delta F_6'\tvert_{\L^2}+
C~\delta^{5/2}~\rho~\tvert\mu_1-\mu_2\tvert_{\sigma}~.
\label{eqn:drdldmain}
\end{equs}
\end{lemma}
\begin{corollary}\label{cor:onprogressedeux}
The map $\mu\mapsto r_2(\mu)$ satisfies
\begin{equs}
\epsilon^2~\tvert
\chi~{\cal L}_{\mu,r}~{\cal L}_{v}^{-1}~\bigl(r_2(\mu_1)-r_2(\mu_2)\bigr)'
\tvert_{\L^2}&\leq
\epsilon^2~C~(\delta^{5/2}~\rho+\delta^4)~\tvert\mu_1-\mu_2\tvert_{\sigma}
\label{eqn:drdldmaincor}
\end{equs}
for some constant $C$.
\end{corollary}
\begin{proof}
This is again a statement on $F_6$. For the proof, see Section
\ref{sec:boundfsix}.
\end{proof}
Again, choosing $c_{\epsilon}$ sufficiently small, the
r.h.s. of (\ref{eqn:drdldmaincor}) will be bounded by $\tvert\mu_1-\mu_2\tvert_{\sigma}$.
Thus the hypotheses of the next lemma can be fulfilled.
\begin{lemma}\label{lem:brdeuxdiffwmain}
Let $r_2(\mu_i)$ be the solution of (\ref{eqn:therdeqrapun}) with
$\mu=\mu_i$, and define $\Delta r_2=r_2(\mu_1)-r_2(\mu_2)$
and $\Delta F_6=F_6(s(\mu_1),\mu_1)-F_6(s(\mu_2),\mu_2)$. Assume that
$\mu_i\in{\cal B}_{\sigma}(c_{\mu}~\rho)$, and that
\begin{equs}
\epsilon^2~\tvert\chi~{\cal L}_{\mu,r}~{\cal L}_{v}^{-1}~r_2(\mu_i)'\tvert_{\L^2}+
\epsilon^2~\tvertb
\frac{\chi~{\cal L}_{\mu,r}}{\Lepsilon }~{r_2(\mu_i)}'
\tvertb_{{\cal W},\sigma}&\leq c_{\mu}~\rho~,
\label{eqn:assumebrdeuxmainrap}\\
\epsilon^2~\tvert\chi~{\cal L}_{\mu,r}~{\cal L}_{v}^{-1}~\Delta r_2(\mu_i)'\tvert_{\L^2}
&\leq \tvert\mu_1-\mu_2\tvert_{\sigma}~.
\end{equs}
Then there exists a constant $C$ such that
\begin{equs}
\tvertb
\frac{\chi~{\cal L}_{\mu,r}}{\Lepsilon }~\Delta r_2'
\tvertb_{{\cal W},\sigma}&\leq
\frac{
\tvertb
\frac{{\cal L}_{\mu,r}}{\Lepsilon ~\Gepsilon ~{\cal L}_r}~
\Delta F_6'
\tvertb_{{\cal W},\sigma}
+C~\balpha~\tvert\mu_1-\mu_2\tvert_{\sigma}
}{1-\epsilon^2~2~\balpha~\Cm ~c_{\mu}~c_{\delta}^{-1/2}}~,
\label{eqn:drdsmain}
\end{equs} 
for all $\epsilon$ satisfying
$\epsilon^2~2~\balpha~\Cm ~c_{\mu}~c_{\delta}^{-1/2}<1$.
\end{lemma}
\begin{proof}   
See Section \ref{sec:varboundsrd}.
\end{proof}
\begin{corollary}\label{cor:ontermine}
There exists a constant $C$ such that
\begin{equs}
\epsilon^2~
\tvertb\frac{\chi{\cal L}_{\mu,r}}{\Lepsilon }~
\Delta r_2'\tvertb_{{\cal W},\sigma}
&\leq 
\frac{
\max\bigl(\frac{1}{3},\frac{\alpha^2}{1-\alpha^2}\bigr)
\left(1
+\epsilon^2~C~(1+\delta^{5/2}~\rho)\right)
}{1-\epsilon^2~2~\balpha~\Cm ~c_{\mu}~c_{\delta}^{-1/2}}
~\tvert\mu_1-\mu_2\tvert_{\sigma}
~,
\label{eqn:finished}
\end{equs}
for all $\mu\in{\cal B}_{\sigma}(c_{\mu}~\rho)$, and for all
$\epsilon$ satisfying
$\epsilon^2~2~\balpha~\Cm ~c_{\mu}~c_{\delta}^{-1/2}<1$.
\end{corollary}
\begin{proof}
This is also a statement on $F_6$. For the proof, see Section
\ref{sec:boundfsix}.
\end{proof}
We are now in position to prove that the hypotheses
(\ref{eqn:lundefrap})--(\ref{eqn:lundefdiffrap}) are satisfied if
$\epsilon$ is sufficiently small. 

\begin{theorem}\label{thm:thecondissat}
Condition \ref{cond:condonF} is satisfied if
$\epsilon_0\leq c_{\epsilon}~\sqrt{1-2~\alpha^2}~\rho^{-4}$ with 
$c_{\epsilon}$ sufficiently small and if the initial data $\mu_0$ and $s_0$
are in the class $\class$.
\end{theorem}
\begin{proof}
By Definition \ref{cond:thecondition} (see also (\ref{eqn:encorerdeuxdiff})), 
if the initial data $\mu_0$ and $s_0$ are in the class $\class$, then
\begin{equs}
256~\epsilon^2~\|r_{2,0}\|_{\sigma-1}\leq
\lambda_{2}~\Bigl(\frac{\epsilon}{\epsilon_0}\Bigr)^2~c_{\mu}~\rho~,
\end{equs}
with
$\lambda_2<\min\bigl(\frac{2}{3},\frac{1-2~\alpha^2}{1-\alpha^2}\bigr)$.
This means that the hypotheses of Corollaries \ref{cor:firstpartrd}, 
\ref{cor:onprogresse}, \ref{cor:onprogressedeux} and \ref{cor:ontermine}
on $r_{2,0}$ are satisfied with $\lambda_{2,2}+\lambda_{2,{\cal
W}}\leq\lambda_2~\bigl(\frac{\epsilon}{\epsilon_0}\bigr)^2$.
Collecting the results of these corollaries, and using
(\ref{eqn:lundefrapzero})--(\ref{eqn:lundefdiffrapzero}), we have
\begin{equs}
\epsilon^4~\tvert r_2(\mu_i)\tvert_{\L^2}+\epsilon^2~\tvert{\cal L}_{v}^{-1}~F(\mu_i)'\tvert_{\L^2}
&\leq 
(\frac{1}{4}+\epsilon^2)~\lambda_{2,2}~c_{\mu}~\rho
+\epsilon^2~C\bigl(\delta^{5/2}~\rho^2+\delta^4~\rho\bigr)\\
&\leq\Bigl(\frac{\epsilon}{\epsilon_0}\Bigr)^2~c_{\mu}~\rho~
\tilde{\lambda}_2(\lambda_2,\epsilon_0,\rho,\delta)
\label{eqn:premcond}
~,
\end{equs}  
where
\begin{equs}
\tilde{\lambda}_2(\lambda_2,\epsilon_0,\rho,\delta)
&\equiv\Bigl(\frac{\epsilon}{\epsilon_0}\Bigr)^2~c_{\mu}~\rho~\Bigl(
\lambda_{2}~
+\epsilon_0^2~C\bigl(\delta^{5/2}~\rho+\delta^4\bigr)\Bigr)~.
\end{equs}
In the same way, using $\epsilon\leq\epsilon_0$, we have
\begin{equs}
4~\epsilon^2~\tvert{\cal L}_{v}^{-1}~F(\mu_i)'\tvert_{\L^2}
+\epsilon^2~
\tvertb\frac{F(\mu_i)'}{\Lepsilon }\tvertb_{{\cal W},\sigma}
&\leq\tilde{\lambda}_1(\alpha,\epsilon_0,\rho,\delta)~c_{\mu}~\rho~,
\label{eqn:seccond}\\
4~\epsilon^2~\tvert{\cal L}_{v}^{-1}~\Delta F'\tvert_{\L^2}
+\epsilon^2~
\tvertb\frac{\Delta F'}{\Lepsilon }\tvertb_{{\cal
W},\sigma}
&\leq 
\tilde{\lambda}_1(\alpha,\epsilon_0,\rho,\delta)~
\tvert\mu_1-\mu_2\tvert_{\sigma}~,
\label{eqn:tercond}
\end{equs}
where
\begin{equs}
\tilde{\lambda}_1(\alpha,\epsilon_0,\rho,\delta)\equiv
\frac{
\max\Bigl(\frac{1}{3},\frac{\alpha^2}{1-\alpha^2}\Bigr)
\left(1
+\epsilon_0^2~C~\bigl(\delta^{5/2}~\rho+\delta^4+1\bigr)\right)+\lambda_{2}
}{1-\epsilon^2~2~\balpha~\Cm ~c_{\mu}~c_{\delta}^{-1/2}}~.
\end{equs}
Now, let $\delta=c_{\delta}~\rho^2$,
$\epsilon_0=c_{\epsilon}~\sqrt{1-2~\alpha^2}~\rho^{-4}$. Since
$\lambda_2<\min\bigl(\frac{2}{3},\frac{1-2~\alpha^2}{1-\alpha^2}\bigr)$,
we have
\begin{equs}
\tilde{\lambda}_1(\alpha,\epsilon_0,\rho,\delta)&\leq
\frac{
\max\Bigl(\frac{1}{3},\frac{\alpha^2}{1-\alpha^2}\Bigr)
\left(1+c_\epsilon^2~(1-2~\alpha^2)~C_1\right)+\lambda_{2}
}{1-c_{\epsilon}^2~(1-2~\alpha^2)~C_2}~,\\
\tilde{\lambda}_2(\lambda_2,\epsilon_0,\rho,\delta)&\leq
\frac{2}{3}+C_3~c_{\epsilon}^2~(1-2~\alpha^2)\leq\frac{2}{3}+C_3~c_{\epsilon}^2~,
\end{equs}
for some (positive) constants $C_1,C_2$ and $C_3$. We now choose
\begin{equs}
c_{\epsilon}<
\min_{\alpha^2\in[0,1/2]}\left(
\frac{1}{C_3}~,~
\frac{1}{1-2~\alpha^2}~\left(\frac{\min\bigl(\frac{2}{3},\frac{1-2~\alpha^2}{1-\alpha^2}\bigr)-\lambda_2}
{C_1~
\max\bigl(\frac{1}{3},\frac{\alpha^2}{1-\alpha^2}\bigr)
+C_2}\right)
\right)^{1/2}
~,
\label{eqn:cechoice}
\end{equs}
and get
$\tilde{\lambda}_1(\alpha,\epsilon_0,\rho,\delta)<1$ and 
$\tilde{\lambda}_2(\lambda_2,\epsilon_0,\rho,\delta)<1$ as requested
for Condition \ref{cond:condonF} to hold. Note that there always exists a
$c_{\epsilon}>0$ satisfying inequality (\ref{eqn:cechoice}), since
$\lambda_2<\min\bigl(\frac{2}{3},\frac{1-2~\alpha^2}{1-\alpha^2}\bigr)$.
\end{proof}

\subsection*{Acknowledgements}   
The author would like to express his gratitude to Jean-Pierre Eckmann and
Pierre Collet for proposing the problem. Jean--Pierre Eckmann's suggestions
and advises during the elaboration of the results and the redaction of the
paper were invaluable. Finally, the author would also like to thank Pierre
Collet, Martin Hairer, Emmanuel Zabey and Serge\"i Kuksin for helpful
discussions.

\appendix
\newappendix{Coercive functional for the phase}\label{app:coercive}
In this section, we prove the Proposition \ref{prop:coerciveness}. This is
a nondegeneracy result on the operator $\Lepsilon$ based on a similar
result of \cite{Collet} for ${\cal L}_{\mu,c}=\partial_x^4+\partial_x^2$.
We will need some technical alterations of their proof to take into account
that $\Lepsilon$ is of lower order than ${\cal L}_{\mu,c}$. However, since
the two operators are equal in the limit $\epsilon\to0$ ($\Lepsilon=\Gepsilon~{\cal L}_{\mu,c}$ and
${\displaystyle\lim_{\epsilon\to0}\Gepsilon}={\rm Id}$ by (\ref{eqn:gedef})
and (\ref{eqn:lmudef})), we will recover their result as a particular case.
\begin{proposition}\label{prop:coercivenessrappel}
For all $L\geq2\pi$, there exist a constant $K$ and an antisymmetric
periodic function $\phi$ such that for all $\gamma\in[\frac{1}{4},1]$, all
$\epsilon\leq(\pi L^{2/5})^{-1}$ and any antisymmetric periodic function $v$, one has
\begin{equs}
\frac{3}{4}({\cal L}_{v}~v,{\cal L}_{v}~v)
\leq
(v,v)_{\gamma\phi}&\leq
\|\phi'\|_{\infty}~(v,v)+(v'',v'')~,\\
(\phi,\phi)_{\gamma\phi}&\leq K~L^{16/5}~,\\
(\phi,\phi)&\leq{\textstyle\frac{4}{3}}~L^3~,
\end{equs}
where the inner products $(\cdot,\cdot)$ and $(\cdot,\cdot)_{\gamma\phi}$ are
defined by
\begin{equs}
(v,v)&=
\int_{-L/2}^{L/2}
\hspace{-5mm}
\text{d}x~
v(x)^2~,~~~
(v,v)_{\gamma\phi}
=
\int_{-L/2}^{L/2}
\hspace{-5mm}  
\text{d}x~
v(x)~(\Lepsilon +\gamma\phi'(x))~v(x)~,
\end{equs}
and ${\cal L}_{v}$ is multiplicative in Fourier space with symbol
${\cal
L}_{v}(k)=\sqrt{\frac{1}{3}\frac{1+k^4}{1+\frac{\epsilon^2~k^2}{2}}}$.
\end{proposition}
\begin{remark}\label{rem:striepsi}
The restriction $\epsilon\leq(\pi L^{2/5})^{-1}$ is a convenient one because then we
can use the same function $\phi$ as that defined in \cite{Collet}. We will
see later that we need a much stronger restriction ($\epsilon\leq
{\cal O}(L^{-32/5})$) anyway.
\end{remark}
\begin{proof}[proof of Proposition \ref{prop:coercivenessrappel}]
The proof really amounts to construct the function $\phi$. Let
$q\equiv\frac{2\pi}{L}\leq1$ and $M$ the smallest integer (strictly) larger
than $\frac{1}{2}~L^{7/5}$. We define $\phi$ by
\begin{equs}
\phi(x)=\sum_{n\in{\bf Z}}\ed^{iqnx}~\phi_n~,
\end{equs}
where the Fourier coefficients $\phi_n$ are given by
\begin{equs}
\phi_{n}=\left\{
\begin{array}{ll}
0~,~~&~n=0\\[0.3cm]
\frac{4~i}{qn}~,~~&~1\leq|n|\leq~2M\\[0.3cm]
\frac{4~i~f(|n|/2M-1)}{qn}~,&~\text{otherwise}
\end{array}
\right.~,
\end{equs}
where $f$ is a non-increasing ${\cal C}^1$ function satisfying $f(0)=1$,
$f'(0)=0$ and
\begin{equs}
f\geq0,~~\sup|f'|<1,~~~
\int_0^{\infty}
\hspace{-2mm}{\rm d}k~(1+k)^2~|f(k)|^2<\infty~.
\end{equs}
The proof then follows from the three technical lemmas below.
\end{proof}
\begin{lemma}\label{lem:rzerophi}
There exists a constant $K$ such that the function $\phi$ defined above
satisfies
\begin{equs}
(\phi,\phi)&\leq
{\textstyle\frac{4}{3}}~L^3~,\\
(\phi,\phi)_{\gamma\phi}&\leq K~L^{16/5}~,\\
(v,v)_{\gamma\phi}&\leq
K~L^{7/5}~\|v\|_{\L^2}^2+\|v''\|_{\L^2}^2~.
\end{equs}
for all periodic antisymmetric functions $v$.
\end{lemma}
\begin{proof}
For the first inequality, we have
\begin{equs}
(\phi,\phi)=\frac{4\pi}{q}\sum_{n=1}^{\infty}|\phi_n|^2
\leq \frac{4^3~\pi}{q^3}\sum_{n=1}^{\infty}\frac{1}{n^2}
=\frac{1}{6}\left(\frac{4\pi}{q}\right)^3
=\frac{4}{3}~L^3~.
\end{equs}
For the second inequality, we use that $\phi$ is periodic, so that
$\int\phi^2\phi'=0$, giving
\begin{equs}
(\phi,\phi)_{\gamma\phi}
=(\phi,\Lepsilon \phi)
=\frac{4\pi}{q}\sum_{n=1}^{\infty}
\Lepsilon (qn)~|\phi_n|^2~,
\end{equs}
where
\begin{equs}
\Lepsilon (qn)
=
\frac{(qn)^4-(qn)^2}{1+\frac{\epsilon^2(qn)^2}{2}}~.
\end{equs}
Since $\Lepsilon (qn)\leq(qn)^4$ and $M<L^{7/5}$, we get
\begin{equs}
(\phi,\phi)_{\gamma\phi}
&=(\phi,\Lepsilon \phi)
=\frac{4\pi}{q}\sum_{n=1}^{\infty}
\Lepsilon (qn)~|\phi_n|^2\\
&\leq
4^3\pi~q\left(
~\sum_{n=1}^{2M}n^2
+
(2M)^2
\sum_{n=1}^{\infty}
\Bigl(1+\frac{n}{2M}\Bigr)^2
f\Bigl(\frac{n}{2M}\Bigr)^2
\right)
\\
&\leq
C~L^{16/5}~\left(1+\int_0^{\infty}\hspace{-2mm}{\rm d}k~
(1+k)^2~f(k)^2\right)
~.
\end{equs}
Finally, using again $\Lepsilon (qn)\leq (qn)^4$, we have
\begin{equs}
(v,v)_{\gamma\phi}\leq
\|\phi'\|_{\L^{\infty}}~\|v\|_{\L^2}^2+\|v''\|_{\L^2}^2~.
\end{equs}
Using the Cauchy--Schwartz inequality, we have
\begin{equs}
\|\phi'\|_{\L^{\infty}}&\leq
2\sum_{n=1}^{\infty}
|qn|~|\phi_n|
\leq
16M+2\sum_{n=1}^{\infty}\left|
f\Bigl(\frac{n}{2M}\Bigr)
\right|
\\&\leq
16M+4M\int_{0}^{\infty}\hspace{-2mm}{\rm d}k 
~\Bigl(\frac{1+k}{1+k}\Bigr)~|f(k)|\\
&\leq C~L^{7/5}
\left(
1+\sqrt{\int_{0}^{\infty}\hspace{-2mm}{\rm d}k 
~(1+k)^2~|f(k)|^2}
\right)~.
\end{equs}
This completes the proof of the Lemma.
\end{proof}
\begin{lemma}\label{lem:positivity}
For all $L\geq2\pi$, for all $\gamma\in[\frac{1}{4},1]$ and for all
$\epsilon\leq\frac{1}{Mq}$, one has
\begin{equs}
(v,v)_{\gamma\phi}\geq
\frac{3}{4}({\cal L}_{v}~v,{\cal L}_{v}~v)~.
\label{eqn:inthelemme}
\end{equs}
\end{lemma}
\begin{proof}
Following \cite{Collet} one shows first that 
\begin{equs}
(v,v)_{\gamma\phi}=
2~L~\left[
\sum_{n>0}
(\Lepsilon (qn)+\gamma\psi_{2n})v_n^2
+2\gamma\sum_{k>m>0}
v_k~v_m
(\psi_{|k+m|}-\psi_{|k-m|})
\right]~,
\end{equs}
where $\psi_n=-iqn~\phi_n$. Then one notices that for 
$0\leq\epsilon\leq1$, one has
\begin{equs}
\Lepsilon (qn)+\gamma\psi_{2n}\geq
\tau(qn)^2\equiv
\frac{1}{2}\frac{1+(qn)^4}{1+\frac{\epsilon^2(qn)^2}{2}}
\geq \tau_1(qn)^2\equiv
\frac{1}{2}\frac{(qn)^4}{1+\frac{\epsilon^2(qn)^2}{2}}~.
\end{equs}
The definition of $\tau$ here is different from that of
\cite{Collet}, except in the $\epsilon=0$ limit. Set now $w_n=v_n\tau_n$
(in particular $w=\sqrt{\frac{3}{2}}~{\cal L}_{v}~v$), so that
\begin{equs}
(v,v)_{\gamma\phi}\geq
2~L~
\left[
\sum_{n>0}
w_n^2+2\gamma
\sum_{0<m<k}
w_k\frac{\psi_{|k+m|}-\psi_{|k-m|}}{\tau_k\tau_m}w_m
\right]
\equiv (w~,~({\rm Id}+2\gamma\Gamma)~w)~.
\end{equs}
In Lemma \ref{lem:hilbertschmidt} below, we prove that the Hilbert--Schmidt
norm of $2\gamma\Gamma$ is less than $\frac{1}{2}$, hence
\begin{equs}
(v,v)_{\gamma\phi}\geq\frac{1}{2}~(w,w)=
\frac{3}{4}~({\cal L}_{v}~v,{\cal L}_{v}~v)~,
\end{equs}
which proves inequality (\ref{eqn:inthelemme}).
\end{proof}
\begin{lemma}\label{lem:hilbertschmidt}
For all $L\geq2\pi$, for all $\gamma\in[\frac{1}{4},1]$ and for all
$\epsilon\leq\frac{1}{Mq}$, the Hilbert--Schmidt norm of $2\gamma\Gamma$ is
smaller than $\frac{1}{2}$, where the operator $\Gamma$ is defined by
\begin{equs}
(w,\Gamma w)=
L~
\sum_{0<m<k}
w_k\frac{\psi_{|k+m|}-\psi_{|k-m|}}{\tau_k\tau_m}w_m~.
\end{equs}
\end{lemma}
\begin{proof}
Note again that $\epsilon\leq\frac{1}{Mq}=(\pi L^{2/5})^{-1}$
is only a convenient restriction (see remark \ref{rem:striepsi}). 
To prove this lemma, it is sufficient (see \cite{Collet}) to show that
\begin{equs}
\|\Gamma\|^2_{\rm HS}\equiv
\sum_{0<m<k}
\left|
\frac{\psi_{|k+m|}-\psi_{|k-m|}}{\tau_k~\tau_m}
\right|^2<\frac{1}{16}~.
\label{eqn:hs}
\end{equs}
By definition of $\phi$, for all $k>m>0$, we have
\begin{equs}
|\psi_{k-m}-\psi_{k+m}|=0~,~~\text{if}~k+m\leq 2M~,
\end{equs}
and
\begin{equs}
|\psi_{k-m}-\psi_{k+m}|\leq4\min\left\{1,\frac{m}{M}\right\}~,~~\text{for all}~k>m~.
\end{equs}
We distinguish two sets of summation indices 
$S=S_{\rm I}\cup S_{\rm II}$ in the sum (\ref{eqn:hs}),
\begin{equs}
S_{\rm I}&=\left\{
(m,k)\in{\bf N}^2~{\rm s.t.}~M+1\leq m~{\rm and}~m+1\leq k\right\}~,\\
S_{\rm II}&=\left\{
(m,k)\in{\bf N}^2~{\rm s.t.}~
1\leq m\leq M~{\rm and}~2M-m+1\leq k\right\}~,
\end{equs}
and write $\|\Gamma\|^2_{\rm HS}=T_{\rm I}+T_{\rm II}$ accordingly. 

In the region I, we have $m>M$, and using $\epsilon\leq\frac{1}{Mq}$ and
$\frac{1}{\tau(k)}\leq\frac{1}{\tau_1(k)}$, we get
\begin{equs}
T_{\rm I}\leq16
\sum_{m=M+1}^{\infty}\frac{1}{\tau(qm)^{2}}
\sum_{k=m+1}^{\infty}\frac{1}{\tau(qk)^{2}}
\leq16
\int_{M}^{\infty}\hspace{-2mm}{\rm d}m~
\frac{1}{\tau_1(qm)^{2}}
\int_{m}^{\infty}\hspace{-2mm}{\rm d}k~
\frac{1}{\tau_1(qk)^{2}}
\leq \frac{200}{9}\frac{1}{q^8~M^6}~,
\end{equs}
whereas in the region II, we have $m\leq M$ and $k\geq M+1$, and using again 
$\epsilon\leq\frac{1}{Mq}$ and $\frac{1}{\tau(k)}\leq\frac{1}{\tau_1(k)}$, we get
\begin{equs}
T_{\rm II}&\leq
\frac{16}{M^2}
\sum_{m=1}^{M}\frac{m^2}{\tau(qm)^{2}}
\sum_{k=2M-m+1}^{\infty}\frac{1}{\tau(qk)^{2}}
\leq
\frac{16}{M^2}
\sum_{m=1}^{M}\frac{m^2}{\tau(qm)^{2}}
\int_{M}^{\infty}\hspace{-2mm}{\rm d}k~
\frac{1}{\tau_1(qk)^{2}}\\
&\leq\frac{160}{3}\frac{1}{M^5~q^4}
\sum_{m=1}^{M}
\left(
\frac{1}{q^2}
\frac{q^2~m^2}{1+m^4~q^4}+
\frac{1}{2~M^2~q^4}
\frac{q^4~m^4}{1+m^4~q^4}
\right)
\\
&\leq
\frac{160}{3}\frac{1}{M^5~q^4}
\left(
\frac{1}{q^2}
\int_0^{\infty}
\hspace{-2mm}
\frac{{\rm d}m}{1+q^2~m^2}
+
\frac{1}{2~M~q^4}
\right)~.
\end{equs}
Collecting these results, we get
\begin{equs}
\|\Gamma\|^2_{\rm HS}\leq
\frac{80~\pi}{3}~\frac{1}{q^7~M^5}
+\frac{440}{9}~\frac{1}{q^8~M^6}~.
\end{equs}
Note that this bound is worse than that of \cite{Collet} by numerical factors
only (in their bound $\frac{80~\pi}{3}$ is replaced by $\frac{128}{3}$ and
$\frac{440}{9}$ by $\frac{16}{3}$), but is uniform in $\epsilon\leq\frac{1}{Mq}$.
This motivates the restriction $\epsilon\leq\frac{1}{Mq}$. The proof is
then completed using $M>\frac{1}{2}~L^{7/5}$.
\end{proof}

\newappendix{Proofs for Section \ref{sec:highfreq}}\label{app:highk}
\begin{lemma}\label{lem:ldliestimerep}
Let $\sigma\geq\frac{3}{2}$. There exists a constant $C$ such that
for all $n\leq\sigma-\frac{3}{2}$ and for all $m\leq\sigma-1$, we have
\begin{equs}[3]
\|f^{(m)}\|_{\sigma-m}&+\|\Gepsilon~ f^{(m)}\|_{\sigma-m}
&~\leq~& c_{\infty}~\delta^{m}~&\|f\|_{\sigma}~,
\label{eqn:sdestimrap}
\\
\|f^{(n)}\|_{\L^{\infty}}&+
\|\Gepsilon~ f^{(n)}\|_{\L^\infty}
&~\leq~&
c_{\infty}~\delta^{n+\frac{1}{2}}~&\|f\|_{\sigma}~,
\label{eqn:Linfestimrap}
\end{equs}
where $f^{(m)}$ is the $m$--th order spatial
derivative of $f$.
\end{lemma}
\begin{proof}
Throughout the proof, we use that $\Gepsilon$ acts multiplicatively in
Fourier space, $(\Gepsilon~ f)_n=\Gepsilon(qn)~f_n$ with
\begin{equs}
\Gepsilon(k)=
\frac{1}{1+\frac{\epsilon^2~k^2}{2}}\leq1~,
\end{equs}
so that $\Gepsilon$ is a bounded operator in the $l^{p}$ and
$\|\cdot\|_{\sigma}$ norms. Using that
$\|f\|_{\L^{\infty}}\leq\|f\|_{l^1}$, and that the space derivative commutes
with $\Gepsilon $, we see that we need only prove (\ref{eqn:sdestimrap})
and (\ref{eqn:Linfestimrap}) for the terms without $\Gepsilon$, and with
$\L^{\infty}$ replaced by $l^1$ in (\ref{eqn:Linfestimrap}). In the sequel,
we denote by $K$ the operator with symbol $K(k)=|k|$. 

For (\ref{eqn:sdestimrap}), we use that $|x|\leq\sqrt{1+x^2}$ and that
$\|\cdot\|_{\L^2}=\sqrt{L}\|\cdot\|_{l^2}$ to show that
\begin{equs}
\|f^{(m)}\|_{\sigma-m}&=
\|f^{(m)}\|_{{\cal W},\sigma-m}+\|f^{(m)}\|_{\L^{2}}
\leq \delta^m~\|f\|_{{\cal W},\sigma}+\|f^{(m)}\|_{\L^{2}}\\
&\leq
\delta^m~\|f\|_{{\cal W},\sigma}+
\sqrt{L}~\|K^{m}~P_{<}f\|_{l^2}+
\sqrt{L}~\|K^{m}~P_{>}f\|_{l^2}
\\
&\leq
\delta^m~\|f\|_{{\cal W},\sigma}+
\delta^{m}~\sqrt{L}~\|P_{<}f\|_{l^2}+
\sqrt{L}~\delta^m\|
(1+(K/\delta)^2)^{m/2}~P_{>}f\|_{l^2}\\
&\leq
\delta^m~\|f\|_{{\cal W},\sigma}+
\delta^{m}~\|f\|_{\L^2}+
\delta^{m}~
\sqrt{
\int_{-\infty}^{\infty}
\frac{2\pi~{\rm d}x}{(1+x^2)^{\sigma-m}}}
~\|f\|_{{\cal W},\sigma}\\
&\leq \delta^m~\|f\|_{\sigma}
\left(
1+\sqrt{\int_{-\infty}^{\infty}\frac{2\pi~{\rm d}x}{(1+x^2)^{\sigma-m}}}
\right)~.
\end{equs}
For (\ref{eqn:Linfestimrap}), using the Cauchy--Schwartz inequality, we have
\begin{equs}
\|P_{<}f\|_{l^1}\leq
\sqrt{\frac{2~\delta}{q}}~\|P_{<}f\|_{l^2}
\leq\sqrt{\frac{\delta}{\pi}}~
\sqrt{L}~\|P_{<}f\|_{l^2}
\leq\sqrt{\frac{\delta}{\pi}}~\|P_{<}f\|_{\L^2}~,
\label{eqn:cauchyschwartz}
\end{equs}
so that
\begin{equs}
\|f^{(n)}\|_{l^1}&\leq
\|K^{n}~P_{<}f\|_{l^1}+
\|K^{n}~P_{>}f\|_{l^1}
\\
&\leq
\frac{\delta^{n+\frac{1}{2}}}{\sqrt{\pi}}~\|f\|_{\L^2}+
\|K^{n}~P_{>}f\|_{l^1}\\
&\leq
\delta^{n+\frac{1}{2}}~\|f\|_{\L^2}+
\delta^{n+\frac{1}{2}}
\int_{-\infty}^{\infty}
\frac{{\rm d}x}{(1+x^2)^{\frac{\sigma-n}{2}}}
~\|f\|_{{\cal W},\sigma}\\
&\leq
\delta^{n+\frac{1}{2}}~\|f\|_{\sigma}
\left(1+\int_{-\infty}^{\infty}\frac{{\rm d}x}{(1+x^2)^{\frac{\sigma-n}{2}}}\right)~.
\end{equs}
Since $\sigma-n\geq\frac{3}{2}$ and $\sigma-m\geq1$, setting
\begin{equs}
c_{\infty}=1+\max\left\{
\sqrt{\int_{-\infty}^{\infty}\frac{2\pi~{\rm d}x}{1+x^2}}~~,~~
\int_{-\infty}^{\infty}\frac{{\rm d}x}{(1+x^2)^{\frac{3}{4}}}
\right\}~,
\end{equs}
the proof is completed.
\end{proof}
Before proving Propositions \ref{prop:key} and \ref{prop:division}, we
prove a simpler lemma (see also \cite{Bricmont}).

\begin{lemma}\label{lem:keyestimbricmontrap}
Let $\sigma_1,\sigma_2\geq\frac{3}{2}$ and
$\sigma=\min(\sigma_1,\sigma_2)\geq\frac{3}{2}$, then there exists a
constant $c_b$ depending only on $\sigma$ such that
\begin{equs}
\|uv\|_{{\cal N},\sigma}\leq
c_b~\sqrt{\delta}~\|u\|_{{\cal N},\sigma_1}~\|v\|_{{\cal N},\sigma_2}~,
\label{eqn:unebornerap}
\end{equs}
and if $\sigma<1$, we have the two particular cases
\begin{equs}
\|uv\|_{{\cal N},\frac{1}{2}}&\leq
c_b~\sqrt{\delta}~\|u\|_{{\cal N},1}~\|v\|_{{\cal N},1}~,
\label{eqn:ununbornerap}\\
\|uv\|_{{\cal N},0}&\leq
c_b~\sqrt{\delta}~\|u\|_{\L^2}~\|v\|_{\L^2}~.
\label{eqn:zeroineqldld}
\end{equs}
\end{lemma}
\begin{proof}
We begin with (\ref{eqn:zeroineqldld}). We have
\begin{equs}
\|uv\|_{{\cal N},0}=\frac{\sqrt{\delta}}{q}\sup_{n\in{\bf Z}}
\sum_{m\in{\bf Z}}|u_n|~|v_{m-n}|
\leq
\frac{\sqrt{\delta}~L}{2\pi}~\|u\|_{l^2}~\|v\|_{l^2}
\leq\frac{\sqrt{\delta}}{2\pi}~\|u\|_{\L^2}~\|v\|_{L^2}~.
\end{equs}
For the other inequalities, we proceed as follows. Let
$p=\frac{q}{\delta}$, then we have
\begin{equs}
\|uv\|_{{\cal N},\sigma}
&\leq
\frac{1}{\sqrt{\delta}}
\sup_{n\in{\bf Z}}
\frac{(1+(pn)^2)^{\frac{\sigma}{2}}}{p}
\sum_{m\in{\bf Z}}
~|u_{m}||v_{n-m}|\\
&\leq
\left(\sup_{n\in{\bf Z}}
\sum_{m\in{\bf Z}}
\frac{p}{(1+(pm)^2)^{\frac{\sigma_1}{2}}}
\frac{(1+(pn)^2)^{\frac{\sigma}{2}}}{(1+(p(m-n))^2)^{\frac{\sigma_2}{2}}}\right)~
\sqrt{\delta}~
\|u\|_{{\cal N},\sigma_1}~
\|v\|_{{\cal N},\sigma_2}\\
&\leq
\left(\sup_{x\in{\bf R}}
\int_{-\infty}^{\infty}
\hspace{-4mm}\text{d}y~
\frac{1}{(1+y^2)^{\frac{\sigma}{2}}}
\frac{(1+x^2)^{\frac{\sigma}{2}}}{(1+(x-y)^2)^{\frac{\sigma}{2}}}\right)~
\sqrt{\delta}~\|u\|_{{\cal N},\sigma_1}~
\|v\|_{{\cal N},\sigma_2}~,\\
\|uv\|_{{\cal N},\frac{1}{2}}
&\leq
\left(\sup_{x\in{\bf R}}
\int_{-\infty}^{\infty}
\hspace{-4mm}\text{d}y~
\frac{1}{(1+y^2)^{\frac{1}{2}}}
\frac{(1+x^2)^{\frac{1}{4}}}{(1+(x-y)^2)^{\frac{1}{2}}}\right)~
\sqrt{\delta}~\|u\|_{{\cal N},1}~\|v\|_{{\cal N},1}~.
\end{equs}
Thus we define
\begin{equs}
c_b&=\max\left\{\frac{1}{2\pi}~~,~~S\Bigl(1,\frac{1}{2}\Bigr)
~~,~~S(\sigma,\sigma)\right\}
~,
\end{equs}
where
\begin{equs}
S(\sigma,\sigma')=\sup_{x\in{\bf R}}
\int_{-\infty}^{\infty}
\hspace{-4mm}\text{d}y~
\frac{1}{(1+y^2)^{\frac{\sigma}{2}}}
\frac{(1+x^2)^{\frac{\sigma'}{2}}}{(1+(x-y)^2)^{\frac{\sigma}{2}}}~.
\end{equs}
To see that $c_b<\infty$, we can assume without loss of generality that
$x\geq0$, and split the $y$ integration into two pieces,
$y\in(-\infty,x/2]$ and $y\in[x/2,\infty)$. We then have
\begin{equs}[2]
y\in(-\infty,x/2] &~~\Rightarrow~~& 
\frac{1+x^2}{1+(x-y)^2}\leq4~~&,~\frac{1}{1+(x-y)^2}\leq\frac{1}{1+y^2}~,\\
y\in[x/2,\infty) &~~\Rightarrow~~& 
\frac{1+x^2}{1+y^2}\leq4~~&,~~\frac{1}{1+y^2}\leq\frac{1}{1+(x-y)^2}~,
\end{equs}
from which we get
\begin{equs}
S(\sigma,\sigma)&\leq\int_{-\infty}^{x/2}
\hspace{-4mm}\text{d}y~
\frac{2^{\sigma}~}{(1+y^2)^{\frac{\sigma}{2}}}+
\int_{x/2}^{\infty}
\hspace{-4mm}\text{d}y~
\frac{2^{\sigma}~}{(1+(x-y)^2)^{\frac{\sigma}{2}}}
\leq\int_{-\infty}^{\infty}
\hspace{-4mm}\text{d}y~
\frac{2~2^{\sigma}~}{(1+y^2)^{\frac{\sigma}{2}}}\leq2^{\sigma}~15~,
\\
S\Bigl(1,\frac{1}{2}\Bigr)&\leq
\int_{-\infty}^{x/2}
\hspace{-4mm}\text{d}y~
\frac{2^{1/2}~}{(1+y^2)^{\frac{3}{4}}}+
\int_{x/2}^{\infty}
\hspace{-4mm}\text{d}y~
\frac{2^{1/2}~}{(1+(x-y)^2)^{\frac{3}{4}}}
\leq
\int_{-\infty}^{\infty}
\hspace{-4mm}\text{d}y~
\frac{2~2^{1/2}~}{(1+y^2)^{\frac{3}{4}}}
\leq2^{\sigma}~15~.
\end{equs}
The proof of the Lemma is completed.
\end{proof}
\begin{proposition}\label{prop:keyrap}
Let $\|u\|_{\sigma_1}<\infty$, $\|v\|_{\sigma_2}<\infty$ and
$\sigma=\min(\sigma_1,\sigma_2)\geq\frac{3}{2}$, then there exists a
constant $\Cm$ depending only on $\sigma$ such that
\begin{equs}
\|uv\|_{\sigma}\leq
\Cm~\sqrt{\delta}~
\|u\|_{\sigma_1}~
\|v\|_{\sigma_2}~,
\label{eqn:uneborneproprap}
\end{equs}
and if $\sigma<1$, we have the two particular cases
\begin{equs}
\|uv\|_{{\cal W},\frac{1}{2}}&\leq
\Cm~\sqrt{\delta}~\|u\|_{1}~\|v\|_{1}~,
\label{eqn:bricmontundeuxx}\\
\|uv\|_{{\cal W},0}&\leq
\Cm~\sqrt{\delta}~\|u\|_{\L^2}~\|v\|_{\L^2}~.
\label{eqn:bricmontzerox}
\end{equs}
\end{proposition}
\begin{proof}   
We first note that if $\sigma=\min(\sigma_1,\sigma_2)\geq\frac{3}{2}$, 
by Lemma \ref{lem:ldliestimerep}, we have
\begin{equs}
\|uv\|_{\L^2}\leq \|u\|_{\L^{\infty}}~\|v\|_{\L^2}\leq
\Cinfty ~\sqrt{\delta}~
\|u\|_{\sigma_1}~
\|v\|_{\sigma_2}~.
\end{equs}
So the $\L^2$ part of (\ref{eqn:uneborneproprap}) is proved. For the
$\|\cdot\|_{{\cal W},\sigma}$ part of (\ref{eqn:uneborneproprap}),
for (\ref{eqn:bricmontundeuxx}) and for (\ref{eqn:bricmontzerox}), we write $u=u_{<}+u_{>}$, where
$u_{<}=P_{<}u$ and $u_{>}=P_{>}u$ and the same for $v$. Then we have
\begin{equs}
\|uv\|_{{\cal W},\sigma}\leq
\|uv\|_{{\cal N},\sigma}\leq
 \|u_{<}~v_{<}\|_{{\cal N},\sigma}
+\|u_{<}~v_{>}\|_{{\cal N},\sigma}
+\|u_{>}~v_{<}\|_{{\cal N},\sigma}
+\|u_{>}~v_{>}\|_{{\cal N},\sigma}~.
\end{equs}
Clearly, $\|P_{>}f\|_{{\cal N},\sigma}\leq\|f\|_{{\cal
W},\sigma} \leq\|f\|_{\sigma}$, and we can apply directly
Lemma \ref{lem:keyestimbricmontrap} to the last term. The first three terms
are bounded using Lemmas \ref{lem:bricmontweird} and
\ref{lem:bricmontweirdweird} below.
\end{proof}
\begin{lemma}\label{lem:bricmontweird}
Let $\sigma\geq0$, then there exists a constant $C$ depending only on
$\sigma$ such that
\begin{equs}
\|(P_{>}u)~(P_{<}v)\|_{{\cal N},\sigma}\leq
C~\sqrt{\delta}~\|u\|_{\sigma}~\|v\|_{\sigma}~.
\label{eqn:uneautreborne}
\end{equs}
\end{lemma}
\begin{proof}   
Let $p=\frac{q}{\delta}$. By the Cauchy--Schwartz inequality, we have
$\|P_{<}v\|_{l^1}\leq\sqrt{\delta}~\|v\|_{\L^2}$ (see
(\ref{eqn:cauchyschwartz}) above), so that
\begin{equs}
\|(P_{>}u)~(P_{<}v)\|_{{\cal N},\sigma}&\leq
\frac{1}{\sqrt{\delta}}
\sup_{n\in{\bf Z}}
{\textstyle\frac{(1+(pn)^2)^{\frac{\sigma}{2}}}{p}}
\sum_{|m|\leq\frac{1}{p}}
|v_{m}|~|u_{n-m}|
\\
&\leq \|u\|_{{\cal W},\sigma}~
\|P_{<}v\|_{l^1}
\sup_{n\in{\bf Z}}
\sup_{|m|\leq1}
\left(\frac{1+n^2}{1+(m-n)^2}\right)^{\frac{\sigma}{2}}\\
&\leq
C~\sqrt{\delta}~\|u\|_{{\cal W},\sigma}~\|v\|_{\L^2}
\leq C~\sqrt{\delta}~\|u\|_{\sigma}~\|v\|_{\sigma}~.
\end{equs}
This completes the proof of the Lemma.
\end{proof}
\begin{lemma}\label{lem:bricmontweirdweird}
Let $\sigma\geq0$, then there exists a constant $C$ depending only on
$\sigma$ such that
\begin{equs}
\|(P_{<}u)~(P_{<}v)\|_{{\cal N},\sigma}\leq
C~\sqrt{\delta}~\|u\|_{\sigma}~\|v\|_{\sigma}~.
\label{eqn:unetroisborne}
\end{equs}
\end{lemma}
\begin{proof} 
Let $p=\frac{q}{\delta}$, we have
\begin{equs}
\|(P_{<}u)(P_{<}v)\|_{{\cal N},\sigma}&\leq
\frac{1}{\sqrt{\delta}}
\sup_{n\in{\bf Z}}
{\textstyle\frac{(1+(pn)^2)^{\frac{\sigma}{2}}}{p}}
\sum_{\stackrel{|m|\leq\frac{1}{p}}{|m-n|\leq\frac{1}{p}}}
|u_m|~|v_{n-m}|
\\&\leq
\frac{5^{\frac{\sigma}{2}}}{2\pi}~\sqrt{\delta}~L~
\sum_{\stackrel{|m|\leq\frac{1}{p}}{|m-n|\leq\frac{1}{p}}}
|u_m|~|v_{n-m}|
\leq
5^{\frac{\sigma}{2}}~\sqrt{\delta}~L~
\|u\|_{l^2}~
\|v\|_{l^2}
\\&\leq 5^{\frac{\sigma}{2}}~\sqrt{\delta}~
~\|u\|_{\L^2}
~\|v\|_{\L^2}
\leq
C~\sqrt{\delta}~\|u\|_{\sigma}~\|v\|_{\sigma}~.
\end{equs}
This completes the proof.
\end{proof}
\begin{proposition}
Let $\|u\|_{\sigma_1}<\infty$, $\|v\|_{\sigma_2}<\infty$ and
$\sigma=\min(\sigma_1,\sigma_2)\geq\frac{3}{2}$, then
\begin{equs}
\left\|\frac{u}{1+v}\right\|_{\sigma}
\leq 
\frac{\|u\|_{\sigma_1}}{1-\Cm~\sqrt{\delta}~\|v\|_{\sigma_2}}
~.
\end{equs}
for all $v$ satisfying $\Cm~\sqrt{\delta}~\|v\|_{\sigma_2}<1$, where 
$\Cm$ is the constant of Proposition \ref{prop:keyrap}.
\end{proposition}
\begin{proof}
The idea is to write a geometric series for $\frac{1}{1+v}$, and to use
that since $\Cm~\sqrt{\delta}~\|v\|_{\sigma_2}<1$, the series is
convergent. Indeed, using Proposition \ref{prop:keyrap} inductively, we have
\begin{equs}
\left\|\frac{u}{1+w}\right\|_{\sigma}\leq
\sum_{m\geq0}\|uv^m\|_{\sigma}
\leq \|u\|_{\sigma_1}
\sum_{m\geq0}
\Bigl(\Cm~\sqrt{\delta}~\|v\|_{\sigma_2}\Bigr)^m
\leq 
\frac{\|u\|_{\sigma_1}}{1-\Cm~\sqrt{\delta}~\|v\|_{\sigma_2}}~.
\label{eqn:geom}
\end{equs}
This completes the proof.
\end{proof}

\begin{proposition}
Let $\delta\geq2$, then
\begin{equs}
\left\|
\ed^{-\Lepsilon t}~
f(\cdot)
\right\|_{{\cal W},\sigma}
&\leq
\ed^{-4t}~
\|f(\cdot)\|_{{\cal W},\sigma}
~,\\
\left\|
\int_{0}^{t}
\hspace{-2mm}{\rm d}s~
\ed^{-\Lepsilon (t-s)}~
g'(\cdot,s)
\right\|_{{\cal W},\sigma}
&\leq
\sup_{0\leq s\leq t}\left\|\frac{g'(\cdot,s)}{\Lepsilon }\right\|_{{\cal W},\sigma}~,
\end{equs}
where $\ed^{-\Lepsilon t}$ is the propagation Kernel associated with
$\partial_t f=-\Lepsilon f$.
\end{proposition}
\begin{proof}
In Fourier space, the propagation Kernel $\ed^{-\Lepsilon t}$ acts as
\begin{equs}
\left(\ed^{-\Lepsilon t}~f\right)_n=
\ed^{-\Lepsilon (qn)t}~f_n~,~~~\text{with}~~~
\Lepsilonk=\frac{k^4-k^2}{1+\frac{\epsilon^2k^2}{2}}~.
\end{equs}
For $\epsilon\leq1$, $\delta\geq2$ and $|k|\geq\delta$, one has 
$\Lepsilonk\geq\Lepsilond\geq4$, 
which gives
\begin{equs}
\sup_{t\geq0}
\left\|
\ed^{-\Lepsilon t}~
f(\cdot)
\right\|_{{\cal W},\sigma}
&\leq
\ed^{-\Lepsilon (\delta)~t}~
\|f(\cdot)\|_{{\cal W},\sigma}
\leq
\ed^{-4t}~
\|f(\cdot)\|_{{\cal W},\sigma}~.
\end{equs}
Next, we use that for $q|n|\geq\delta$, we have
\begin{equs}
\left|\int_{0}^{t}
\hspace{-2mm}{\rm d}s
\left(\ed^{-\Lepsilon (t-s)}~
g'(\cdot,s)\right)_n\right|&\leq
\int_{0}^{t}\hspace{-2mm}{\rm d}s
~\ed^{-\Lepsilon (qn)~(t-s)}~
~|qn|~|g_n(s)|\\
&\leq
\frac{q}{\sqrt{\delta}}~\Bigl(1+\bigl(\frac{qn}{\delta}\bigr)^2\Bigr)^{-\sigma/2}
~\Lepsilon (qn)
\int_{0}^{t}\hspace{-2mm}{\rm d}s
~\ed^{-\Lepsilon (qn)~(t-s)}~
\left\|
\frac{g'(\cdot,s)}{\Lepsilon }
\right\|_{{\cal W},\sigma}
\\
&\leq
\frac{q}{\sqrt{\delta}}~\Bigl(1+\bigl(\frac{qn}{\delta}\bigr)^2\Bigr)^{-\sigma/2}
\left(
1-\ed^{-\Lepsilon (qn)~t}
\right)~
\sup_{0\leq s\leq t}
\left\|
\frac{g'(\cdot,s)}{\Lepsilon }
\right\|_{{\cal W},\sigma}~.
\end{equs}
Since $1-\ed^{-\Lepsilon (qn)~t}\leq1$ for
$qn\geq\delta\geq2$, the proof is completed.
\end{proof}

\begin{lemma}\label{lem:unsurLerap}
Let $\delta\geq2$, then
\begin{equs}
\left\|\frac{g'}{\Lepsilon }\right\|_{{\cal W},\sigma}
&\leq\frac{\sqrt{2}}{\delta}\left\|g\right\|_{{\cal W},\sigma-1}~,\\
\left\|\frac{\Gepsilon ~ g'}{\Lepsilon }\right\|_{{\cal W},\sigma}
&\leq\frac{2^{7/2}}{3~\delta^3}\left\|g\right\|_{{\cal W},\sigma-3}~.
\end{equs}
\end{lemma}
\begin{proof}   
We have
\begin{equs}
\left\|\frac{g'}{\Lepsilon }\right\|_{{\cal W},\sigma}
&\leq
\frac{\sqrt{\delta}}{q}
\sup_{|n|\geq\frac{\delta}{q}}
\Bigl(1+\bigl(\frac{qn}{\delta}\bigr)^2\Bigr)^{\sigma/2}
\Bigl(1+\frac{\epsilon^2(qn)^2}{2}\Bigr)~
\frac{|qn|~|g_n|}{(qn)^4-(qn)^2}\\
&\leq
\frac{1}{\delta}
\frac{\sqrt{\delta}}{q}
\sup_{|n|\geq\frac{\delta}{q}}
\frac{\Bigl(1+\bigl(\frac{qn}{\delta}\bigr)^2\Bigr)^{1/2}}{\frac{|qn|}{\delta}}
\frac{\Bigl(1+\frac{(qn)^2}{2}\Bigr)}{(qn)^2-1}
\Bigl(1+\bigl(\frac{qn}{\delta}\bigr)^2\Bigr)^{\frac{\sigma-1}{2}}~|g_n|\\
&\leq
\frac{1}{\delta}~\|g\|_{{\cal W},\sigma-1}~
\Bigl(\sup_{x\geq1}\frac{\sqrt{1+x^2}}{x}\Bigr)~
\Bigl(\sup_{x\geq2}\frac{1+\frac{x^2}{2}}{x^2-1}\Bigr)
\leq\frac{\sqrt{2}}{\delta}~\|g\|_{{\cal W},\sigma-1}~,
\end{equs}
and similarly
\begin{equs}
\left\|\frac{\Gepsilon ~ g'}{\Lepsilon }\right\|_{{\cal W},\sigma}
&\leq
\frac{\sqrt{\delta}}{q}
\sup_{|n|\geq\frac{\delta}{q}}
\Bigl(1+\bigl(\frac{qn}{\delta}\bigr)^2\Bigr)^{\sigma/2}
\frac{|qn|~|g_n|}{(qn)^4-(qn)^2}\\
&\leq
\frac{1}{\delta^3}~\|g\|_{{\cal W},\sigma-3}~
\Bigl(\sup_{x\geq1}\frac{\sqrt{1+x^2}}{x}\Bigr)^3~
\Bigl(\sup_{x\geq2}\frac{x^4}{x^4-x^2}\Bigr)
\leq\frac{2^{7/2}}{3~\delta^3}~\|g\|_{{\cal W},\sigma-3}~.
\end{equs}
This completes the proof.
\end{proof}

\newappendix{Bounds on nonlinear terms}\label{app:thefis}
We begin by recalling that 
\begin{equs}
F_0(s,\mu)&=
\alpha^2~\chi\Bigl(
\bigl(2+\epsilon^2(1+\alpha^2)\bigr)~s^2+
\frac{s'~\mu}{1+\epsilon^4~\alpha^2~s}
-\frac{2~\epsilon^2~\alpha^2~s~s''}{1+\epsilon^4~\alpha^2~s}
\Bigr)\\
&\phantom{=}~
-\frac{1}{4}~\Gepsilon ~\mu^2
-\frac{1}{4}~\Gepsilon ~(\mu^2)''~.
\end{equs}
Let $r=s-\frac{\epsilon^2}{2}s''$. We will now prove that
we can write $F_0(s,\mu)$ as
\begin{equs}
F_0(s,\mu)=F_{1}(s,\mu)+\Gepsilon ~ F_2(s,r,\mu)~,
\label{eqn:decdecdec}
\end{equs}
with
\begin{equs}
F_1(s,\mu)&=
\chi~\alpha^2~\Bigl(
\bigl(2+\epsilon^2(1+\alpha^2)\bigr)~s^2-
\frac{\alpha^2~(\epsilon^4~s')~s~\mu}{1+\epsilon^4~\alpha^2~s}
-\frac{2~\alpha^2~s~(\epsilon^2~s'')}{1+\epsilon^4~\alpha^2~s}
\Bigr)
~,\\
F_{2}(s,r,\mu)&=
-\frac{1}{4}~\mu^2-\frac{1}{4}~(\mu^2)''
+
\frac{\chi~\alpha^2}{2}\Bigl(
2~\mu~r'+4~\mu'~r-4~\mu'~s-
\mu''~(\epsilon^2~s')
\Bigr)~.
\end{equs}
To prove (\ref{eqn:decdecdec}) it is sufficient to show
that
\begin{equs}
s'\mu=\Gepsilon ~
\left(
\mu~r'+2\mu'~r-2\mu'~s-\frac{1}{2}\mu''~(\epsilon^2~s')
\right)~.
\end{equs}
But this is true because acting on both sides of this equation with
$\Bigl(1-\frac{\epsilon^2}{2}\partial_x^2\Bigr)$ gives
\begin{equs}
\Bigl(1-\frac{\epsilon^2}{2}\partial_x^2\Bigr)(\mu~s')&=
\mu~s'-\frac{\epsilon^2}{2}
\Bigl(
\mu~s'''+2\mu'~s''+\mu''~s'
\Bigr)~,
\end{equs}
and using $\epsilon^2~s''=2s-2r$ and
$\epsilon^2~s'''=2s'-2r'$, we get the desired
result.

\subsection{Bounds on $r_1$}
In this subsection we prove bounds on $r_1(\mu)$ and
$r_1(\mu_1)-r_1(\mu_2)$ in terms of $\|\mu\|_{\sigma}$.
We recall that $r_1(\mu)$ is defined by
\begin{equs}
r_1(\mu)=-\frac{1}{32}~(4\mu'+\epsilon^2~\mu^2)~,
\label{eqn:defsrzero}
\end{equs}
and assume that the following holds
\begin{equs}
\|\mu\|_{\sigma}&\leq c_{\mu}~\rho~,~~~
\|\mu_i\|_{\sigma}\leq c_{\mu}~\rho~,~~~
\delta=c_{\delta}~\rho^2~,~~~
\epsilon\leq1~,
\label{eqn:asumunnew}
\end{equs}
where $c_{\delta}>1$. 
\begin{proposition}\label{prop:run}
Assume that (\ref{eqn:asumunnew}) hold, and that $r_1$ is defined by
(\ref{eqn:defsrzero}). Then there exists a constant $c_{r_1}$ such that
\begin{equs}
\|r_1(\mu)\|_{\sigma-1}
&\leq c_{r_1}~\delta~\rho~,
\label{eqn:szerosdo}  \\
\|r_1(\mu_1)-r_1(\mu_2)\|_{\sigma-1}
&\leq c_{r_1}~\delta~\|\mu_1-\mu_2\|_{\sigma-1}~.
\label{eqn:szerosdoa}
\end{equs}
\end{proposition}
\begin{proof}
Using Lemma \ref{lem:ldliestime}, Proposition \ref{prop:key} and the assumptions
(\ref{eqn:asumunnew}), we have
\begin{equs}
\|r_1(\mu)\|_{\sigma-1}&\leq
\frac{\delta}{8}~\|\mu\|_{\sigma}+\frac{\epsilon^2}{32}~\|\mu^2\|_{\sigma}
\leq \frac{\delta}{8}~\|\mu\|_{\sigma}
\bigl(1+\frac{\Cm}{4~\sqrt{\delta}}~\|\mu\|_{\sigma}\bigr)
\leq\frac{1}{8}\Bigl(1+\frac{\Cm~c_{\mu}}{4~\sqrt{c_{\delta}}}\Bigr)~\delta~\rho~,
\end{equs}
and since $\mu_1^2-\mu_2^2=(\mu_1-\mu_2)(\mu_1+\mu_2)$, we have
\begin{equs}
\|r_1(\mu_1)-r_1(\mu_2)\|_{\sigma}&\leq
\frac{\delta}{8}~\|\mu_1-\mu_2\|_{\sigma}+
\frac{\epsilon^2}{32}~\|\mu_1^2-\mu_2^2\|_{\sigma}\\
&\leq \frac{\delta}{8}~\|\mu_1-\mu_2\|_{\sigma}
\bigl(1+\frac{\Cm}{4~\sqrt{\delta}}~\|\mu_1+\mu_2\|_{\sigma}\bigr)\\
&\leq 
\frac{1}{8}\Bigl(1+\frac{\Cm~c_{\mu}}{2~\sqrt{c_{\delta}}}\Bigr)~
\delta~\|\mu_1-\mu_2\|_{\sigma}~.
\end{equs}
Setting
$c_{r_1}=\frac{1}{8}\Bigl(1+\frac{\Cm~c_{\mu}}{2~\sqrt{c_{\delta}}}\Bigr)$
completes the proof.
\end{proof}

\subsection{Bounds on $F_0$, $F_1$ and $F_2$}
In this subsection, we define
\begin{equs}
F_0(\mu)&=F_0(\Gepsilon ~ r(\mu),\mu)~,\\
F_1(\mu)&=F_1(\Gepsilon ~ r(\mu),\mu)~,\\
F_2(\mu)&=F_2(\Gepsilon ~ r(\mu),r(\mu),\mu)~,
\end{equs}
and we suppose that for all $\mu$, $\mu_1$ and $\mu_2$ in
${\cal B}_{\sigma}(c_{\mu}~\rho)$, we have
\begin{equs}
\|r(\mu)\|_{\sigma-1}&\leq c_{r}~\delta~\rho~,\\
\|r(\mu_1)-r(\mu_2)\|_{\sigma-1}&\leq c_{r}~\delta~
\|\mu_1-\mu_2\|_{\sigma}~,
\end{equs}
for some constant $c_r$ of order $c_{r_1}$, where $c_{r_1}$ is given by
Proposition \ref{prop:run}. See also Section \ref{sec:ampli}.

\begin{theorem}\label{thm:lesfonctions}
Let $\delta=c_{\delta}~\rho^2$. There exist
constants $c_{\epsilon}$ and $c_{F_0}$ such that
\begin{equs}
\|F_0(\mu)\|_{\L^2}\leq c_{F_0}~\delta^{5/2}~\rho^2~,
\label{eqn:boundfld}
\\
\|F_0(\mu)\|_{\sigma-2}\leq c_{F_0}~\delta^{5/2}~\rho^2~,
\label{eqn:boundfsmdeux}
\\
\left\|
\frac{F_0(\mu)'}{\Lepsilon }
\right\|_{{\cal W},\sigma}
\leq c_{F_0}~\delta^{5/2}~\rho^2~,
\label{eqn:boundfun}
\end{equs}
for all $\epsilon\leq c_{\epsilon}~\delta^{-3/8}~\rho^{-1/4}$ and
for all $\mu\in{\cal B}_{\sigma}(c_{\mu}~\rho)$.
\end{theorem}
\begin{proof}
We recall that, defining $s(\mu)=\Gepsilon ~ r(\mu)$, we have
\begin{equs}
F_0(\mu)&=
\chi~\alpha^2~\Bigl(
\bigl(2+\epsilon^2(1+\alpha^2)\bigr)~s^2+
\frac{s'~\mu}{1+\epsilon^4~\alpha^2~s}
-\frac{2~\epsilon^2~\alpha^2~s~s''}{1+\epsilon^4~\alpha^2~s}
\Bigr)
-\frac{1}{4}~\Gepsilon ~\mu^2-\frac{1}{4}~\Gepsilon ~(\mu^2)''\\
&=F_1(\mu)+\Gepsilon ~ F_2(\mu)~,\\
F_1(\mu)&=
\chi~\alpha^2~\Bigl(
\bigl(2+\epsilon^2(1+\alpha^2)\bigr)~s^2-
\frac{\alpha^2~(\epsilon^4~s')~s~\mu}{1+\epsilon^4~\alpha^2~s}
-\frac{2~\alpha^2~s~(\epsilon^2~s'')}{1+\epsilon^4~\alpha^2~s}
\Bigr)
~,\\
F_{2}(\mu)&=
-\frac{1}{4}~\mu^2-\frac{1}{4}~(\mu^2)''
+
\frac{\chi~\alpha^2}{2}\Bigl(
2~\mu~r'+4~\mu'~r-4~\mu'~s-
\mu''~(\epsilon^2~s')
\Bigr)~,
\end{equs}
where we omitted the $\mu$ dependence of $s$ and $r$ for
concision. Note that from the definition of $\Gepsilon $, we have
\begin{equs}
\left\|s\right\|_{\sigma-1}\leq\left\|r\right\|_{\sigma-1}~,~~~
\left\|\epsilon s'\right\|_{\sigma-1}\leq\sqrt{2}\left\|r\right\|_{\sigma-1}
~~~\mbox{and}~~~
\left\|\epsilon^2 s''\right\|_{\sigma-1}\leq2\left\|r\right\|_{\sigma-1}~,
\end{equs}
since $s=\Gepsilon ~ r$. Using these inequalities and
Propositions \ref{prop:key} and \ref{prop:division}, we have
\begin{equs}
\left\|s^2\right\|_{\sigma-1}
&\leq \Cm ~\sqrt{\delta}~
\|r\|_{\sigma-1}^2
\label{eqn:ffline}
~,\\
\left\|
\frac{s~(\epsilon^2 s)''}{1+\epsilon^4~\alpha^2~s}
\right\|_{\sigma-1}
&\leq 
\frac{C~\sqrt{\delta}~
\|r\|_{\sigma-1}^2}
{1-\Cm ~\epsilon^4~\alpha^2~\sqrt{\delta}~\|s\|_{\sigma-1}}
~,\label{eqn:secline}\\
\Bigl\|\frac{\mu~s'}{1+\epsilon^4~\alpha^2~s}\Bigr\|_{\sigma-2}&\leq
\frac{C~\delta^{3/2}~\|\mu\|_{\sigma}~\|r\|_{\sigma-1}}
{1-\Cm~\epsilon^4~\alpha^2~\sqrt{\delta}~\|r\|_{\sigma-1}}
~,\label{eqn:mligne}\\
\left\|
\frac{\mu~s~(\epsilon^4~s)'}{1+\epsilon^4~\alpha^2~s}
\right\|_{\sigma-1}
&\leq 
\frac{C~\epsilon^3~\delta~
\left\|\mu\right\|_{\sigma-1}
\left\|r\right\|_{\sigma-1}^2}
{1-C ~\epsilon^4~\alpha^2~\sqrt{\delta}~\|r\|_{\sigma-1}}
~,\label{eqn:lastline}\\
\left\|
\mu''~(\epsilon^2~s')
\right\|_{\sigma-3}&\leq 
\epsilon~C~\delta^{5/2}
\left\|\mu\right\|_{\sigma}~
\left\|r\right\|_{\sigma-1}
~,\label{eqn:itsnottrue}
\\
\|\mu^2\|_{\sigma-3}&
\leq C~\sqrt{\delta}~\|\mu\|_{\sigma}^2
~,\label{eqn:pourlesdiff}\\
\|(\mu^2)''\|_{\sigma-3}&\leq
C~\delta^{5/2}~\|\mu\|_{\sigma}^2
\label{eqn:jun}
~,\\
\|\mu~r'\|_{\sigma-3}&\leq
C~\delta^{3/2}~\|\mu\|_{\sigma}
~\|r\|_{\sigma-1}
\label{eqn:jdeux}
~,\\
\|\mu'~r\|_{\sigma-3}+
\|\mu'~s\|_{\sigma-3}&\leq
C~\delta^{3/2}~\|\mu\|_{\sigma}
\|r\|_{\sigma-1}
~.
\label{eqn:llline}
\end{equs}
By hypothesis, we have $\|r\|_{\sigma-1}\leq c_r~\delta~\rho$. We now choose 
$c_{\epsilon}=\bigl(\frac{1}{2~\Cm~c_r~\alpha^2}\bigr)^{1/4}$, and get that for all 
$\mu\in{\cal B}_{\sigma}(c_{\mu}~\rho)$, and for all $\epsilon\leq
c_{\epsilon}~\delta^{-3/8}~\rho^{-1/4}\leq c_{\epsilon}~c_{\delta}^{-3/8}~\rho^{-1}$,
\begin{equs}
\frac{1}{1-\Cm ~\epsilon^4~\alpha^2~\sqrt{\delta}~\|r\|_{\sigma-1}}&\leq2~,\\
\epsilon^3~\sqrt{\delta}~\|\mu\|_{\sigma}\leq
c_{\mu}~\epsilon^3~\sqrt{\delta}~\rho
&=c_{\mu}~\sqrt{c_{\delta}}~\epsilon^3~\rho^2
\leq c_{\mu}~\sqrt{c_{\delta}}~c_{\epsilon}^{3/2}~c_{\delta}^{-9/8}~.
\end{equs}
Hence the r.h.s. of the inequalities (\ref{eqn:ffline})--(\ref{eqn:llline})
are all bounded by some constant times $\delta^{5/2}~\rho^2$, except
(\ref{eqn:itsnottrue}) which is bounded by a constant times $\delta^{7/2}~\rho^2$.
From (\ref{eqn:ffline})--(\ref{eqn:lastline}) and Lemma
\ref{lem:ldliestime} for the two last terms of $F_0(\mu)$, we see that 
(\ref{eqn:boundfsmdeux}) holds, then (\ref{eqn:boundfld}) also holds
because $\|F_0(\mu)\|_{\L^2}\leq\|F_0(\mu)\|_{\sigma-2}$. For
(\ref{eqn:boundfun}), we use first Lemma \ref{lem:unsurLe} which gives
\begin{equs}
\left\|\frac{F_0(\mu)'}{\Lepsilon }
\right\|_{{\cal W},\sigma}
\leq
\frac{\sqrt{2}}{\delta}\left\|F_1(\mu)\right\|_{{\cal W},\sigma-1}+
\frac{2^{7/2}}{3~\delta^3}\left\|F_2(\mu)\right\|_{{\cal W},\sigma-3}~.
\end{equs}
Using (\ref{eqn:itsnottrue})--(\ref{eqn:llline}) for the $F_2$--term
and (\ref{eqn:ffline}), (\ref{eqn:secline}) and (\ref{eqn:lastline}) for the $F_1$--term
completes the proof.
\end{proof}

\begin{theorem}\label{thm:lesdifferences}
Let $\delta= c_{\delta}~\rho^2$, and let $c_{\epsilon}$ be given by Theorem
\ref{thm:lesfonctions}. There exists a constant $c_{F_0}$ such that
\begin{equs}
\|F_0(\mu_1)-F_0(\mu_2)\|_{\L^2}
\leq c_{F_0}~\delta^{5/2}~\rho~\|\mu_1-\mu_2\|_{\sigma}~,
\label{eqn:diffun}
\\
\|F_0(\mu_1)-F_0(\mu_2)\|_{\sigma-2}
\leq c_{F_0}~\delta^{5/2}~\rho~\|\mu_1-\mu_2\|_{\sigma}~,
\label{eqn:diffsun}
\\
\left\|
\frac{F_0(\mu_1)'-F_0(\mu_2)'}{\Lepsilon }
\right\|_{{\cal W},\sigma}
\leq c_{F_0}~\delta^{5/2}~\rho~\|\mu_1-\mu_2\|_{\sigma}~,
\label{eqn:diffdeux}
\end{equs}
for all $\epsilon\leq c_{\epsilon}^{1/4}~\rho^{-1/4}~\delta^{-3/8}$ and
for all $\mu_i\in{\cal B}_{\sigma}(c_{\mu}~\rho)$.
\end{theorem}
\begin{proof}
We use the three following equalities
\begin{equs}
a_1b_1-a_2b_2&=
(a_1-a_2)b_1+(b_1-b_2)a_2~,
\label{eqn:sillicone}
\\
a_1b_1c_1-a_2b_2c_2&=
(a_1-a_2)b_1c_1+(b_1-b_2)a_2c_1+(c_1-c_2)a_2b_2~,
\label{eqn:silliconetwo}\\
\frac{f(\mu_1)}{1+\epsilon^4~s(\mu_1)}-\frac{f(\mu_2)}{1+\epsilon^4~s(\mu_2)}
&=
\frac{f(\mu_1)-f(\mu_2)}{1+\epsilon^4~s(\mu_1)}+
\frac{f(\mu_2)}{1+\epsilon^4~s(\mu_2)}
~\frac{\epsilon^4~\Delta s}{1+\epsilon^4~s(\mu_1)}~,
\label{eqn:simpliflyer}
\end{equs}
where $\Delta s=s(\mu_2)-s(\mu_1)$. Then we proceed exactly as we did in
the proof of Theorem \ref{thm:lesfonctions}. For example, in
(\ref{eqn:pourlesdiff}), we used that
\begin{equs}
\|\mu^2\|_{\sigma-3}\leq \Cm~\sqrt{\delta}~\|\mu\|_{\sigma}^2
\leq \Cm~c_{\mu}^2~\sqrt{\delta}~\rho^2~,
\label{eqn:tutun}
\end{equs}
then here, this bound is replaced by
\begin{equs}
\|\mu_1^2-\mu_2^2\|_{\sigma-3}&\leq
\Cm~\sqrt{\delta}~\|\mu_1+\mu_2\|_{\sigma}~\|\mu_1-\mu_2\|_{\sigma}
\leq 2~\Cm~c_{\mu}~\sqrt{\delta}~\rho~\|\mu_1-\mu_2\|_{\sigma}~.
\label{eqn:tutdeux}
\end{equs}
All other estimates are similar.
\end{proof}
\begin{corollary}\label{cor:onFsept}
Let $\delta=c_{\delta}~\rho$, and define
$F_7(\mu)=\frac{\epsilon^2}{8}~\bigl(\partial_x+\frac{\epsilon^2~\mu}{2}\bigr)
F_0(\mu)'$. Then there exist constants $c_{\epsilon}$ and
$c_{F_7}$ such that for all $\epsilon\leq c_{\epsilon}~\rho^{-2}$ and
$\mu_i\in{\cal B}_{\sigma}(c_{\mu}~\rho)$ the following bounds hold
\begin{equs}
\left\|F_7(\mu_i)\right\|_{\L^2}+
\left\|{\cal L}_{\mu,r}~{\cal L}_{v}^{-1}~F_7(\mu_i)'\right\|_{\L^2}
&\leq
c_{F_{7}}~\delta^{5/2}~\rho^2~,\label{eqn:ldlikeun}\\
\left\|
{\cal L}_{\mu,r}~{\cal L}_{v}^{-1}~
\Delta F_7'
\right\|_{\L^2}&\leq
c_{F_{7}}~\delta^{5/2}~\rho~\|\mu_1-\mu_2\|_{\sigma}~,
\label{eqn:ldlikedeux}\\
\left\|\frac{{\cal L}_{\mu,r}}{{\cal
L}_{\epsilon}~\Gepsilon ~{\cal L}_r}~F_7(\mu_i)'\right\|_{{\cal W},\sigma}&\leq
c_{F_{7}}~\max\Bigl(2,\frac{\alpha^2}{1-\alpha^2}\Bigr)~\delta^{5/2}~\rho^2~,\label{eqn:wlikeun}\\
\left\|\frac{{\cal L}_{\mu,r}}{{\cal
L}_{\epsilon}~\Gepsilon ~{\cal L}_r}~\Delta F_7'\right\|_{{\cal W},\sigma}&\leq 
c_{F_{7}}~\max\Bigl(2,\frac{\alpha^2}{1-\alpha^2}\Bigr)~\delta^{5/2}~\rho~\|\mu_1-\mu_2\|_{\sigma}~,\label{eqn:wlikedeux}
\end{equs}
where $\Delta F_7=\bigl(F_7(\mu_1)-F_7(\mu_2)\bigr)$.
\end{corollary}
\begin{proof}
We first use that $\|{\cal L}_{\mu,r}~{\cal L}_{v}^{-1}~f'\|_{\L^2}\leq
16\|f\|_{\L^2}$ (see Lemma \ref{lem:someproperties} in Appendix
\ref{app:therdeuxmap}), so that for the $\L^2$ bounds
(\ref{eqn:ldlikeun})--(\ref{eqn:ldlikedeux}), we need only bound
$\|F_7(\mu_i)\|_{\L^2}$ and $\|F_7(\mu_1)-F_7(\mu_2)\|_{\L^2}$.
Then we have also $\|f\|_{\L^2}\leq\|f\|_{\sigma'}$ for any $\sigma'>0$,
from which we get
\begin{equs}
\|F_7(\mu_i)\|_{\L^2}&\leq
\epsilon^2\|F_0(\mu_i)''\|_{\L^2}+
\frac{\epsilon^4}{2}
\|\mu_i~F_0(\mu_i)'\|_{\L^2}\\
&\leq 
\epsilon^2~\|F_0(\mu_i)''\|_{\sigma-4}+
\frac{\epsilon^4~\Cm~\sqrt{\delta}}{2}
\|\mu_i\|_{\sigma-3}~\|F_0(\mu_i)'\|_{\sigma-3}\\
&\leq
\epsilon^2~\delta^2~\Bigl(
\Cinfty ^2+\frac{\epsilon^2~\Cm ~\Cinfty ~c_{\mu}}{\sqrt{c_{\delta}}}
\Bigr)~\|F_0(\mu_i)\|_{\sigma-2}
\leq C~\delta^{5/2}~\rho^2~,
\end{equs}
since $\|F_0(\mu_i)\|_{\sigma-2}\leq c_{F_0}~\delta^{5/2}~\rho^2$ and
$\epsilon\leq c_{\epsilon}~\rho^{-2}$. Similarly, since
\begin{equs}
\mu_1~F_0(\mu_1)'-\mu_2~F_0(\mu_2)'=
\frac{1}{2}\Delta\mu~\bigl(F_0(\mu_1)+F_0(\mu_2)\bigr)'+
\frac{1}{2}(\mu_1+\mu_2)~\Delta F_0'~,
\label{eqn:ptittrux}
\end{equs}
where $\Delta\mu=\mu_1-\mu_2$ and  $\Delta F_0=F_0(\mu_1)-F_0(\mu_2)$,
we also have
\begin{equs}
\|\Delta F_7\|_{\L^2}&\leq
C_1~\|\Delta F_0\|_{\sigma-2}
+C_2~\epsilon^2~\|F_0(\mu_1)+F_0(\mu_2)\|_{\sigma-2}~\|\mu_1-\mu_2\|_{\sigma}~.
\end{equs}
The proof of (\ref{eqn:ldlikedeux}) is completed noting that
$\|\Delta F_0\|_{\sigma-2}\leq
c_{F_0}~\delta^{5/2}~\rho~\|\mu_1-\mu_2\|_{\sigma}$, and using again
$\epsilon\leq c_{\epsilon}~\rho^{-2}$.

For the proof of (\ref{eqn:wlikeun}) and (\ref{eqn:wlikedeux}), we define
$\balpha=\max(2,\frac{\alpha^2}{1-\alpha^2})$, then (see Lemma
\ref{lem:someproperties} of Appendix \ref{app:therdeuxmap})
\begin{equs}
\epsilon^2~\left\|
\frac{{\cal L}_{\mu,r}}{\Gepsilon ~{\cal L}_r}~f''
\right\|_{{\cal W},\sigma}
&\leq 
8~
\max\Bigl(\frac{1}{3},\frac{\alpha^2}{1-\alpha^2}\Bigr)
~\|f\|_{{\cal W},\sigma}
\leq
C~\balpha~\|f\|_{{\cal W},\sigma}~,
\\
\left\|
\frac{{\cal L}_{\mu,r}}{\Lepsilon ~\Gepsilon ~{\cal L}_r}~f'
\right\|_{{\cal W},\sigma}
&\leq
C~\balpha~\delta^{-3}~\|f\|_{{\cal W},\sigma-3}~
\end{equs}
for some constant $C$. Using this, we conclude that 
\begin{equs}
\left\|\frac{{\cal L}_{\mu,r}}{{\cal
L}_{\epsilon}~\Gepsilon ~{\cal L}_r}~F_7(\mu_i)'\right\|_{\L^2}&\leq 
C~\balpha\left(
\left\|\frac{F_0(\mu_i)'}{\Lepsilon }\right\|
+\frac{\epsilon^4~\delta^{-3}}{2}
\|\mu_i~F_0(\mu_i)'\|_{\sigma-3}
\right)\\
&\leq 
C~\balpha\left(
\left\|\frac{F_0(\mu_i)'}{\Lepsilon }\right\|
+\frac{\epsilon^4~\delta^{-3/2}}{2}
\|\mu_i\|_{\sigma}~\|F_0(\mu_i)\|_{\sigma-2}
\right)\\
&\leq 
C'~\balpha\left(
\left\|\frac{F_0(\mu_i)'}{\Lepsilon }\right\|
+\|F_0(\mu_i)\|_{\sigma-2}
\right)\leq C''~\balpha~\delta^{5/2}~\rho^2~.
\end{equs}
The proof of (\ref{eqn:wlikedeux}) is very similar (use e.g.
(\ref{eqn:ptittrux}) and proceed as for (\ref{eqn:wlikeun}) above).
\end{proof}

\subsection{Bounds on $F_3$}
We begin by recalling that
\begin{equs}
F_3(s,\mu)&=
-\alpha^2~\chi~
\Bigl({\textstyle\frac{3}{2}}~s^2
+\epsilon^2~{\textstyle\frac{1}{32}}~s~\mu^2
+{\textstyle\frac{\alpha^2}{2}}~\epsilon^4~s^3\Bigr)~.
\end{equs}
The map $F_3$ satisfies the two following propositions. The first one is
used for the properties of $s$, and the second one for those of $r_2$.
\begin{proposition}\label{prop:onFtrois}
Let $c_{\mu},c_{r_1},c_{s_0},\rho>0$, $\delta>2$, and
$c_{s}>2(c_{r_1}+c_{s_0})$. There exists a constant $c_{\epsilon}$ such
that for all $\epsilon\leq c_{\epsilon}~\delta^{-5/4}~\rho^{-1/2}$, for all
$\mu_i\in{\cal B}_{\sigma}(c_{\mu}~\rho)$ and for all $s_i\in{\cal
B}_{\sigma-1}(c_{s}~\delta~\rho)$ the following
bounds hold
\begin{equs}
\frac{2~\epsilon^4}{\chi}~\|F_3(s_i,\mu_i)\|_{\sigma-1}&
<\Bigl(
1-\frac{2(c_{r_1}+c_{s_0})}{c_s}
\Bigr)~c_{s}~\delta~\rho~,
\label{eqn:sansdiff}
\\
\frac{2~\epsilon^4}{\chi}~\|F_3(s_1,\mu_i)-F_3(s_2,\mu_i)\|_{\sigma-1}&\leq
\Bigl(
1-\frac{2(c_{r_1}+c_{s_0})}{c_s}
\Bigr)~\|s_1-s_2\|_{\sigma-1}~,
\label{eqn:avecdiffun}
\\
\frac{2~\epsilon^4}{\chi}~\|F_3(s_i,\mu_1)-F_3(s_i,\mu_2)\|_{\sigma-1}&\leq
2~c_{s_0}~\delta~\|\mu_1-\mu_2\|_{\sigma-1}~.
\label{eqn:avecdiffdeux}
\end{equs}
\end{proposition}
\begin{proof}
We have
\begin{equs}
\epsilon^4\|s_i^2\|_{\sigma-1}&\leq
\epsilon^4~\Cm~\sqrt{\delta}~(c_s~\delta~\rho)^2=
\bigl(\epsilon^4~\delta^{3/2}~\rho~\Cm~c_s\bigr)~c_s~\delta~\rho~,\\
\epsilon^6\|s_i~\mu_i^2\|_{\sigma-1}&\leq
\bigl(\epsilon^6~\delta~\rho^2~\Cm^2~c_{\mu}^2\bigr)~c_s~\delta~\rho~,\\
\epsilon^8\|s_i^3\|_{\sigma-1}&\leq
\epsilon^8~\Cm^2~\delta~(c_s~\delta~\rho)^3=
\bigl(\epsilon^8~\delta^3~\rho^2~\Cm^2~c_s^2\bigr)~c_s~\delta~\rho~,\\
\epsilon^4\|s_1^2-s_2^2\|_{\sigma-1}&\leq
\epsilon^4~\Cm~\sqrt{\delta}~\bigl(\|s_1\|_{\sigma-1}+\|s_2\|_{\sigma-1}\bigr)
\|s_1-s_2\|_{\sigma-1}\\
&\leq
\bigl(\epsilon^4~\delta^{3/2}~\rho~2~\Cm~c_s\bigr)~\|s_1-s_2\|_{\sigma-1}~,\\
\epsilon^8\|s_1^3-s_2^3\|_{\sigma-1}&\leq
\epsilon^8~\Cm^2~\delta~\bigl(\|s_1\|_{\sigma-1}+\|s_2\|_{\sigma-1}\bigr)^2
\|s_1-s_2\|_{\sigma-1}\\&\leq
\bigl(\epsilon^8~\delta^{3}~\rho^2~4~\Cm^2~c_s^2\bigr)~\|s_1-s_2\|_{\sigma-1}~,\\
\epsilon^6\|s_i~\mu_1^2-s_i~\mu_2^2\|_{\sigma-1}&\leq
\epsilon^6~\Cm^2~\delta~\|s_i\|_{\sigma-1}~
\bigl(\|\mu_1\|_{\sigma}+\|\mu_2\|_{\sigma}\bigr)
\|\mu_1-\mu_2\|_{\sigma}\\
&\leq\bigl(
\epsilon^4~\delta~\rho^2~\Cm^2~c_{\mu}~c_s
\bigr)~\delta~\|\mu_1-\mu_2\|_{\sigma}~.
\end{equs}
Choosing $c_{\epsilon}$ independent of $\delta$ and $\rho$ and
sufficiently small, we can satisfy
(\ref{eqn:sansdiff})--(\ref{eqn:avecdiffdeux}).
\end{proof}

\begin{proposition}\label{prop:onFtroisrd}
Let $\delta=c_{\delta}~\rho^2>2$, $c_{\mu},c_{s},\rho>0$ and ${\cal
L}_s=1-\frac{\epsilon^2}{2}\partial_x^2$. There exist constants
$c_{\epsilon}$ and $m_{\epsilon}$ such that for all $\epsilon\leq
c_{\epsilon}~\rho^{-m_{\epsilon}}$, for all $\mu_i\in{\cal
B}_{\sigma}(c_{\mu}~\rho)$ and for all maps $s$ satisfying
$\|s(\mu_i)\|_{\sigma-1}\leq c_{s}~\delta~\rho$ and
$\|s(\mu_1)-s(\mu_2)\|_{\sigma-1}\leq c_s~\delta~\|\mu_1-\mu_2\|$
the following bounds hold
\begin{equs}
\|{\cal L}_s~F_3(s(\mu_i),\mu_i)\|_{\sigma-3}&\leq
c_{F_3}~\delta^{5/2}~\rho^2~,
\label{eqn:onFtroisrd}
\\
\|{\cal L}_s~F_3(s(\mu_1),\mu_1)
 -{\cal L}_s~F_3(s(\mu_2),\mu_2)\|_{\sigma-3}&\leq
c_{F_3}~\delta^{5/2}~\rho~\|\mu_1-\mu_2\|_{\sigma}
\label{eqn:onFtroisrddiff}
~,
\end{equs}
for some constant $c_{F_3}$.
\end{proposition}
\begin{proof}
We first note that since ${\cal L}_s=1-\frac{\epsilon^2}{2}\partial_x^2$,
we have
\begin{equs}
\|{\cal L}_s~f\|_{\sigma-3}\leq
\Bigl(
1+\frac{\Cinfty ~\epsilon^2~\delta^2}{2}
\Bigr)~\|f\|_{\sigma-1}\leq
\Bigl(
1+\frac{\Cinfty ~c_{\epsilon}^2~c_{\delta}^2}{2}
\Bigr)~\|f\|_{\sigma-1}
\leq C~\|f\|_{\sigma-1}
~,
\end{equs}
if $m_{\epsilon}\geq2$. Let $s(\mu_i)=s_i$, then, as in the proof of
Proposition \ref{prop:onFtrois}, we have
\begin{equs}
\|s_i^2\|_{\sigma-1}
+\epsilon^2\|s_i~\mu_i^2\|_{\sigma-1}
+\epsilon^4\|s_i^3\|_{\sigma-1}
&\leq\delta^{5/2}~\rho^2~\Bigl(
\Cm~c_s^2+
\frac{\epsilon^2~\Cm^2~c_{\mu}^2~c_s~\rho}{\sqrt{\delta}}+
\epsilon^4~\delta^{3/2}~\rho~\Cm^2~c_s^3\Bigr)\\
&\leq
\delta^{5/2}~\rho^2~\Bigl(
\Cm~c_s^2+
\frac{c_{\epsilon}^2~\Cm^2~c_{\mu}^2~c_s}{\sqrt{c_{\delta}}}+
c_{\epsilon}^4~c_{\delta}^{3/2}~\Cm^2~c_s^3\Bigr)~,
\end{equs}
if $m_{\epsilon}\geq1$. The proof of
(\ref{eqn:onFtroisrddiff}) is similar, we omit the details.
\end{proof}

\subsection{Bounds on $F_4$}
We begin by recalling that
\begin{equs}
F_4(s,\mu)&=-{\textstyle\frac{\alpha^2~\chi}{8}}(2s'\mu+s\mu')~,
\end{equs}
then we have the
\begin{proposition}\label{prop:onFquatrerd}
Let $\delta=c_{\delta}~\rho^2>2$, $c_{\mu},c_{s},\rho>0$ and ${\cal
L}_s=1-\frac{\epsilon^2}{2}\partial_x^2$. There exist constants
$c_{\epsilon}$ and $m_{\epsilon}$ such that for all $\epsilon\leq
c_{\epsilon}~\rho^{-m_{\epsilon}}$, for all $\mu_i\in{\cal
B}_{\sigma}(c_{\mu}~\rho)$ and for all maps $s$ satisfying
$\|{\cal L}_s~s(\mu_i)\|_{\sigma-1}\leq c_{s}~\delta~\rho$ and $\|{\cal
L}_s~s(\mu_1)-{\cal L}_s~s(\mu_2)\|_{\sigma-1}\leq
c_s~\delta~\|\mu_1-\mu_2\|$ the following bounds hold
\begin{equs}
\|{\cal L}_s~F_4(s(\mu_i),\mu_i)\|_{\sigma-3}&\leq
c_{F_4}~\delta^{5/2}~\rho^2~,
\label{eqn:onFquatrerd}
\\
\|{\cal L}_s~F_4(s(\mu_1),\mu_1)
 -{\cal L}_s~F_4(s(\mu_2),\mu_2)\|_{\sigma-3}&\leq
c_{F_4}~\delta^{5/2}~\rho~\|\mu_1-\mu_2\|_{\sigma}
\label{eqn:onFquatrerddiff}
~,
\end{equs}
for some constant $c_{F_4}$.
\end{proposition}
\begin{proof}   
We first note that $\|f\|_{\sigma}\leq\|{\cal L}_s~f\|_{\sigma}$, and that
(see Proposition \ref{prop:onFtroisrd})
$\|{\cal L}_s~f\|_{\sigma-3}\leq C~\|f\|_{\sigma-1}$ if
$m_{\epsilon}\geq2$. We thus have
\begin{equs}
\|{\cal L}_s~(s(\mu_i)~\mu_i')\|_{\sigma-3}&\leq
C~\sqrt{\delta}~\|s(\mu_i)\|_{\sigma-1}~\|\mu_i'\|_{\sigma-1}
\leq C~\delta^{5/2}~\rho^2~,
\end{equs}
and for the other term, we use
\begin{equs}
{\cal L}_s (s'~\mu)=
\mu~({\cal L}_s s')
+2~\mu'~({\cal L}_s s)
+s'~({\cal L}_s~\mu)
-2~s~\mu'-s'~\mu~,
\end{equs}
which gives
\begin{equs}
\|{\cal L}_s~(s(\mu_i)'~\mu_i)\|_{\sigma-3}&\leq
C~\delta^{5/2}~\rho^2~.
\end{equs}
The proof of (\ref{eqn:onFquatrerddiff}) is similar, we omit the details.
\end{proof}

\subsection{Bounds on $F_{6}$}\label{sec:boundfsix}

We recall that
\begin{equs}
F_6(s,\mu)&={\cal L}_s~\Bigl(F_3(s,\mu)+F_4(s,\mu)
\Bigr)+F_7(\mu)+F_8(\mu)~,\\
F_7(\mu)&=\frac{\epsilon^2}{8}~\bigl(\partial_x+\frac{\epsilon^2~\mu}{2}\bigr)~F_0(\mu)'~,\\
F_8(\mu)&=
-\frac{1}{8}~\bigl(
\partial_x+\frac{\epsilon^2~\mu}{2}
\bigr)
\Bigl(\Lepsilon ~\mu+\mu~\mu'\Bigr)~.
\end{equs}
We will now prove the estimates of Corollaries \ref{cor:firstpartrd}, 
\ref{cor:onprogresse}, \ref{cor:onprogressedeux} and \ref{cor:ontermine}.
These estimates follow immediately from the bounds on $F_6$ of the next
proposition. 

\begin{proposition}
Let $\delta=c_{\delta}~\rho$ then there exist constants $c_{\epsilon}$ and
$c_{F_6}$ such that for all $\epsilon\leq c_{\epsilon}~\rho^{-2}$ and for
all $\mu_i\in{\cal B}_{\sigma}(c_{\mu}~\rho)$ the following bounds hold
\begin{equs}
\epsilon^2~\left\|\frac{{\cal L}_{\mu,r}}{{\cal
L}_{\epsilon}~\Gepsilon ~{\cal L}_r}~F_6(\mu_i)'\right\|_{{\cal W},\sigma}&\leq
\max\Bigl(\frac{1}{3},\frac{\alpha^2}{1-\alpha^2}\Bigr)
\Bigl(c_{\mu}~\rho+\epsilon^2~c_{F_{6}}~\delta^{5/2}~\rho^2\Bigr)~,\label{eqn:huitwlikeun}\\
\epsilon^2~\left\|\frac{{\cal L}_{\mu,r}}{{\cal
L}_{\epsilon}~\Gepsilon ~{\cal L}_r}~\Delta F_6'\right\|_{{\cal W},\sigma}&\leq 
\max\Bigl(\frac{1}{3},\frac{\alpha^2}{1-\alpha^2}\Bigr)
\bigl(1+\epsilon^2~c_{F_{6}}~\delta^{5/2}~\rho\bigr)~\|\mu_1-\mu_2\|_{\sigma}~,\label{eqn:huitwlikedeux}
\end{equs}
and 
\begin{equs}
\epsilon^2~\left\|F_6(\mu_i)\right\|_{\L^2}+
\left\|{\cal L}_{\mu,r}~{\cal L}_{v}^{-1}~F_6(\mu_i)'\right\|_{\L^2}
&\leq
c_{F_{6}}~\Bigl(\delta^{4}~\rho+
\delta^{5/2}~\rho^2\Bigr)~,\label{eqn:huitldlikeun}\\
\left\|
{\cal L}_{\mu,r}~{\cal L}_{v}^{-1}~
\Delta F_6'
\right\|_{\L^2}&\leq
c_{F_{6}}~\Bigl(\delta^{4}+\delta^{5/2}~\rho\Bigr)~\|\mu_1-\mu_2\|_{\sigma}~,
\label{eqn:huitldlikedeux}
\end{equs}
where $\Delta F_6=F_6(\mu_1)-F_6(\mu_2)$.
\end{proposition}
\begin{proof}
For the proof of (\ref{eqn:huitwlikeun}) and (\ref{eqn:huitwlikedeux}), we define
$\balpha=\max(2,\frac{\alpha^2}{1-\alpha^2})$, then we have (see Lemma
\ref{lem:someproperties} of Appendix \ref{app:therdeuxmap})
\begin{equs}
\frac{\epsilon^2}{8}~\left\|
\frac{{\cal L}_{\mu,r}}{\Gepsilon ~{\cal L}_r}~f''
\right\|_{{\cal W},\sigma}
&\leq 
\max\Bigl(\frac{1}{3},\frac{\alpha^2}{1-\alpha^2}\Bigr)
~\|f\|_{{\cal W},\sigma}~,
\\
\left\|
\frac{{\cal L}_{\mu,r}}{\Lepsilon ~\Gepsilon ~{\cal L}_r}~f'
\right\|_{{\cal W},\sigma}
&\leq
C~\balpha~\delta^{-3}~\|f\|_{{\cal W},\sigma-3}~,
\end{equs}
for some constant $C$. On the other hand, we have
\begin{equs}
\left\|
\frac{\epsilon^2}{2}\mu_i\Lepsilon ~\mu_i
\right\|_{{\cal W},\sigma-3}&\leq
\Cm~\sqrt{\delta}~\|\mu_i\|_{\sigma}~
\|\epsilon^2\Lepsilon ~\mu_i\|_{\sigma-2}\leq
C~\delta^{5/2}~\rho^2~,\\
\left\|
\bigl(\partial_x+\frac{\epsilon^2~\mu_i}{2}\bigr)\mu_i\mu_i'
\right\|_{{\cal W},\sigma-3}&\leq
C_1~\delta^{5/2}~\rho^2+
\epsilon^2~C_2~\delta^2~\rho^3
\leq C~\delta^{5/2}~\rho^2~,
\end{equs}
from which we get, using Corollary \ref{cor:onFsept}, Propositions \ref{prop:onFtroisrd} and
\ref{prop:onFquatrerd} above for the contributions of $F_0$, $F_3$ and $F_4$,
\begin{equs}
\epsilon^2~\left\|\frac{{\cal L}_{\mu,r}}{{\cal
L}_{\epsilon}~\Gepsilon ~{\cal L}_r}~F_6(\mu_i)'\right\|_{{\cal W},\sigma}&\leq
\max\Bigl(\frac{1}{3},\frac{\alpha^2}{1-\alpha^2}\Bigr)~
\|\mu_i\|_{{\cal W},\sigma}
+\epsilon^2~C~\balpha~\delta^{5/2}~\rho^2
~.
\end{equs}
The proof of (\ref{eqn:huitwlikeun}) follows since $\balpha\leq
6\max\bigl(\frac{1}{3},\frac{\alpha^2}{1-\alpha^2}\bigr)$. The proof of
(\ref{eqn:huitwlikedeux}) being very similar, we omit the details.

For the proof of (\ref{eqn:huitldlikeun}) and (\ref{eqn:huitldlikedeux}), we
first show that 
\begin{equs}
\frac{1}{8}\|{\cal L}_{\mu,r}~{\cal L}_{v}^{-1}~
\Lepsilon ~\mu''\|_{\L^2}&\leq 2\|(1+\partial_x^4)~\mu\|_{\L^2}
\leq C~\delta^4~\|\mu\|_{\sigma}
\leq C~\delta^4~\rho~,\\
\frac{\epsilon^2}{8}~\|\Lepsilon ~\mu'\|_{\L^2}&\leq
\frac{\epsilon^2}{16}\|(1-\partial_x^2)^{5/2}~\mu\|_{\L^2}
\leq\frac{\epsilon^2~\delta^5}{16}\|\mu\|_{\sigma}
\\&\leq
\delta^{5/2}~\rho^2~\Bigl(\frac{c_{\mu}~\epsilon^2~\delta^{5/2}}{16~\rho}\Bigr)
\leq C~\delta^{5/2}~\rho^2~,
\end{equs}
since by hypothesis $\delta=c_{\delta}~\rho^2$ and $\epsilon\leq c_{\epsilon}~\rho^{-2}$.
To complete the proof of (\ref{eqn:huitldlikeun}), we first note that
$\|{\cal L}_{\mu,r}~{\cal L}_{v}^{-1}~f'\|_{\L^2}\leq 16\|f\|_{\L^2}$, then
using also Corollary \ref{cor:onFsept}, Propositions \ref{prop:onFtroisrd}
and \ref{prop:onFquatrerd} above, we see that it is in fact sufficient to
show that $\|F_6(\mu_i)-\frac{1}{8}~\Lepsilon ~\mu_i'\|_{\L^2}\leq
C~\delta^{5/2}~\rho^2$, but we have
\begin{equs}
\|F_6(\mu_i)-\frac{1}{8}~\Lepsilon ~\mu_i'\|_{\L^2}&\leq
\frac{\epsilon^2}{16}\|\mu_i\Lepsilon ~\mu_i\|_{\L^2}+
\frac{1}{8}\|(\mu_i~\mu_i')'\|_{\L^2}+
\frac{\epsilon^2}{8}\|\mu_i^2~\mu_i'\|_{\L^2}\\
&\leq
C_1~\sqrt{\delta}~\|\mu_i\|_{\sigma}~\|\epsilon^2~\Lepsilon ~\mu_i\|_{\sigma-2}+
C_2~\delta^{5/2}~\|\mu_i\|_{\sigma}^2+
\epsilon^2~C_3~\delta^2~\|\mu_i\|_{\sigma}^3\\
&\leq C~\delta^{5/2}~\|\mu_i\|_{\sigma}^2~\Bigl(
1+\frac{\|\mu_i\|_{\sigma}}{\sqrt{\delta}}
\Bigr)\leq C~\delta^{5/2}~\rho^2~.
\end{equs}
The proof of (\ref{eqn:huitldlikedeux}) follows easily using Corollary
\ref{cor:onFsept}, Propositions \ref{prop:onFtroisrd} and
\ref{prop:onFquatrerd}, and equalities like
\begin{equs}
\mu_1~f(\mu_1)-\mu_2~f(\mu_2)=
\frac{1}{2}(\Delta\mu)~\bigl(f(\mu_1)+f(\mu_2)\bigr)+
\frac{1}{2}(\mu_1+\mu_2)~\Delta f~,
\label{eqn:ptittruxhuit}
\end{equs}
for $\Delta\mu=\mu_1-\mu_2$ and $\Delta f=f(\mu_1)-f(\mu_2)$.
\end{proof}

\newappendix{Proof of Proposition \ref{prop:contraun}}\label{app:contra}
Before proving Proposition \ref{prop:contraun}, we prove a simpler Lemma.
\begin{lemma}\label{lem:contratrois}
Let $\delta$ and $\epsilon_0$ be given by Proposition \ref{prop:contraun},
and let ${\cal F}(\tilde{\mu},\mu_{0})$ be the solution of
\begin{equs}   
\partial_t\mu=-\Lepsilon ~\mu-\mu~\mu'+\epsilon^2~F(\tilde{\mu})'~,~~~~\mu(x,0)=\mu_{0}(x)~.
\end{equs}
Assume that 
\begin{equs}
\tvert{\cal F}(\tilde{\mu},\mu_{0})\tvert_{\sigma}\leq c_{\mu}~\rho~,
\end{equs}
and that (\ref{eqn:lundefdiff}) holds with $\lambda_1<1$ for all $\epsilon\leq\epsilon_0$, for
all $\tilde{\mu}\in{\cal B}_{\sigma}(c_{\mu}~\rho)$,
and for all $\mu_0\in{\cal B}_{0,\sigma}(c_{\mu}~\rho)$. 
Then for all $0<c_{\lambda}<1$, there exists a $t_1>0$ such that
\begin{equs}
\sup_{0\leq t\leq t_1}~
\|{\cal F}(\tilde{\mu}_1,\mu_{0})(\cdot,t)
 -{\cal F}(\tilde{\mu}_2,\mu_{0})(\cdot,t)\|_{\L^2}
\leq c_{\lambda}~\lambda_1~
\sup_{0\leq t\leq t_1}
\|\tilde{\mu}_1(\cdot,t)-\tilde{\mu}_2(\cdot,t)\|_{\sigma}~,
\end{equs}
for all $\tilde{\mu}_{i}\in{\cal B}_{\sigma}(c_{\mu}~\rho)$.
\end{lemma}
\begin{proof}
Let $\mu_i={\cal F}(\tilde{\mu}_i,\mu_0)$, $i=1,2$ and
\begin{equs}
\mu_{\pm}={\cal F}(\tilde{\mu}_1,\mu_0)\pm{\cal F}(\tilde{\mu}_2,\mu_0)~.
\end{equs}
We have
\begin{equs}   
\partial_t\mu_1&=-\Lepsilon ~\mu_1-\frac{1}{2}(\mu_1^2)'+\epsilon^2~F(\tilde{\mu}_1)'~,
~~~~\mu_1(x,0)=\mu_0(x)~,\\
\partial_t\mu_2&=-\Lepsilon ~\mu_2-\frac{1}{2}(\mu_2^2)'+\epsilon^2~F(\tilde{\mu}_2)'~,
~~~~\mu_2(x,0)=\mu_0(x)~.
\end{equs}
Subtracting these two equations, we get
\begin{equs}   
\partial_t\mu_{-}&=
-\Lepsilon ~\mu_{-}-\frac{1}{2}(\mu_{+}\mu_{-})'
+\epsilon^2~\bigl(F(\tilde{\mu}_1)-F(\tilde{\mu}_2)\bigr)'~,
~~~~\mu_{-}(x,0)=0~.
\label{eqn:difference}
\end{equs}
Let $\Delta F=F(\tilde{\mu}_1)-F(\tilde{\mu}_2)$, then multiplying
(\ref{eqn:difference}) by $\mu_{-}$, integrating over $[-L/2,L/2]$ and using Young's
inequality, we get
\begin{equs}   
\partial_t(\mu_{-},\mu_{-})&=
-2(\mu_{-},\Lepsilon ~\mu_{-})
-\frac{1}{2}
(\mu_{-},\mu_{+}'~\mu_{-})
+2~\epsilon^2~(\mu_{-},\Delta F')\\
&\leq
(\mu_{-},({\cal L}_{v}^2-2\Lepsilon )\mu_{-})
+\frac{1}{2}\|\mu_{+}'\|_{\L^{\infty}}(\mu_{-},\mu_{-})
+\epsilon^4~\|{\cal L}_{v}^{-1}~\Delta F'\|_{\L^2}^2\\
&\leq
(1+\frac{1}{2}\|\mu_{+}'\|_{\L^{\infty}})~(\mu_{-},\mu_{-})
+\epsilon^4~\|{\cal L}_{v}^{-1}~\Delta F'\|_{\L^2}^2~.
\end{equs}
Now we use that
\begin{equs}
1+\frac{1}{2}\|\mu_{+}'\|_{\L^{\infty}}
\leq
1+\frac{\|\mu_{1}'\|_{\L^{\infty}}+\|\mu_{2}'\|_{\L^{\infty}}}{2}\leq 
1+\Cinfty ~c_{\mu}~\delta^{3/2}~\rho\equiv\zeta~,
\end{equs}
and that by (\ref{eqn:lundefdiff}), for all $\epsilon\leq\epsilon_0$
\begin{equs}
\epsilon^2~\|{\cal L}_{v}^{-1}~\bigl(F(\tilde{\mu}_1)-F(\tilde{\mu}_2)\bigr)'\|_{\L^2}\leq
\lambda_1~
\|\tilde{\mu}_1-\tilde{\mu}_2\|_{\sigma}~,
\end{equs}
with $\lambda_1<1$ to conclude that
\begin{equs}
\|{\cal F}(\tilde{\mu}_1,\mu_0)(\cdot,t)
 -{\cal F}(\tilde{\mu}_2,\mu_0)(\cdot,t)\|_{\L^2}
&\leq
\lambda_1~
\sqrt{\frac{\ed^{\zeta t}-1}{\zeta}}~
\sup_{0\leq s\leq t}
\|\tilde{\mu}_1(\cdot,s)-\tilde{\mu}_2(\cdot,s)\|_{\sigma}~.
\end{equs}
Setting
\begin{equs}
t_1=\frac{1}{\zeta}\ln\Bigl(1+c_{\lambda}^2~\zeta\Bigr)
\end{equs}
completes the proof.
\end{proof}
Proposition \ref{prop:contraun} is then an easy consequence of the following
proposition.
\begin{proposition}\label{prop:contraquatre}
There exist constants $c_{\delta}$ sufficiently large and $c_{\lambda}$
sufficiently small such that if $t_1$ is given by Lemma \ref{lem:contratrois}, and 
${\cal F}(\tilde{\mu},\mu_{0})$, the solution of
\begin{equs}   
\partial_t\mu=-\Lepsilon ~\mu-\mu~\mu'+\epsilon^2~F(\tilde{\mu})'~,
~~~~\mu(x,0)=\mu_{0}(x)~,
\end{equs}
satisfies
\begin{equs}
\tvert{\cal F}(\tilde{\mu},\mu_{0})\tvert_{\sigma}\leq c_{\mu}~\rho~,
\end{equs}
and (\ref{eqn:lundefdiff}) holds with $\lambda_1<1$ for all
$\epsilon\leq\epsilon_0$, for all $\tilde{\mu}\in{\cal
B}_{\sigma}(c_{\mu}~\rho)$, and for all $\mu_0\in{\cal
B}_{0,\sigma}(c_{\mu}~\rho)$, then
there exists a constant $0<\lambda<1$ such that
\begin{equs}
\sup_{0\leq t\leq t_1}~
\|{\cal F}(\tilde{\mu}_1,\mu_{0})(\cdot,t)
 -{\cal F}(\tilde{\mu}_2,\mu_{0})(\cdot,t)\|_{\sigma}
\leq\lambda~\sup_{0\leq t\leq t_1}
\|\tilde{\mu}_1(\cdot,t)-\tilde{\mu}_2(\cdot,t)\|_{\sigma}~,
\label{eqn:desdiffforF}
\end{equs}
for all $\tilde{\mu}_{i}\in{\cal B}_{\sigma}(c_{\mu}~\rho)$.
\end{proposition}
\begin{proof}   
We will use the same definitions as in Lemma \ref{lem:contratrois} above,
and $\Delta F=F(\tilde{\mu}_1)-F(\tilde{\mu}_2)$. We first note that we have 
\begin{equs}
\sup_{0\leq t\leq t_1}~\|\mu_{-}(\cdot,t)\|_{\sigma}&=
\sup_{0\leq t\leq t_1}~
\|{\cal F}(\tilde{\mu}_1,\mu_{0})(\cdot,t)
 -{\cal F}(\tilde{\mu}_2,\mu_{0})(\cdot,t)\|_{\sigma}
\leq2~c_{\mu}~\rho<\infty~,\\
\sup_{0\leq t\leq t_1}~\|\mu_{+}(\cdot,t)\|_{\sigma}&=
\sup_{0\leq t\leq t_1}~
\|{\cal F}(\tilde{\mu}_1,\mu_{0})(\cdot,t)
 +{\cal F}(\tilde{\mu}_2,\mu_{0})(\cdot,t)\|_{\sigma}
\leq2~c_{\mu}~\rho<\infty~.
\end{equs}
To prove (\ref{eqn:desdiffforF}), the idea is to use Duhamel's representation formula
\begin{equs}
\mu_{-}(x,t)=
-\frac{1}{2}\int_{0}^{t}\hspace{-2mm}{\rm d}s~
\ed^{-\Lepsilon (t-s)}~(\mu_{-}\mu_{+})'(x,s)
+\epsilon^2~\int_{0}^{t}
\hspace{-2mm}{\rm d}s~
\ed^{-\Lepsilon (t-s)}~
\Delta F'(x,s)~,
\label{eqn:repdiff}
\end{equs}
for the solution of (\ref{eqn:difference}). 
We have
\begin{equs}
\epsilon^2~\left\|\int_{0}^{t}
\hspace{-2mm}{\rm d}s~
\ed^{-\Lepsilon (t-s)}~\Delta F'(\cdot,s)
\right\|_{{\cal W},\sigma}
\leq
\epsilon^2~
\sup_{0\leq s\leq t_1}
\left\|
\frac{\Delta F'(\cdot,s)}{\Lepsilon }
\right\|_{{\cal W},\sigma}
\leq\lambda_1
\sup_{0\leq t\leq t_1}~
\|\tilde{\mu}_{-}(\cdot,t)\|_{{\cal W},\sigma}
~,
\end{equs}
with $\lambda_1<1$. Then, from (\ref{eqn:repdiff}), Proposition
\ref{prop:propagation} and Lemma \ref{lem:contratrois}, we have
\begin{equs}
\sup_{0\leq t\leq t_1}
\|\mu_{-}(\cdot,t)\|_{\sigma}
&\leq \lambda_1~(1+c_{\lambda})
\sup_{0\leq t\leq t_1}~
\|\tilde{\mu}_{-}(\cdot,t)\|_{{\cal W},\sigma}
+\frac{\sqrt{2}}{\delta}~
\sup_{0\leq t\leq t_1}~
\|\mu_{-}(\cdot,t)~\mu_{+}(\cdot,t)\|_{{\cal W},\sigma-1}\\
&\leq
\lambda_1~(1+c_{\lambda})
\sup_{0\leq t\leq t_1}~
\|\tilde{\mu}_{-}(\cdot,t)\|_{{\cal W},\sigma}
+\frac{\sqrt{2}~\Cm }{\sqrt{\delta}}~
\sup_{0\leq t\leq t_1}~
\|\mu_{-}(\cdot,t)\|_{\sigma}~\|\mu_{+}(\cdot,t)\|_{\sigma}\\
&\leq
\lambda_1~(1+c_{\lambda})
\sup_{0\leq t\leq t_1}~
\|\tilde{\mu}_{-}(\cdot,t)\|_{{\cal W},\sigma}
+\frac{2^{3/2}~\Cm ~c_{\mu}~\rho}{\sqrt{\delta}}~
\sup_{0\leq t\leq t_1}~
\|\mu_{-}(\cdot,t)\|_{\sigma}~.
\end{equs}
Since ${\displaystyle\sup_{0\leq t\leq t_1}}~\|\mu_{-}(\cdot,t)\|_{\sigma}<\infty$, and
$\delta= c_{\delta}~\rho^2$, with $c_{\delta}>8~\Cm ~c_{\mu}$, we have
\begin{equs}
\sup_{0\leq t\leq t_1}
\|\mu_{-}(\cdot,t)\|_{\sigma}
&\leq
\lambda_1~\frac{(1+c_{\lambda})}{1-\frac{2^{3/2}~\Cm~c_{\mu}}{\sqrt{c_{\delta}}}}
\sup_{0\leq t\leq t_1}~
\|\tilde{\mu}_{-}(\cdot,t)\|_{{\cal W},\sigma}~.
\end{equs}
Hence, we finally get
\begin{equs}
\sup_{0\leq t\leq t_1}
\|{\cal F}(\tilde{\mu}_{1},\mu_0)(\cdot,t)
 -{\cal F}(\tilde{\mu}_{2},\mu_0)(\cdot,t)
\|_{\sigma}
\leq 
\lambda~
\sup_{0\leq t\leq t_1}
\|\tilde{\mu}_{1}(\cdot,t)-\tilde{\mu}_{2}(\cdot,t)\|_{\sigma}~,
\end{equs}
with
\begin{equs}
\lambda=
\lambda_1~\frac{1+c_{\lambda}}{1-\frac{2^{3/2}~\Cm~c_{\mu}}{\sqrt{c_{\delta}}}}~.
\end{equs}
Choosing $c_{\delta}$ sufficiently large and $c_{\lambda}$ sufficiently small we
can certainly make $\lambda$ arbitrarily close to $\lambda_1<1$, in particular, we can
make $\lambda<1$, which completes the proof.
\end{proof}

\newappendix{Further properties of the amplitude equation}\label{app:amplitude}
\begin{corollary}
Assume that $\|r_0\|_{\sigma-1}\leq c_{r_1}~\delta~\rho$. Then
$\mu\mapsto r(\mu)$ satisfies
\begin{equs}
\tvert r(\mu)\tvert_{\sigma-1}&\leq 8~c_{r_1}~\delta~\rho~,
\label{eqn:existerbrap}
\\
\tvert r(\mu_1)-r(\mu_2)\tvert_{\sigma-1}&\leq 8~c_{r_1}~\delta~
\tvert\mu_1-\mu_2\tvert_{\sigma}~,
\label{eqn:existerbdiffrap}
\end{equs}
if the conditions of Theorem \ref{thm:onr} are satisfied.
\end{corollary}
\begin{proof}
As a first step, we note that $\tvert r(\mu)\tvert_{\sigma-3}$ is
finite, because
\begin{equs}
\tvert r(\mu)\tvert_{\sigma-3}\leq
\tvert s(\mu)\tvert_{\sigma-3}+
\frac{\epsilon^2}{2}\tvert s(\mu)''\tvert_{\sigma-3}
\leq
\bigl(1+\frac{\epsilon^2~\delta^2}{2}\bigr)\tvert s(\mu)\tvert_{\sigma-1}~.
\end{equs}
On the other hand, we have
\begin{equs}
\tvert r(\mu)\tvert_{\sigma-1}\leq
\tvert r(\mu)\tvert_{\sigma-3}+
\tvert r(\mu)\tvert_{{\cal W},\sigma-1}~,
\label{eqn:desboundsforr}
\end{equs}
and with the same arguments as the proof of Proposition \ref{prop:bs}, we
have that for all $\sigma'\leq\sigma-1$,
\begin{equs}
\tvert r(\mu)\tvert_{{\cal W},\sigma'}&\leq
\|r_0\|_{{\cal W},\sigma-1}
+\frac{\epsilon^4}{\chi}\tvert F_3(s,\mu)\tvert_{{\cal W},\sigma-1}
+\frac{\epsilon^4~\alpha^2}{8}\tvert(s~\mu')\tvert_{{\cal W},\sigma-1}
+\frac{\epsilon^4~\alpha^2}{4}\tvert(s~\mu)'\tvert_{{\cal W},\sigma'}\\
&\leq \frac{3}{2}~c_{r_1}~\delta~\rho
+\frac{\epsilon^4~\alpha^2~\Cm~\delta^{3/2}}{8}
\tvert s\tvert_{\sigma-1}
~\tvert\mu\tvert_{\sigma}
+\frac{\epsilon^4~\alpha^2}{4}\tvert(s~\mu)'\tvert_{{\cal
W},\sigma'}\\
&\leq 2~c_{r_1}~\delta~\rho+
\frac{\epsilon^4~\alpha^2}{4}\tvert(s~\mu)'\tvert_{{\cal W},\sigma'}~,
\end{equs}
since $\epsilon\leq c_{\epsilon}~\delta^{-5/4}~\rho^{-1/2}$. And now, we
use that
\begin{equs}
s\mu=\Gepsilon ~\left(
r\mu-\epsilon^2s'\mu'-\frac{\epsilon^2}{2}s\mu''
\right)~,
\end{equs}
from which we get
\begin{equs}
\tvert r(\mu)\tvert_{{\cal W},\sigma'}&\leq
2~c_{r_1}~\delta~\rho
+\frac{\epsilon^6~\alpha^2}{8}\tvert \Gepsilon ~(2s'~\mu'+s~\mu'')'\tvert_{{\cal W},\sigma-1}
+\frac{\epsilon^4~\alpha^2}{4}\tvert \Gepsilon ~(r(\mu)~\mu)'\tvert_{{\cal W},\sigma'}\\
&\leq
2~c_{r_1}~\delta~\rho
+\frac{\epsilon^4~\alpha^2}{4}\left(
2\tvert s'~\mu'\tvert_{{\cal W},\sigma-2}
+\tvert s~\mu''\tvert_{{\cal W},\sigma-2}
\right)
+\frac{\epsilon^4~\alpha^2}{4}\tvert \Gepsilon ~(r(\mu)~\mu)'\tvert_{{\cal W},\sigma'}\\
&\leq
2~c_{r_1}~\delta~\rho
+\frac{\epsilon^4~\alpha^2~\Cm~\delta^{5/2}~3}{4}
\tvert s\tvert_{\sigma-1}
~\tvert\mu\tvert_{\sigma}
+\frac{\epsilon^4~\alpha^2}{4}\tvert \Gepsilon ~(r(\mu)~\mu)'\tvert_{{\cal W},\sigma'}~.
\end{equs}
Using this inequality and (\ref{eqn:desboundsforr}), we finally have
\begin{equs}
\tvert r(\mu)\tvert_{\sigma'}&\leq
7~c_{r_1}~\delta~\rho
+\frac{\epsilon^4~\alpha^2}{4}\tvert \Gepsilon ~(r(\mu)~\mu)'\tvert_{{\cal W},\sigma'}~.
\label{eqn:dautresboundsforr}
\end{equs}
Since $\tvert \Gepsilon ~(r(\mu)~\mu)'\tvert_{{\cal W},\sigma'}\leq
2\tvert r(\mu)~\mu\tvert_{{\cal W},\sigma'-1}$, we use
(\ref{eqn:dautresboundsforr}) with $\sigma'=\sigma-2$, and then with $\sigma'=\sigma-1$
to conclude that $\tvert r(\mu)\tvert_{{\cal W},\sigma-1}$ is
finite, and then we have
\begin{equs}
\tvert r(\mu)\tvert_{\sigma-1}&
\leq7~c_{r_1}~\delta~\rho
+\frac{\epsilon^3~\alpha^2~\sqrt{2}}{4}\tvert r(\mu)~\mu\tvert_{{\cal W},\sigma-1}\\
&\leq7~c_{r_1}~\delta~\rho
+\bigl(\epsilon^3~\sqrt{\delta}~\alpha^2~\Cm~\tvert\mu\tvert_{\sigma}\bigr)
\tvert r(\mu)\tvert_{\sigma-1}~.
\end{equs}
Since $\epsilon\leq c_{\epsilon}~\delta^{-5/4}~\rho^{-1/2}$, this last
parenthesis is smaller than $\frac{1}{8}$, and the proof of
(\ref{eqn:existerbrap}) is completed. The proof of 
(\ref{eqn:existerbdiffrap}) is similar, we omit the details.
\end{proof}

\newappendix{The $\mu\mapsto r_2(\mu)$ map.}\label{app:therdeuxmap}
We begin with a preliminary lemma.
\begin{lemma}\label{lem:someproperties}
We have
\begin{equs}
\frac{\epsilon^2}{8}~\left\|
\frac{{\cal L}_{\mu,r}}{\Gepsilon ~{\cal L}_r}~f''
\right\|_{{\cal W},\sigma}
&\leq\max\Bigl(\frac{1}{3},\frac{\alpha^2}{1-\alpha^2}\Bigr)~\|f\|_{{\cal W},\sigma}~,
\label{eqn:withfsec}
\\
\|{\cal L}_{\mu,r}~f\|_{\sigma}&\leq 8~\|f\|_{\sigma}~,
\label{eqn:forlmur}\\
\|{\cal L}_{v}^{-1}~f'\|_{\L^2}&\leq 2~\|f\|_{\L^2}~,
\label{eqn:forlvundeux}\\
\|(1-\partial_x^2)^{-1}~{\cal L}_{v}~f\|_{\L^2}&\leq \|f\|_{\L^2}~,
\label{eqn:forlvundeuxsdx}\\
\left\|
\frac{{\cal L}_{\mu,r}}{\Lepsilon ~\Gepsilon ~{\cal L}_r}~f'
\right\|_{{\cal W},\sigma}
&\leq
11~\max\Bigl(2,\frac{\alpha^2}{1-\alpha^2}\Bigr)~\delta^{-3}~\|f\|_{{\cal W},\sigma-3}~,
\label{eqn:withfp}
\end{equs}
for all $\epsilon^2\leq1$ and $\alpha^2<1/2$.
\end{lemma}
\begin{proof}
In terms of the Fourier coefficients, we have
\begin{equs}
\frac{\epsilon^2}{8}
\left(\frac{{\cal L}_{\mu,r}}{\Gepsilon ~{\cal L}_r}~f''\right)_n=
-\left(\frac{\epsilon^2~(qn)^2}{8}~
\frac{{\cal L}_{\mu,r}(qn)}{\Gepsilon (qn)~{\cal L}_r(qn)}\right)~f_n~,
\end{equs}
and
\begin{equs}
\frac{\epsilon^2~k^2}{8}~
\frac{{\cal L}_{\mu,r}(k)}{\Gepsilon (k)~{\cal L}_r(k)}
&=
\Bigl(\frac{\epsilon^2~k^2}{2}\Bigr)~
\frac{1+\epsilon^2~\bigl(\frac{1+\alpha^2}{2}\bigr)-\alpha^2~\bigl(\frac{\epsilon^2~k^2}{2}\bigr)
}{
1+\bigl(
3+\epsilon^2~\frac{1+\alpha^2}{2}
\bigr)~\bigl(\frac{\epsilon^2~k^2}{2}\bigr)
+(1-\alpha^2)~\bigl(\frac{\epsilon^4~k^4}{4}\bigr)
}\\
&=
\frac{\xi^2~(\lambda^2-\alpha^2~\xi^2)}{1+(2+\lambda^2)~\xi^2+(1-\alpha^2)~\xi^4}~,
\end{equs}
with $\xi=\frac{\epsilon^2~k^2}{2}$ and
$\lambda^2=1+\epsilon^2~\bigl(\frac{1+\alpha^2}{2}\bigr)$. Then as a
function of $\xi$, we have
\begin{equs}
-\frac{\alpha^2}{1-\alpha^2}\leq
\frac{\xi^2~(\lambda^2-\alpha^2~\xi^2)}{1+(2+\lambda^2)~\xi^2+(1-\alpha^2)~\xi^4}
\leq
\frac{\lambda^4}{\lambda^4+4~\lambda^2+4\alpha^2}\leq\frac{1}{3}~,
\end{equs}
where the last inequality comes from from the fact that 
$\epsilon^2\leq1$ and $\alpha^2<1$ imply that $1\leq\lambda^2\leq2$. This
proves (\ref{eqn:withfsec}). For (\ref{eqn:forlmur}), we have
$\left({\cal L}_{\mu,r}~f\right)_n={\cal L}_{\mu,r}(qn)~f_n$, and with the
above notations,
\begin{equs}
|{\cal L}_{\mu,r}(k)|=4~\frac{|\lambda^2-\alpha^2~\xi^2|}{1+\xi^2}
\leq4~\max(\alpha^2,\lambda^2)\leq8~,
\end{equs}
while for (\ref{eqn:forlvundeux}) and (\ref{eqn:forlvundeuxsdx}), we use that
\begin{equs}
|ik~{\cal L}_{v}(k)^{-1}|
&\leq
\sqrt{\frac{3~k^2\bigl(1+\frac{k^2}{2}\bigr)}{1+k^4}}
\leq2~,\\
\left|\frac{{\cal L}_{v}(k)}{1+k^2}\right|
&\leq1~.
\end{equs}
For (\ref{eqn:withfp}), we have
\begin{equs}
\left(\frac{{\cal L}_{\mu,r}}{\Lepsilon ~\Gepsilon ~{\cal L}_r}~f'\right)_n=
i~qn~\left(
\frac{{\cal L}_{\mu,r}(qn)}{\Lepsilon (qn)~\Gepsilon (qn)~{\cal L}_r(qn)}\right)~f_n~,
\end{equs}
then for $|qn|=|k|\geq\delta\geq2$, we have
\begin{equs}
\left|\frac{{\cal L}_{\mu,r}(k)}{\Lepsilonk~\Gepsilon (k)~{\cal L}_r(k)}\right|
&\leq
\frac{8}{k^4}
\sup_{|k|\geq\delta}
\Bigl|\frac{k^4}{k^4-k^2}\Bigr|
~\sup_{|\xi|\geq0}
\Bigl|
\frac{(1+\xi^2)~(\lambda^2-\alpha^2~\xi^2)}{1+(2+\lambda^2)~\xi^2+(1-\alpha
^2)~\xi^4}\Bigr|
\\
&\leq\frac{32}{3~k^4}~\max\Bigl(\lambda^2~,~\frac{\alpha^2}{1-\alpha^2}\Bigr)
\leq\frac{11}{k^4}~\max\Bigl(2~,~\frac{\alpha^2}{1-\alpha^2}\Bigr)
~.
\end{equs}
The proof of (\ref{eqn:withfp}) is completed noting that
$\|K^{-4}~f'\|_{{\cal W},\sigma}=\delta^{-3}~\|f\|_{{\cal W},\sigma-3}$,
if $(K^{-4}~f)_n\equiv (qn)^{-4}~f_n$.
\end{proof}

\subsection{Coercive functionals for the amplitude}\label{app:coercrdeux}
\begin{proposition}\label{prop:coercrdeux}
Let $c_{\mu}>0$ and $\alpha^2<1$. There exists a constant $c_{\epsilon}$ such that
\begin{equs}
\int r_2~\Gepsilon ~{\cal L}_r~r_2-
\frac{\epsilon^4}{16}\int r_2~\mu~{\cal L}_{\mu,r}~r_2'
\geq\frac{3}{4}\int r_2^2~,
\label{eqn:corcampun}
\\
\int r_4~\Gepsilon ~{\cal L}_r~r_4-
\frac{\epsilon^4}{16}\int r_4~{\cal L}_{\mu,r}~{\cal L}_{v}^{-1}\Bigl(\mu~{\cal L}_{v}~r_4\Bigr)'
\geq\frac{3}{4}\int r_4^2~.
\label{eqn:corcampdeux}
\end{equs}
for all $\epsilon\leq c_{\epsilon}~\delta^{-5/4}~\rho^{-1/2}$ and for all
$\mu\in{\cal B}_{\sigma}(c_{\mu}~\rho)$.
\end{proposition}
\begin{proof}
We notice first that ${\cal L}_{\mu,r}r_2'=a_1~\Gepsilon ~
r_2'-a_2~\frac{\epsilon^2}{2}~\Gepsilon ~ r_2'''$ with 
$a_1=4+2~\epsilon^2~(1+\alpha^2)$ and 
$a_2=4\alpha^2$. Then we have
\begin{equs}
\left|
\int\mu~r_2~\Gepsilon ~~r_2'
\right|&\leq
\|\mu\|_{\L^{\infty}}~\|r_2\|_{\L^2}~\|r_2'\|_{\L^2}
\leq
\frac{\Cinfty ~c_{\mu}~\rho~\sqrt{\delta}}{2}\Bigl(
\|r_2\|_{\L^2}^2+\|r_2'\|_{\L^2}^2
\Bigr)~,\\
\left|
\int\mu~r_2~\frac{\epsilon^2}{2}~\Gepsilon ~~r_2'''
\right|&\leq
\|\mu\|_{\L^{\infty}}~\|r_2\|_{\L^2}~\|r_2'\|_{\L^2}
\leq
\frac{\Cinfty ~c_{\mu}~\rho~\sqrt{\delta}}{2}\Bigl(
\|r_2\|_{\L^2}^2+\|r_2'\|_{\L^2}^2
\Bigr)~,
\end{equs}
since (by Fourier Transform) we have 
$\|\Gepsilon ~ f\|_{\L^2}\leq\|f\|_{\L^2}$ and
$\|\frac{\epsilon^2}{2}~\Gepsilon ~ f''\|_{\L^2}\leq\|f\|_{\L^2}$.
We thus get
\begin{equs}
\left|
\frac{\epsilon^4}{16}\int r_2~\mu~{\cal L}_{\mu,r}~r_2'
\right|
&\leq
{\textstyle\epsilon^4~\frac{(a_1+a_2)~\Cinfty ~c_{\mu}~\rho~\sqrt{\delta}}{32}}
\left(\int r_2^2
+
\int (r_2')^2
\right)\\
&\leq
{\textstyle\epsilon^2~\frac{(a_1+a_2)~\Cinfty ~c_{\mu}~\rho~\sqrt{\delta}}{16}}
\left(\int r_2^2
+\frac{\epsilon^2}{2}\int (r_2')^2
\right)
~.
\end{equs}
Let now $a_3=3+\epsilon^2\Bigl(\frac{1+\alpha^2}{2}\bigr)$ and
$a_4=1-\alpha^2$. We have
\begin{equs}
\int r_2~\Gepsilon ~{\cal L}_r~r_2-
\frac{\epsilon^4}{16}\int r_2~\mu~{\cal L}_{\mu,r}~r_2'
\geq
\gamma
\int r_2^2~,
\end{equs}
where
\begin{equs}
\gamma=\min_{\xi\in{\bf R}}\left(
\frac{1+a_3~\xi^2+a_4~\xi^4}{1+\xi^2}-
{\textstyle\epsilon^2~\rho~\sqrt{\delta}~
\bigl(\frac{(a_1+a_2)~\Cinfty ~c_{\mu}}{16}\bigr)}
(1+\xi^2)
\right)~.
\end{equs}
Since $a_3\geq3$ and $a_4>0$, choosing $c_{\epsilon}$ sufficiently small
completes the proof of (\ref{eqn:corcampun}). The proof of
(\ref{eqn:corcampdeux}) is similar. We first use
\begin{equs}
\int r_4~{\cal L}_{\mu,r}~{\cal L}_{v}^{-1}\Bigl(\mu~{\cal
L}_{v}^{1/2}~r_4\Bigr)'&=
-\int \Bigl({\cal L}_{\mu,r}~{\cal L}_{v}^{-1}~r_4\Bigr)'~
\mu~{\cal L}_{v}~r_4
=\int f~\mu~(1-\partial_x^2)~g\\
&=\int f~\mu~g+f'~\mu~g'+f~\mu'~g
~,
\end{equs}
where $f={\cal L}_{\mu,r}~{\cal L}_{v}^{-1}~r_4'$ and 
$g=(1-\partial_x^2)^{-1}~{\cal L}_{v}~r_4$. Let $f^{(m)}$ be the
$m$--th order spatial derivative of $f$. Then we have
$\|f^{(m)}\|_{\L^2}\leq16~\|r_4^{(m)}\|_{\L^2}$ and
$\|g^{(m)}\|_{\L^2}\leq\|r_4^{(m)}\|_{\L^2}$. Furthermore, we have
$\|\mu'\|_{\L^{\infty}}\leq \Cinfty ~c_{\mu}~\delta^{3/2}~\rho$ and
$\|\mu\|_{\L^{\infty}}\leq \Cinfty ~c_{\mu}~\delta^{1/2}~\rho\leq
\Cinfty ~c_{\mu}~\delta^{3/2}~\rho$. Using these inequalities, we have
\begin{equs}
\frac{\epsilon^4}{16}
\left|
\int r_4~{\cal L}_{\mu,r}~{\cal L}_{v}^{-1}\Bigl(\mu~{\cal
L}_{v}^{1/2}~r_4\Bigr)'\right|
&\leq
\epsilon^4~\Cinfty ~c_{\mu}~\delta^{3/2}~\rho~\bigl(
\|r_4\|_{\L^2}^2+
\|r_4\|_{\L^2}~\|r_4'\|_{\L^2}+
\|r_4'\|_{\L^2}^2
\bigr)\\
&\leq 
3~\epsilon^2~\Cinfty ~c_{\mu}~\delta^{3/2}~\rho
\left(
\int r_4^2
+\frac{\epsilon^2}{2}\int (r_4')^2\right)~.
\end{equs}
As above, choosing $c_{\epsilon}$ sufficiently small completes the proof of
(\ref{eqn:corcampdeux}).
\end{proof}

\subsection{Various bounds on $r_2$}\label{sec:varboundsrd}
\begin{lemma}\label{lem:ldrdrap}
Let $r_2$ be the solution of (\ref{eqn:therdeqrapun}). Then $r_2$
satisfies
\begin{equs}
\tvert\chi{\cal L}_{\mu,r}~{\cal L}_{v}^{-1}~r_2'\tvert_{\L^2}&\leq
64~\|r_{2,0}\|_{\L^2}+
\sqrt{2}~\tvert{\cal L}_{\mu,r}~{\cal L}_{v}^{-1}~F_6(s(\mu),\mu)'\tvert_{\L^2}
\label{eqn:rdeuxsansdiffwrap}~.
\end{equs}
\end{lemma}
\begin{proof}
Let $r_4=\chi~{\cal L}_{\mu,r}~{\cal L}_{v}^{-1}~r_2'$, then $r_4$
satisfies
\begin{equs}
\partial_t r_4=-\frac{\chi}{\epsilon^4}
\left(
\Gepsilon ~{\cal L}_{r}~r_4
+\frac{\epsilon^4}{16}
{\cal L}_{\mu,r}~{\cal L}_{v}^{-1}~
(\mu~{\cal L}_{v}~r_4)'\right)+\frac{\chi}{\epsilon^4}~
{\cal L}_{\mu,r}~{\cal L}_{v}^{-1}~F_6(s(\mu),\mu)'~,
\label{eqn:stilltherdeq}
\end{equs}
with initial condition $r_4(x,0)=\chi~{\cal L}_{\mu,r}~{\cal
L}_v^{-1/2}~r_{2,0}(x)'\equiv r_{4,0}(x)$. Using Proposition
\ref{prop:coercrdeux}, we then have
\begin{equs}
\partial_t (r_4,r_4)&\leq
-\frac{\chi}{\epsilon^4}~\frac{3}{2}~(r_4,r_4)+
\frac{2~\chi}{\epsilon^4}(r_4,{\cal L}_{\mu,r}~{\cal L}_{v}^{-1}~F_6(s(\mu),\mu)')\\
&\leq
-\frac{\chi}{\epsilon^4}~(r_4,r_4)+
\frac{2~\chi}{\epsilon^4}\|{\cal L}_{\mu,r}~{\cal L}_{v}^{-1}~F_6(s(\mu),\mu)'\|_{\L^2}^2~.
\end{equs}
Integrating this differential inequality and using that 
$\|r_{4,0}\|_{\L^2}\leq 64~\|r_{2,0}\|_{\L^2}$
completes the proof.
\end{proof}

\begin{lemma}\label{lem:brdeux}
Let $r_2$ be the solution of (\ref{eqn:therdeqrapun}) with
$\mu\in{\cal B}_{\sigma}(c_{\mu}~\rho)$, and assume that $r_2$
satisfies
\begin{equs}
\epsilon^2~\tvert\chi~{\cal L}_{\mu,r}~{\cal L}_{v}^{-1}~r_2'\tvert_{\L^2}
\leq c_{\mu}~\rho~.
\end{equs}
Then there exists a constant $C$ such that
\begin{equs}
\tvertb
\frac{\chi~{\cal L}_{\mu,r}}{\Lepsilon }~r_2'
\tvertb_{{\cal W},\sigma}&\leq
\frac{
\frac{4~\chi}{\delta}~
\|r_{2,0}\|_{{\cal W},\sigma-1}+
\tvertb
\frac{{\cal L}_{\mu,r}}{\Lepsilon }~
\frac{F_6(s(\mu),\mu)'}{\Gepsilon ~{\cal L}_r}
\tvertb_{{\cal W},\sigma}+
C~\epsilon^2~\max\bigl(2~,~\frac{\alpha^2}{1-\alpha^2}\bigr)~c_{\mu}~\rho}
{1-\epsilon^2~2~\Cm ~c_{\mu}~c_{\delta}^{-1/2}~\max(2,\frac{\alpha^2}{1-\alpha^2})}~,
\end{equs}
for all $\epsilon$ satisfying
$\epsilon^2~2~\Cm ~c_{\mu}~c_{\delta}^{-1/2}~\max(2,\frac{\alpha^2}{1-\alpha^2})<1$.
\end{lemma}
\begin{proof}   
We define
\begin{equs}
r_3=P_{>}~\Bigl(\frac{\chi~{\cal L}_{\mu,r}}{\Lepsilon }~r_2'\Bigr)~,
\label{eqn:rtroisdef}
\end{equs}
and we note that $\tvert r_3\tvert_{{\cal W},\sigma}<\infty$, because
\begin{equs}
\tvert r_3\tvert_{{\cal W},\sigma}=
\tvertb\frac{\chi~{\cal L}_{\mu,r}}{\Lepsilon }~r_2'\tvertb_{\sigma}
\leq \frac{4~\chi}{\delta}~\tvert r_2\tvert_{\sigma-1}
\leq \frac{4~\chi}{\delta~\epsilon^4}~
\Bigl(\tvert r(\mu)\tvert_{\sigma-1}+\tvert r_1(\mu)\tvert_{\sigma-1}
\Bigr)~.
\end{equs}
On the other hand, $r_3$ satisfies
\begin{equs}
\partial_t r_3&=-\frac{\chi}{\epsilon^4}~\Gepsilon ~{\cal L}_{r}~r_3
+
\frac{\chi}{16}~P_{>}~\Bigl(\frac{{\cal L}_{\mu,r}}{\Lepsilon }~
\bigl(\mu~\chi~{\cal L}_{\mu,r}~r_2'\bigr)'\Bigr)
+\frac{\chi~}{\epsilon^4}~P_{>}~\Bigl(\frac{{\cal L}_{\mu,r}}{\Lepsilon }~
F_6(s(\mu),\mu)'\Bigr)\\
&=-\frac{\chi}{\epsilon^4}~\Gepsilon ~{\cal L}_{r}~r_3
+\frac{\chi}{16}~P_{>}~\Bigl(\frac{{\cal L}_{\mu,r}}{\Lepsilon }~
\bigl(\mu~\Lepsilon ~r_3\bigr)'\Bigr)+
\frac{\chi}{\epsilon^4}~
P_{>}~F_9(\mu,P_{<}~r_2)~,
\end{equs}
with $r_3(x,0)=r_{3,0}(x)$ and
\begin{equs}
F_9(\mu,P_{<}~r_2)=\frac{\chi~\epsilon^4}{16}\Bigl(\frac{{\cal L}_{\mu,r}}{\Lepsilon }~
\bigl(\mu~{\cal L}_{\mu,r}~P_{<}~r_2'\bigr)'\Bigr)
+\frac{1}{\epsilon^4}\Bigl(\frac{{\cal L}_{\mu,r}}{\Lepsilon }~
F_6(s(\mu),\mu)'\Bigr)~.
\end{equs}
Then we use Duhamel's formula for the solution to conclude that
\begin{equs}
\tvert r_3\tvert_{{\cal W},\sigma}&\leq
\|r_{3,0}\|_{{\cal W},\sigma}+
\frac{\epsilon^4}{16}~
\tvertb
\frac{{\cal L}_{\mu,r}}{\Lepsilon ~\Gepsilon ~{\cal L}_r}~
\bigl(\mu~\Lepsilon ~r_3\bigr)'
\tvertb_{{\cal W},\sigma}
+
\tvertb\frac{F_9(\mu,P_{<}~r_2)}{\Gepsilon ~{\cal L}_r}\tvertb_{{\cal W},\sigma}\\
&\leq
\frac{4~\chi}{\delta}
\|r_{2,0}\|_{{\cal W},\sigma-1}+
\epsilon^2~
\frac{11}{16}~
\max\Bigl(2~,~\frac{\alpha^2}{1-\alpha^2}\Bigr)~
\epsilon^2~\delta^{-3}~
\tvert
\mu~\Lepsilon ~r_3
\tvert_{{\cal W},\sigma-3}
\\&\phantom{=}~+
\tvertb\frac{F_9(\mu,P_{<}~r_2)}{\Gepsilon ~{\cal L}_r}\tvertb_{{\cal W},\sigma}~,
\end{equs}
where we used Lemma \ref{lem:someproperties}. Then we have
\begin{equs}
\epsilon^2~\delta^{-3}~
\tvert
\mu~\Lepsilon ~r_3
\tvert_{{\cal W},\sigma-3}\leq
\Cm ~\delta^{-5/2}~
\tvert\mu\tvert_{\sigma}~
\tvert\epsilon^2~\Lepsilon ~r_3\tvert_{\sigma-3}
\leq
2~\Cm ~c_{\mu}~c_{\delta}^{-1/2}~\tvert r_3\tvert_{\sigma}~,
\end{equs}
from which we get
\begin{equs}
\tvert r_3\tvert_{{\cal W},\sigma}&\leq
\frac{
\frac{4~\chi}{\delta}
\|r_{2,0}\|_{{\cal W},\sigma-1}
+
\tvert
\frac{F_9(\mu,P_{<}~r_2)}{\Gepsilon ~{\cal L}_r}~
\tvert_{{\cal W},\sigma}
}{
1-\epsilon^2~2~\Cm~c_{\mu}~c_{\delta}^{-1/2}~\max(2~,~\frac{\alpha^2}{1-\alpha^2})}~,
\end{equs}
since by hypothesis
$\epsilon^2~2~\Cm~c_{\mu}~c_{\delta}^{-1/2}~\max(2~,~\frac{\alpha^2}{1-\alpha^2})<1$.
Using twice Lemma \ref{lem:someproperties}, we also have
\begin{equs}
\frac{\chi~\epsilon^4}{16}
\tvertb
\frac{{\cal L}_{\mu,r}}{\Lepsilon ~\Gepsilon ~{\cal L}_r}~
\bigl(\mu~{\cal L}_{\mu,r}~P_{<}~r_2'\bigr)'\tvertb_{{\cal W},\sigma}
&\leq
\epsilon^4~
\frac{11}{16}
~m(\alpha)~\delta^{-3}~
\tvert\mu~\bigl(\chi~P_{<}~{\cal L}_{\mu,r}~r_2'\bigr)\tvert_{{\cal W},\sigma-3}
\\
&\leq
\epsilon^4~
\frac{11}{2}~
m(\alpha)~\Cm ~c_{\mu}~\delta^{-5/2}~
\rho~\tvert\chi~P_{<}~{\cal L}_{\mu,r}~r_2'\tvert_{\sigma}~,
\end{equs}
where $m(\alpha)=\max\Bigl(2~,~\frac{\alpha^2}{1-\alpha^2}\Bigr)$.
But we have $\|P_{<}~{\cal L}_{v}~f\|_{\sigma}=
\|P_{<}~{\cal L}_{v}~f\|_{\L^2}\leq4~\delta^2~\|f\|_{\L^2}$, so that
\begin{equs}
\tvert\chi~P_{<}~{\cal L}_{\mu,r}~r_2'\tvert_{\sigma}\leq
\tvert\chi~P_{<}~{\cal L}_{v}~{\cal L}_{\mu,r}~{\cal L}_{v}^{-1}~r_2'\tvert_{\L^2}
\leq 4~\delta^2~\tvert\chi~{\cal L}_{\mu,r}~{\cal L}_{v}^{-1}~r_2'\tvert_{\L^2}~.
\end{equs}
Hence we have
\begin{equs}
\frac{\chi~\epsilon^4}{16}
\tvertb
\frac{{\cal L}_{\mu,r}}{\Lepsilon ~\Gepsilon ~{\cal L}_r}~
\bigl(\mu~{\cal L}_{\mu,r}~P_{<}~r_2'\bigr)'\tvertb_{{\cal W},\sigma}
&\leq
\epsilon^2~\left(
22~m(\alpha)~
\frac{\Cm ~c_{\mu}}{\sqrt{c_{\delta}}}\right)~
\epsilon^2~\tvert\chi~{\cal L}_{\mu,r}~{\cal L}_{v}^{-1}~r_2'\tvert_{\L^2}\\
&\leq
\epsilon^2~\left(
22~m(\alpha)~
\frac{\Cm ~c_{\mu}}{\sqrt{c_{\delta}}}\right)~c_{\mu}~\rho~.
\end{equs}
This completes the proof.
\end{proof}

\begin{lemma}\label{lem:brdeuxdiff}
Let $r_2(\mu_i)$ be the solution of (\ref{eqn:therdeqrapun}) with
$\mu=\mu_i\in{\cal B}_{\sigma}(c_{\mu}~\rho)$, and assume that 
\begin{equs}
\epsilon^2~\tvert\chi~{\cal L}_{\mu,r}~{\cal L}_{v}^{-1}~r_2(\mu_i)'\tvert_{\L^2}+
\epsilon^2~\tvertb
\frac{\chi~{\cal L}_{\mu,r}}{\Lepsilon }~{r_2(\mu_i)}'
\tvertb_{{\cal W},\sigma}\leq c_{\mu}~\rho~.
\label{eqn:assumebrdeux}
\end{equs}
Let $\Delta r_2=r_2(\mu_1)-r_2(\mu_2)$ and $\Delta
F_6=F_6(s(\mu_1),\mu_1)-F_6(s(\mu_2),\mu_2)$. Then there exists a constant
$C$ such that
\begin{equs}
\tvert
\chi~{\cal L}_{\mu,r}~{\cal L}_{v}^{-1}~\Delta r_2'
\tvert_{\L^2}&\leq
\tvert\chi~{\cal L}_{\mu,r}~{\cal L}_{v}^{-1}~\Delta F_6'\tvert_{\L^2}+
C~\delta^{5/2}~\rho~\tvert\mu_1-\mu_2\tvert_{\sigma}~.
\label{eqn:drdld}
\end{equs}
\end{lemma}
\begin{proof}
The proof relies on the fact that $\Delta r_2$ satisfies the same equation
as $r_2$, but with initial data $\Delta r_2(x,0)=0$, $F_6(s(\mu),\mu)$
replaced by $\Delta F_6=F_6(s(\mu_1),\mu_1)-F_6(s(\mu_2),\mu_2)$ and
$\mu~{\cal L}_{\mu,r}~r_2'$ replaced by
\begin{equs}
\mu_1~{\cal L}_{\mu,r}~r_2(\mu_1)'-\mu_2~{\cal L}_{\mu,r}~r_2(\mu_2)'=
\Delta\mu~{\cal L}_{\mu,r}~{r_2^{+}}'+
\Bigl(\frac{\mu_1+\mu_2}{2}\Bigr)~{\cal L}_{\mu,r}~\Delta r_2'~,
\label{eqn:somedifferences}
\end{equs}
where $\Delta\mu=\mu_1-\mu_2$ and $r_2^{+}=\frac{r_2(\mu_1)+r_2(\mu_2)}{2}$. 
Since $\frac{\mu_1+\mu_2}{2}$ satisfies the same bound as $\mu$ in Lemma
\ref{lem:ldrdrap}, we see that the conclusion of this Lemma holds with the
replacements $r_{2,0}=0$, $r_2\leftrightarrow\Delta r_2$ and
$f_6\leftrightarrow\Delta F_6$ and an additional term given by
$\frac{\epsilon^4}{8}~\|\chi~{\cal L}_{\mu,r}~{\cal
L}_{v}^{-1/2}~(\Delta\mu~{\cal L}_{\mu,r}~{r_2^{+}}')'\|_{\L^2}$, on which
we have
\begin{equs}
\frac{\epsilon^4}{8}~\|\chi~{\cal L}_{\mu,r}~{\cal
L}_{v}^{-1/2}~(\Delta\mu~{\cal L}_{\mu,r}~{r_2^{+}}')'\|_{\L^2}\leq
8~\epsilon^4~\|\Delta\mu~{\cal L}_{\mu,r}~{r_2^{+}}'\|_{\L^2}~.
\end{equs}
Using (\ref{eqn:assumebrdeux}), defining $r_3$ as in (\ref{eqn:rtroisdef})
and writing $r_2$ instead of $r_2^{+}$ to simplify the notation, we have
\begin{equs}
\epsilon^4~\|\Delta\mu~{\cal L}_{\mu,r}~r_2'\|_{\L^2}&\leq
\epsilon^4~\|\Delta\mu~{\cal L}_{\mu,r}~P_{<}~{\cal L}_{v}~{\cal L}_{v}^{-1}~r_2'\|_{\L^2}+
\epsilon^4~\|\Delta\mu~\Lepsilon ~r_3\|_{\L^2}\\
&\leq
(3~\epsilon^2~\Cinfty ~\delta^{5/2})~
\|\Delta\mu\|_{\sigma}~(\epsilon^2~\|{\cal L}_{v}^{-1}~r_2'\|_{\L^2})+
\epsilon^2~\|\Delta\mu~(\epsilon^2\Lepsilon )~r_3\|_{\sigma-2}\\
&\leq
(3~\epsilon^2~\Cinfty ~\delta^{5/2})~
\|\Delta\mu\|_{\sigma}~c_{\mu}~\rho+
2~\Cm ~\delta^{5/2}~\|\Delta\mu\|_{\sigma}~(\epsilon^2~\|r_3\|_{\sigma})~,
\end{equs}
since $\|P_{<}~{\cal L}_{v}^{-1}~f\|_{\L^2}\leq 3~\|f\|_{\L^2}$ and
$\epsilon^2\|\Lepsilon ~f\|_{\sigma-2}\leq2~\delta^2~\|f\|_{\sigma}$. 
By hypothesis, we have $\epsilon^2~\|r_3\|_{\sigma}\leq c_{\mu}~\rho$ and
the proof is completed.
\end{proof}

\begin{lemma}\label{lem:brdeuxdiffw}
Let $r_2(\mu_i)$ be the solution of (\ref{eqn:therdeqrapun}) with
$\mu=\mu_i$ and define $\Delta r_2=r_2(\mu_1)-r_2(\mu_2)$ and
$\Delta F_6=F_6(s(\mu_1),\mu_1)-F_6(s(\mu_2),\mu_2)$. Assume that
$\mu_i\in{\cal B}_{\sigma}(c_{\mu}~\rho)$, and that
\begin{equs}
\epsilon^2~\tvert\chi~{\cal L}_{\mu,r}~{\cal L}_{v}^{-1}~r_2(\mu_i)'\tvert_{\L^2}+
\epsilon^2~\tvertb
\frac{\chi~{\cal L}_{\mu,r}}{\Lepsilon }~{r_2(\mu_i)}'
\tvertb_{{\cal W},\sigma}&\leq c_{\mu}~\rho~,
\label{eqn:assumebrdeuxrap}\\
\epsilon^2~\tvert\chi~{\cal L}_{\mu,r}~{\cal L}_{v}^{-1}~\Delta r_2(\mu_i)'\tvert_{\L^2}
&\leq \tvert\mu_1-\mu_2\tvert_{\sigma}~.
\end{equs}
Then there exists a constant $C$ such that
\begin{equs}
\tvertb
\frac{\chi~{\cal L}_{\mu,r}}{\Lepsilon }~\Delta r_2'
\tvertb_{{\cal W},\sigma}&\leq
\frac{
\tvert
\frac{{\cal L}_{\mu,r}}{\Lepsilon }~
\frac{\Delta F_6'}{\Gepsilon ~{\cal L}_r}
\tvert_{{\cal W},\sigma}+
C~\max(2,\frac{\alpha^2}{1-\alpha^2})~\tvert\mu_1-\mu_2\tvert_{\sigma}
}{1-\epsilon^2~2~\Cm ~c_{\mu}~c_{\delta}^{-1/2}~\max(2,\frac{\alpha^2}{1-\alpha^2})}~,
\label{eqn:drds}
\end{equs}
for all $\epsilon$ satisfying
$\epsilon^2~2~\Cm ~c_{\mu}~c_{\delta}^{-1/2}~\max(2,\frac{\alpha^2}{1-\alpha^2})<1$.
\end{lemma}
\begin{proof}   
The proof relies again on the fact that $\Delta r_2$ satisfies the same
equation as $r_2$, with initial data $\Delta r_2(x,0)=0$, $F_6(s(\mu),\mu)$
replaced by $\Delta F_6=F_6(s(\mu_1),\mu_1)-F_6(s(\mu_2),\mu_2)$ and
$\mu~{\cal L}_{\mu,r}~r_2'$ replaced by
\begin{equs}
\mu_1~{\cal L}_{\mu,r}~r_2(\mu_1)'-\mu_2~{\cal L}_{\mu,r}~r_2(\mu_2)'=
\Delta\mu~{\cal L}_{\mu,r}~{r_2^{+}}'+
\Bigl(\frac{\mu_1+\mu_2}{2}\Bigr)~{\cal L}_{\mu,r}~\Delta r_2'~.
\label{eqn:yetsomedifferences}
\end{equs}
The proof can be done as that of Lemma \ref{lem:brdeux}, hence we omit the
details.
\end{proof}

\newappendix{Discussion}\label{app:discussion}
The proofs of this section follow from definitions and proofs of Section
\ref{sec:highfreq} which should be read first. By (\ref{eqn:mudef}), we have
\begin{equs}
\eta(x,t)=\frac{\hat{\epsilon}^2}{4}~\int_0^{\hat{\epsilon}~x}
\hspace{-2mm}{\rm d} z~\hat{\mu}(z,\hat{t})~,
\end{equs}
and we get
\begin{equs}
\tvert\eta\tvert_{\L^{\infty}([-L_0/2,L_0/2])}&\leq\hat{\epsilon}^{2}~
\frac{\hat{\epsilon}~L_0}{2}
~\tvert\hat{\mu}\tvert_{\L^{\infty}}
\leq
c_{\mu}~\hat{\epsilon}^{2}~L~\rho
\leq C~\hat{\epsilon}^{2}~\rho^{13/8}
~,\\
\tvert s\tvert_{\L^{\infty}([-L_0/2,L_0/2])}&\leq 
\hat{\epsilon}^{4}~\tvert\hat{s}\tvert_{\L^{\infty}}
\leq C~\epsilon^{4}~\delta^{3/2}~\rho~.
\end{equs}
If $\hat{\epsilon}\leq c_{\epsilon}~\rho^{-m_{\epsilon}}$ with
$m_{\epsilon}\geq4$, we get
\begin{equs}
\tvert\eta\tvert_{\L^{\infty}([-L_0/2,L_0/2])}&\leq C~\epsilon^{2-13/(8~m_{\epsilon})}~,\\
\tvert s\tvert_{\L^{\infty}([-L_0/2,L_0/2])}&\leq C~\epsilon^{4-4/m_{\epsilon}}~,
\label{eqn:boundsphysinfp}
\end{equs}
since $\rho\leq c_{\epsilon}~\hat{\epsilon}^{-1/m_{\epsilon}}$. We also have
\begin{equs}
\tvert\eta'\tvert_{\L^2([-L_0/2,L_0/2])}&\leq 
\hat{\epsilon}^{5/2}~\tvert\hat{\mu}\tvert_{\L^2}\leq
C~\epsilon^{5/2}~\rho\leq
C~\epsilon^{5/2-1/m_{\epsilon}}
\label{eqn:boundetaphysp}
~,\\
\tvert\eta'\tvert_{\L^{\infty}([-L_0/2,L_0/2])}&
\leq\hat{\epsilon}^{3}~\tvert\hat{\mu}\tvert_{\L^{\infty}}
\leq C~\epsilon^{3}~\sqrt{\delta}~\rho
\leq C~\epsilon^{3-2/m_{\epsilon}}
\label{eqn:boundetaphysinfp}~,\\
\tvert s\tvert_{\L^2([-L_0/2,L_0/2])}&
\leq\hat{\epsilon}^{7/2}~\tvert\hat{s}\tvert_{\L^2}
\leq C~\epsilon^{7/2}~\delta~\rho
\leq C~\epsilon^{7/2-3/m_{\epsilon}}~.
\label{eqn:boundsphysp}
\end{equs}
Various other estimates, e.g. on higher order derivatives can be obtained
in a similar way.
\biblio
\end{document}